\documentclass[a4paper,11pt]{article}
\pdfoutput=1

\usepackage{jheppub}
\usepackage[T1]{fontenc}
\usepackage[utf8]{inputenc}
\usepackage{microtype}
\usepackage{amsfonts,mathtools}
\usepackage{mathrsfs}
\usepackage{tikz}
\usetikzlibrary{decorations.markings,decorations.pathmorphing,arrows.meta,patterns}
\usepackage{xcolor}

\allowdisplaybreaks[3]
\hypersetup{
  pdfauthor={Biswajit Sahoo},
  pdftitle={Logarithmic Soft Photon Theorem and Waveform Tails in Higher Dimensions}
}

\def\f{\frac}
\def\p{\partial}
\def\wt{\widetilde}
\def\non{\nonumber}
\newcommand{\bea}{\begin{eqnarray}}
\newcommand{\eea}{\end{eqnarray}}
\newcommand{\be}{\begin{equation}}
\newcommand{\ee}{\end{equation}}

\title{\boldmath Logarithmic soft photon theorem and waveform tails in higher
dimensions}
\author[a,b,1]{Biswajit Sahoo%
\note{ORCID: \href{https://orcid.org/0000-0002-2650-7681}
{0000-0002-2650-7681}.}}
\affiliation[a]{School of Physical Sciences, Indian Association for the
Cultivation of Science (IACS),\\
2A and 2B Raja S.C. Mullick Road, Kolkata 700 032, India}
\affiliation[b]{Department of Mathematics, King's College London,\\
London WC2R 2LS, United Kingdom}
\emailAdd{bisphysahoo@gmail.com}

\abstract{
We derive the leading logarithmic soft-photon theorem in \(d>4\) spacetime
dimensions and its classical radiative counterpart.  A direct one-loop analysis
in massive scalar quantum electrodynamics (QED) yields a factorizing soft term of order
\(\omega^{d-4}\ln\omega\), although the charged-particle \(S\)-matrix is infrared
finite.  The logarithm is generated by the scale-invariant loop-momentum region
\(\omega\ll|\ell|\ll\Lambda\), where \(\Lambda\) denotes a characteristic
hard-particle energy scale.  At leading radiative order, the corresponding
logarithmic contribution to the classical electromagnetic waveform arises from the
long-range acceleration of the asymptotic charged particles.  In even
\(d\geq6\), the straight-line waveform at this radiative order is distributionally
supported in retarded time within an interval whose width is set by the
characteristic size of the hard-scattering region, whereas the logarithmic acceleration term produces
universal early- and late-time radiative tails proportional to
\(|u|^{-(3d-10)/2}\).  In odd \(d\geq5\), straight-line motion already gives a
late-time tail at the same radiative order proportional to \(u^{-(d-4)/2}\).  The acceleration correction adds
universal late- and early-time terms proportional, respectively, to
\(\ln u/u^{(3d-10)/2}\) and \(|u|^{-(3d-10)/2}\).  A comparison of Feynman and
retarded boundary conditions separates the classically radiative contribution
from the intrinsically quantum part of the logarithmic soft factor.
}

\begin{document}
\maketitle
\clearpage
\section{Introduction and summary of results}

The leading soft-photon theorem states that the \(1/\omega\) term in an amplitude
with an additional photon of energy \(\omega\) factorizes into a universal function
of the asymptotic charges and momenta multiplying the amplitude without that
photon.  In the classical limit, the same soft factor determines, at leading
radiative order, the low-frequency waveform produced by the incoming and outgoing asymptotic
trajectories.  In four spacetime dimensions, its inverse Fourier transform gives a
finite difference between the early- and late-retarded-time limits of the
transverse radiative gauge potential at null infinity.  Equivalently, the integral
of the radiative electric field over retarded time is nonzero and gives the
electromagnetic kick memory at this radiative order.  Here and throughout this
section, every statement about an electromagnetic waveform or memory refers to
leading radiative order unless a different radial order is stated explicitly.

Long-range interactions become visible more sharply beyond the leading soft order.
In four dimensions, the Coulomb force accelerates the asymptotic particles for
arbitrarily long times, obstructing an ordinary Taylor expansion in the soft
energy.  This large-time correction produces logarithmic terms in both the quantum
soft expansion and the classical frequency-space electromagnetic waveform at
leading radiative order.  The
inverse Fourier transform of the latter gives power-law electromagnetic tails at
early and late retarded times.  In four-dimensional quantum amplitudes, the
logarithm is usually tied to an infrared singularity, with
\(\omega\) acting as the infrared cutoff that converts the singularity into
\(\ln\omega\).  The corresponding soft theorem also admits asymptotic-symmetry
and celestial-current interpretations.  The logarithmic term therefore probes the
cumulative effect of the long-range force rather than only the localized hard
scattering event.

The central question of this paper is whether an analogous logarithm survives for
\(d>4\), where the massive charged particle \(S\)-matrix in quantum electrodynamics
is infrared finite.  Infrared finiteness does not make the electromagnetic
interaction compactly supported.  The Coulomb field continues to accelerate
charged particles at arbitrarily large separations and causes them to radiate.
Although the force falls more rapidly than in four dimensions, its cumulative
effect produces a non-analytic contribution at the dimension-dependent order
\(\omega^{d-4}\ln\omega\) in both the soft factor and the classical current.  This
is the main result of the paper.  The logarithm occurs after any intervening
analytic powers allowed at the intermediate subleading orders, and its factorized
coefficient is fixed by the asymptotic charges and momenta.  Since no infrared pole
is available from which it could be inferred, establishing this universal
coefficient requires a direct analysis in the long-distance interacting region.  
In the S-matrix calculation, this term comes from the scale-invariant region
\(\omega\ll|\ell|\ll\Lambda\), where \(\Lambda\) denotes a characteristic
hard-particle energy scale, in the one-loop virtual-photon exchange between two
external charged lines, with the logarithm selected by the order-\((d-4)\) term in
the soft-momentum expansion of the charged particle propagator adjacent to the emitted photon.

The classical waveform at leading radiative order gives a complementary motivation
and a physical interpretation of this result.  For even \(d\geq6\), the waveform
generated by the leading \(1/\omega\) term is distributionally supported in retarded
time within an interval of order \(\mathcal R\), set by the hard-scattering region,
and therefore produces no persistent electromagnetic memory at this radiative
order.  The logarithmic term instead generates power-law fields at early and late
retarded times, providing a long-time tail memory at the same radiative order.  In odd dimensions, where the
retarded Green's function has support inside the light cone, straight-line motion
already produces a power-law tail.  The acceleration-induced contribution gives a
logarithmically enhanced subleading correction and also allows an incoming source to act on an
outgoing trajectory.  We derive both the quantum logarithmic soft factor and these
even- and odd-dimensional classical waveforms at leading radiative order.  Throughout, ``tail memory'' denotes
a universal decaying large-\(|u|\) waveform at leading radiative order, rather than the permanent kick
associated with four-dimensional electromagnetic memory at the same radial order.

We now place these results in the context of earlier work.  The soft expansion
originated in the low-frequency analysis of Compton scattering
\cite{Low:1954lowenergy,GellMann:1954goldberger}.  Low subsequently related the
first two terms in radiative scattering to the corresponding non-radiative
process \cite{Low:1958sn}, while Weinberg established the universal leading factor
for general \(S\)-matrix elements
\cite{Weinberg:1964smatrix,Weinberg:1965nx}.  The subleading theorem was developed
further in \cite{Burnett:1968soft,Bell:1969soft}, and its possible
effective-field-theory corrections were classified in \cite{Elvang:2016qvq}.
The classical significance of the same low-frequency behavior emerged through
studies of electromagnetic memory
\cite{Bieri:2013memory,Tolish:2014memory,Susskind:2015memory}.  This led to the
equivalence among the leading soft theorem, electromagnetic memory and the Ward
identity of large \(U(1)\) asymptotic symmetries
\cite{He:2014masslessQED,Kapec:2014evenQED,Pasterski:2015memory,
Campiglia:2015massiveQED,Kapec:2015massiveQED,He:2019higherDleading}, followed by
extensions to further memories and their soft relations
\cite{Mao:2017memory,Hamada:2018memory,Campoleoni:2019memory}.  In parallel, the
connection between soft factors and classical radiative fields developed into
amplitude-based methods for extracting classical radiation
\cite{Goldberger:2016iau,Kosower:2018adc,Laddha:2018rle,
Bautista:2019classical,Laddha:2019yaj,Manu:2020soft,
Akhtar:2025large,Paul:2026logsoft}.

These developments provide the setting for logarithmic corrections to the soft
expansion.  In four dimensions, their relation to long-range acceleration and to
early- and late-time waveform tails was established in
\cite{Laddha:2018myi,Laddha:2018logtail,Sahoo:2018lxl,Saha:2019tub,
Sahoo:2020ryf,Bhatkar:2020conservation,Krishna:2023log,Compere:2025multipole}.
The same logarithmic terms were subsequently connected to asymptotic-symmetry Ward
identities
\cite{Campiglia:2019logward,Bhatkar:2019logward,Choi:2024logsymmetry,
Agrawal:2023logward,Choi:2024triangle} and, in celestial holography, to
higher-spin currents and conformally soft loop operators on the celestial sphere
\cite{Banerjee:2026celestial,Choi:2026celestial}.  The corresponding
higher-dimensional radiation problem has so far revealed two important features:
the absence of persistent memory at leading radiative order from the leading soft term in even
\(d\geq6\) \cite{Garfinkle:2017fre,Laddha:2019yaj}, and the
inside-the-light-cone propagation that produces power-law tails in odd dimensions
\cite{Satishchandran:2017pek,Laddha:2019yaj}.  The present work connects these
threads and extends them by determining the higher-dimensional logarithmic soft factor and its
classical waveform tails at leading radiative order.  We now state the quantum and classical results in a
self-contained form.

We consider \(M\) incoming and \(N\) outgoing massive
charged particles.  Their future-directed physical momenta, masses, charges and
velocities obey
\be\label{eq:intro_scattering_data}
\begin{gathered}
\left(p_a^{\prime\mu},m'_a,q'_a,v_a^{\prime\mu}=\f{p_a^{\prime\mu}}{m'_a}\right),\quad a=1,\ldots,M,
\qquad
\left(p_a^\mu,m_a,q_a,v_a^\mu=\f{p_a^\mu}{m_a}\right),\quad a=1,\ldots,N,
\\
p_a^{\prime2}=-m_a^{\prime2},\quad
v_a^{\prime 2}=-1,
\qquad
p_a^2=-m_a^2,\quad
v_a^2=-1,\qquad 
\sum_{a=1}^Nq_a=\sum_{a=1}^Mq'_a .
\end{gathered}
\ee
We use the mostly-plus metric signature \((-+\cdots+)\), write Lorentz contractions
with a centered dot, and use natural units.  Thus every
physical massive momentum in \eqref{eq:intro_scattering_data} is future-directed
and timelike.
The hard interaction is confined to a spacetime region of characteristic size
\(\mathcal R\).  Outside this region the particles interact only through the
long-range electromagnetic field.  We assume generic massive kinematics, so that
every pair entering the coefficients below has nonzero relative velocity.

We first state the quantum soft-photon theorem.  The emitted photon has
\(k^\mu=\omega\mathbf n^\mu\), where \(\mathbf n^\mu\) is a future-directed null
vector normalized by \(\mathbf n^0=1\), \(\omega>0\), and
\(k.\varepsilon=0\).  The same null direction will be used below for the classical
observer.  Setting \(n=N+M\), we combine the scattering data
\eqref{eq:intro_scattering_data} for the \(n\) hard external lines in the
all-outgoing convention
\be\label{eq:intro_all_outgoing_data}
(P_a,Q_a,\eta_a,\mathfrak m_a)=
\begin{cases}
(p_a,q_a,+1,m_a),&1\leq a\leq N,\\
(-p'_{a-N},-q'_{a-N},-1,m'_{a-N}),&N+1\leq a\leq n.
\end{cases}
\ee
Thus \(P_a^\mu\) and \(Q_a\) are signed all-outgoing quantities, while
\(\eta_aP_a^\mu\) and \(\eta_aQ_a\) are respectively the future-directed physical
momentum and physical charge of line \(a\).  In particular,
\eqref{eq:intro_all_outgoing_data} implies
\(P_a^2=-\mathfrak m_a^2\) and \(\sum_{a=1}^nQ_a=0\).  Momentum conservation for
the radiative amplitude is \(\sum_{a=1}^nP_a^\mu+k^\mu=0\), which reduces to the
hard momentum-conservation condition in the soft limit.
We denote the formal all-orders perturbative amplitudes without and with the soft photon by
\(\mathcal M_n=\sum_{L\geq0}\mathcal M_n^{(L)}\) and
\(\mathcal M_{n+1}=\sum_{L\geq0}\mathcal M_{n+1}^{(L)}\), respectively.  Their
small-\(\omega\) expansion can be organized as
\bea\label{eq:intro_soft_factorization}
\mathcal M_{n+1}(k,\varepsilon)
&=&
\mathcal S_{\rm em}(\varepsilon,k)\mathcal M_n
+\text{non-factorized terms},
\non\\
\mathcal S_{\rm em}(\varepsilon,k)
&=&
\mathcal S_{\rm em}^{(-1)}(\varepsilon,k)
+\mathcal S_{\rm em}^{(0)}(\varepsilon,k)
+\sum_{p=1}^{d-5}\mathcal S_{\rm em}^{(p)}(\varepsilon,k)
+\mathcal S_{\rm em}^{\ln}(\varepsilon,k)+\cdots ,
\non\\
\mathcal S_{\rm em}^{(-1)}(\varepsilon,k)
&=&
\sum_{a=1}^{n}Q_a\f{\varepsilon.P_a}{k.P_a}.
\eea
The logarithmic term in the second line of
\eqref{eq:intro_soft_factorization} is, for \(d>4\),
\bea\label{eq:intro_logarithmic_soft_theorem}
\mathcal S_{\rm em}^{\ln}(\varepsilon,k)
&=&
\ln\omega
\sum_{b=1}^{n}\sum_{\substack{a=1\\a\ne b}}^{n}
Q_aQ_b^2(k.P_b)^{d-4}
\left(\varepsilon.P_a-\f{k.P_a}{k.P_b}\varepsilon.P_b\right)
\non\\
&&\times
\Bigg[
-\f{i(1+\eta_a\eta_b)}{2}\,
\mathcal C_d\,
\f{\mathfrak m_a^{d-2}\mathfrak m_b^2}
{\Delta_{ab}^{(d-1)/2}}
\non\\
&&\hspace{1.4cm}
-\f{(-1)^d\eta_a\eta_b}
{(d-3)2^{d-1}\pi^{d/2}
\Gamma\left(\f{d-2}{2}\right)}
\f{(d-3)\mathcal H_d(s_{ab})
+s_{ab}\mathcal H'_d(s_{ab})}
{\mathfrak m_a\mathfrak m_b^{d-3}}
\Bigg].
\eea
Here, for each ordered pair of distinct hard lines,
\bea\label{eq:intro_quantum_pair_data}
\Delta_{ab}&=&(P_a.P_b)^2-P_a^2P_b^2,
\qquad
s_{ab}=-\eta_a\eta_b\f{P_a.P_b}{\mathfrak m_a\mathfrak m_b}>1,
\non\\
\mathcal H_d(s)&=&
\f{1}{s}\,
\f{\sqrt{\pi}\,\Gamma\left(\f{d-2}{2}\right)}
{\Gamma\left(\f{d-1}{2}\right)}
{}_2F_1\left(1,\f12;\f{d-1}{2};1-\f{1}{s^2}\right),
\non\\
\mathcal C_d&=&
\begin{cases}
\displaystyle
\f{(-1)^{(d-4)/2}}
{(4\pi)^{(d-2)/2}\Gamma\left(\f{d-2}{2}\right)},
&d\ {\rm even},\\[3mm]
\displaystyle
\f{i\,\Gamma\left(\f{4-d}{2}\right)}
{2^{d-2}\pi^{d/2}},
&d\ {\rm odd}.
\end{cases}
\eea
The prime on \(\mathcal H_d\) denotes differentiation with respect to its argument.
Since \(k^\mu=\omega\mathbf n^\mu\), equation
\eqref{eq:intro_logarithmic_soft_theorem} has the advertised order
\(\mathcal O(\omega^{d-4}\ln\omega)\).

For each ordered pair, let \(\Lambda\) be a characteristic hard-energy scale of
order \(\sqrt{-\eta_a\eta_bP_a.P_b}\).  The result
\eqref{eq:intro_logarithmic_soft_theorem} applies for \(\omega\ll\Lambda\).
Because the argument of a logarithm must be dimensionless, the notation
\(\omega^{d-4}\ln\omega\) means
\(\omega^{d-4}\ln(\omega/\Lambda)\).  Changing the frequency-independent reference
scale alters only the non-logarithmic term at order \(\omega^{d-4}\), not the
displayed logarithmic coefficient.  The factor
\(\mathcal S_{\rm em}^{(-1)}=\mathcal O(\omega^{-1})\) in
\eqref{eq:intro_soft_factorization} is the standard universal leading soft factor,
while \(\mathcal S_{\rm em}^{(0)}=\mathcal O(\omega^0)\) denotes Low's subleading
soft operator
\cite{Low:1958sn,Weinberg:1964smatrix,Weinberg:1965nx,
Burnett:1968soft,Bell:1969soft}.  These terms determine the soft expansion through
order \(\omega^0\), apart from specific theory-dependent contributions to the
subleading photon term generated by higher-dimension operators
\cite{Elvang:2016qvq}.  Starting at order \(\omega\), each power
\(\omega^p\) generally contains both a factorizing part and process-dependent
non-factorized terms.  Gauge invariance constrains projections of the former,
giving the hierarchy of partial soft-photon factors discussed in
\cite{Hamada:2018soft,Li:2018soft,Campiglia:2018soft}, but does not determine the
complete unprojected soft factor.  Thus
\(\mathcal S_{\rm em}^{(p)}=\mathcal O(\omega^p)\), \(1\leq p\leq d-5\), denotes
only the factorizing contribution at each intermediate analytic order, while the
undetermined contribution at the same order is included among the non-factorized
terms in the first line of \eqref{eq:intro_soft_factorization}; the sum is empty
for \(d=5\).  The ellipsis in the second line begins with a possible
non-logarithmic factorizing term of order \(\omega^{d-4}\) and includes
factorizing terms at higher soft orders.

The logarithmic factor \eqref{eq:intro_logarithmic_soft_theorem} is gauge invariant
term by term in the ordered-pair sum.  The first term in square brackets is present
only for \(\eta_a\eta_b=1\), equivalently for two incoming or two outgoing lines.
In the Feynman representation of the quantum amplitude, this term is associated
with an on-shell massive-scalar pole, while the second term is associated with an on-shell
virtual photon and is present for every time-orientation assignment.  These labels
do not by themselves determine the classical part; that distinction depends on the
retarded boundary condition and is explained below.  Unlike in four dimensions,
\eqref{eq:intro_logarithmic_soft_theorem} cannot be inferred from an infrared
divergence of the conventional exponentiated QED eikonal factor, in which each
factor \(P_i.\ell\pm i\epsilon\) occurs to the first power.  For \(d>4\), this
factor is infrared finite, and the logarithm instead comes from the scale-invariant
loop-momentum region \(\omega\ll|\ell|\ll\Lambda\).  The \(d=4\) specialization reproduces the
logarithmic soft-photon result of \cite{Sahoo:2018lxl}.

Equation \eqref{eq:intro_soft_factorization} is written for the formal all-orders
perturbative amplitudes.  The derivation in this paper establishes the one-loop
component
\(\mathcal M_{n+1}^{(1)}|_{\omega^{d-4}\ln\omega}
=\mathcal S_{\rm em}^{\ln}\mathcal M_n^{(0)}\) in scalar QED with a
non-derivative gauge-invariant hard interaction.  Since the logarithm is generated
entirely by long-distance photon exchange through the minimal couplings of the
external charged lines, we expect the factor displayed in
\eqref{eq:intro_logarithmic_soft_theorem} to be universal and one-loop exact.  A
proof requires extending the calculation to general spins and gauge-invariant
non-minimal couplings, together with control over higher-loop corrections, in the
spirit of the four-dimensional analysis of \cite{Krishna:2023log}.

\clearpage
The hypergeometric function in \eqref{eq:intro_quantum_pair_data} can be eliminated
in both even and odd dimensions.  Keeping only \(\Delta_{ab}\) as compact notation, the result can
be written directly in terms of the momenta, masses, charges as
\begin{subequations}\label{eq:intro_soft_factors_even_odd_d}
\begin{align}
\begin{aligned}
\left.\mathcal S_{\rm em}^{\ln}(\varepsilon,k)\right|_{d=\mathrm{even}}
&=\ln\omega
\sum_{b=1}^{n}\sum_{\substack{a=1\\a\ne b}}^{n}
Q_aQ_b^2(k.P_b)^{d-4}
\left(\varepsilon.P_a-\f{k.P_a}{k.P_b}\varepsilon.P_b\right)
\\[-1mm]
&\quad\times
\Bigg[
-\f{i(-1)^{(d-4)/2}(1+\eta_a\eta_b)}
{2^{d-1}\pi^{(d-2)/2}\Gamma\left(\f{d-2}{2}\right)}
\f{\mathfrak m_a^{d-2}\mathfrak m_b^2}
{\Delta_{ab}^{(d-1)/2}}
\\[-1mm]
&\qquad
+\f{1}{(d-3)2^{d-1}\pi^{d/2}
\Gamma\left(\f{d-2}{2}\right)}
\Bigg\{
\f{2(-1)^{(d-4)/2}(P_a.P_b)\mathfrak m_a^{d-4}}
{\Delta_{ab}^{(d-2)/2}}
\\[-1mm]
&\qquad
+\sum_{j=1}^{(d-4)/2}
\f{(-1)^{j-1}\sqrt{\pi}\,
\Gamma\left(\f{d-2}{2}-j\right)}
{\Gamma\left(\f{d-1}{2}-j\right)}
\f{(P_a.P_b)\mathfrak m_a^{2j-2}}
{\mathfrak m_b^{d-2j-2}\Delta_{ab}^{j+1}}
\\[-1mm]
&\qquad\quad\times
\left[(d-2-2j)(P_a.P_b)^2
-(d-2)\mathfrak m_a^2\mathfrak m_b^2\right]
\\[-1mm]
&\qquad
+\f{2(-1)^{(d-4)/2}(d-3)\eta_a\eta_b
\mathfrak m_a^{d-2}\mathfrak m_b^2}
{\Delta_{ab}^{(d-1)/2}}
\operatorname{arccosh}\left(
-\eta_a\eta_b\f{P_a.P_b}{\mathfrak m_a\mathfrak m_b}
\right)
\Bigg\}
\Bigg].
\end{aligned}
\label{eq:intro_soft_factor_even_d}
\\[-2mm]
\begin{aligned}
\left.\mathcal S_{\rm em}^{\ln}(\varepsilon,k)\right|_{d=\mathrm{odd}}
&=\ln\omega
\sum_{b=1}^{n}\sum_{\substack{a=1\\a\ne b}}^{n}
Q_aQ_b^2
(k.P_b)^{d-4}
\left(\varepsilon.P_a-\f{k.P_a}{k.P_b}\varepsilon.P_b\right)
\\[-1mm]
&\quad\times
\Bigg[
\f{(1+\eta_a\eta_b)\Gamma\left(\f{4-d}{2}\right)}
{2^{d-1}\pi^{d/2}}
\f{\mathfrak m_a^{d-2}\mathfrak m_b^2}
{\Delta_{ab}^{(d-1)/2}}
\\[-1mm]
&\qquad
+\f{\eta_a\eta_b}
{(d-3)2^{2d-4}\pi^{(d-2)/2}
\Gamma\left(\f{d-2}{2}\right)
\mathfrak m_a\mathfrak m_b^{d-3}}
\Bigg\{
(d-3)\binom{d-3}{\frac{d-3}{2}}
\\[-1mm]
&\qquad
+2\sum_{j=1}^{(d-3)/2}
\binom{d-3}{\frac{d-3}{2}-j}
\left(
\f{\mathfrak m_a\mathfrak m_b+\eta_a\eta_bP_a.P_b}
{\mathfrak m_a\mathfrak m_b-\eta_a\eta_bP_a.P_b}
\right)^j
\\[-1mm]
&\qquad\quad\times
\left[
d-3-\f{2j\eta_a\eta_b(P_a.P_b)\mathfrak m_a\mathfrak m_b}
{\Delta_{ab}}
\right]
\Bigg\}
\Bigg].
\end{aligned}
\label{eq:intro_soft_factor_odd_d}
\end{align}
\end{subequations}
For even \(d\geq6\) and odd \(d\geq5\), the sums terminate at the displayed
dimension-dependent upper limits.  Thus the even-dimensional result is a finite
rational series plus one inverse hyperbolic cosine, while the odd-dimensional
result is a finite rational binomial series.

We next state the classical counterpart using the physical scattering data in
\eqref{eq:intro_scattering_data}.  The quantum theorem uses the positive photon
energy \(\omega>0\), whereas the classical Fourier transform requires real
\(\omega\) with retarded boundary values.  For an observer at
\(\vec x=r\hat x\) in the same null direction introduced above, define
\be\label{eq:intro_observer_data}
\mathbf n^\mu=(1,\hat x),\qquad k^\mu=\omega\mathbf n^\mu,\qquad
u=t-r .
\ee
With the observation data in \eqref{eq:intro_observer_data}, our Fourier convention
and the leading radiative relation for the retarded field in harmonic gauge are
\bea\label{eq:intro_radiative_relation}
\wt A^\mu(\omega,\vec x)
&\equiv&\int_{-\infty}^{\infty}dt\,e^{i\omega t}A^\mu(t,\vec x),
\qquad
A^\mu(t,\vec x)=\int_{-\infty}^{\infty}\f{d\omega}{2\pi}
e^{-i\omega t}\wt A^\mu(\omega,\vec x),
\non\\
\wt A^\mu(\omega,\vec x)
&=&
\f{e^{i\omega r}}{4\pi r^{(d-2)/2}}
\left(\f{\omega+i\epsilon}{2\pi i}\right)^{(d-4)/2}
\widehat J^\mu(k)
+o\left(r^{-(d-2)/2}\right),
\non\\
\widehat J^\mu(k)
&=&
\int d^dy\,e^{-ik.y}J^\mu(y).
\eea
Here \(J^\mu\) is the total classical electromagnetic current and
\(\epsilon\to0^+\) implements the retarded prescription.
The second line of \eqref{eq:intro_radiative_relation} follows from a saddle-point
evaluation of the spatial-momentum integral in the radiation zone
\cite{Laddha:2018rle,Laddha:2019yaj}.  Here
\(o(r^{-(d-2)/2})\) denotes terms that vanish relative to
\(r^{-(d-2)/2}\): multiplying them by \(r^{(d-2)/2}\) gives zero componentwise as
\(r\to\infty\) at fixed \(\hat x\) and \(\omega\).  Combining the
inverse transform in the first line with the outgoing phase \(e^{i\omega r}\) in
the second shows that \(\omega\) is the Fourier variable conjugate to \(u\).

Analogous to the soft-factor expansion in
\eqref{eq:intro_soft_factorization}, the low-frequency waveform at leading radiative order has the
following structure, with the common outgoing phase \(e^{i\omega r}\) suppressed
in the order symbols:
\bea\label{eq:intro_classical_frequency_expansion}
\wt A^\mu(\omega,\vec x)
&\simeq&
\wt A_{\rm st}^\mu(\omega,\vec x)
+\sum_{p=0}^{d-5}
\mathcal O\left(
\f{\omega^{p+(d-4)/2}}{r^{(d-2)/2}}
\right)
+\wt A_{\rm acc,\,log}^\mu(\omega,\vec x)
\non\\
&&+
\mathcal O\left(
\f{\omega^{3(d-4)/2}}{r^{(d-2)/2}}
\right)+\cdots ,
\non\\
\wt A_{\rm st}^\mu(\omega,\vec x)
&=&
\mathcal O\left(
\f{\omega^{(d-6)/2}}{r^{(d-2)/2}}
\right),
\qquad
\wt A_{\rm acc,\,log}^\mu(\omega,\vec x)
=
\mathcal O\left(
\f{\omega^{3(d-4)/2}\ln\omega}{r^{(d-2)/2}}
\right).
\eea
The symbol \(\simeq\) in the first line of
\eqref{eq:intro_classical_frequency_expansion} denotes equality after dropping the
little-\(o\) remainder in \eqref{eq:intro_radiative_relation}, so that only the
coefficient of the leading radiative falloff \(r^{-(d-2)/2}\) is
retained.\footnote{The remainder includes subleading radiative terms beginning at
\(\mathcal O(r^{-d/2})\) and Coulombic terms beginning at
\(\mathcal O(r^{-(d-3)})\).  In \(d=5\) these scale as \(r^{-5/2}\) and \(r^{-2}\),
respectively, but both are \(o(r^{-3/2})\).}
At leading radiative order, the two contributions evaluated here are the universal pieces determined completely
by the asymptotic scattering data: the straight-line field
\(\wt A_{\rm st}^\mu\) and the logarithmic acceleration field
\(\wt A_{\rm acc,\,log}^\mu\), whose soft orders are given in the last line of
\eqref{eq:intro_classical_frequency_expansion}.  The sum in its first line represents
possible contributions analytic in the current $\widehat J^\mu(k)$ at orders
\(\omega^0,\ldots,\omega^{d-5}\), while the non-logarithmic term at order
\(\omega^{d-4}\) in the current has the same power of \(\omega\) as the final term
shown there.  Their coefficients can depend on the details of the scattering event
and are not evaluated here.  At leading radiative order, the waveform powers in
\eqref{eq:intro_classical_frequency_expansion} include the radiative kernel in
\eqref{eq:intro_radiative_relation}; in odd dimensions their retarded boundary
values inherit its branch point.

At leading order in \(|\omega|\mathcal R\), direct integration over the incoming
and outgoing straight-line trajectories, followed by use of
\eqref{eq:intro_radiative_relation} gives for all \(d>4\)
\bea\label{eq:intro_straight_frequency}
\wt A_{\rm st}^\mu(\omega,\vec x)
&\simeq&
\f{i\,e^{i\omega r}}{4\pi r^{(d-2)/2}}
\left(\f{\omega+i\epsilon}{2\pi i}\right)^{(d-4)/2}
\Bigg[
\f{1}{\omega-i\epsilon}
\sum_{b=1}^{M}
\f{q'_bv_b^{\prime\mu}}{\mathbf n.v'_b}
-\f{1}{\omega+i\epsilon}
\sum_{b=1}^{N}
\f{q_bv_b^\mu}{\mathbf n.v_b}
\Bigg].
\eea
Equation \eqref{eq:intro_straight_frequency} follows directly from the leading
soft-photon factor in the last line of
\eqref{eq:intro_soft_factorization}, after converting the all-outgoing variables
back to the physical incoming and outgoing data and applying the radiative relation
\eqref{eq:intro_radiative_relation}
\cite{Low:1958sn,Weinberg:1965nx,Goldberger:2016iau,
Laddha:2018rle,Laddha:2019yaj}.

The leading logarithmic acceleration contribution in
\eqref{eq:intro_classical_frequency_expansion} is
\be\label{eq:intro_acceleration_frequency}
\begin{aligned}
\wt A_{\rm acc,\,log}^\mu(\omega,\vec x)\big|_{d=\mathrm{even}}
&\simeq
\f{e^{i\omega r}}{r^{(d-2)/2}}\,
\f{i^{(d-4)/2}\omega^{3(d-4)/2}}
{(2\pi)^{(d-4)/2}(4\pi)^{d/2}
\Gamma\left(\f{d-2}{2}\right)}
\\[-1mm]
&\quad\times
\left[
\ln(\omega+i\epsilon)\mathcal C_{{\rm out},\,{\rm even}}^\mu(\mathbf n)
+\ln(\omega-i\epsilon)\mathcal C_{{\rm in},\,{\rm even}}^\mu(\mathbf n)
\right],
\\[2mm]
\wt A_{\rm acc,\,log}^\mu(\omega,\vec x)\big|_{d=\mathrm{odd}}
&\simeq
-\f{i\,e^{i\omega r}}{4\pi r^{(d-2)/2}}
\f{(\omega+i\epsilon)^{(d-4)/2}\omega^{d-4}}
{(2\pi i)^{(d-4)/2}(2\pi)^{d-1}}
\\[-1mm]
&\quad\times
\left[
\ln(\omega+i\epsilon)
\left(
\mathcal C_{{\rm out},\,{\rm odd}}^\mu(\mathbf n)
+\mathcal C_{{\rm mix},\,{\rm odd}}^\mu(\mathbf n)
\right)
+\ln(\omega-i\epsilon)
\mathcal C_{{\rm in},\,{\rm odd}}^\mu(\mathbf n)
\right].
\end{aligned}
\ee
The $i\epsilon$ prescriptions in
\eqref{eq:intro_acceleration_frequency} distinguish the outgoing and incoming
asymptotic histories and enforce the retarded boundary condition.  In odd dimensions,
\((\omega+i\epsilon)^{(d-4)/2}\) is also essential because it gives the branch point
of the retarded radiative kernel in
\eqref{eq:intro_radiative_relation}.

The coefficient vectors in \eqref{eq:intro_acceleration_frequency} specify the
angular and kinematic dependence of the acceleration contribution in terms of the
physical data \eqref{eq:intro_scattering_data}.  In each double sum below, \(a\)
labels the accelerated trajectory and \(b\) labels the particle sourcing its
long-range force.  For even \(d\), the vectors are
\bea\label{eq:intro_even_hard_tensors}
\mathcal C_{{\rm out},\,{\rm even}}^\mu(\mathbf n)
&=&
\sum_{a=1}^{N}\sum_{\substack{b=1\\b\ne a}}^N
q_a^2q_b\,
\f{m_a^2m_b^{d-2}(p_a.\mathbf n)^{d-5}}
{\left[(p_a.p_b)^2-p_a^2p_b^2\right]^{(d-1)/2}}
\left(p_b.\mathbf n\,p_a^\mu-p_a.\mathbf n\,p_b^\mu\right),
\non\\
\mathcal C_{{\rm in},\,{\rm even}}^\mu(\mathbf n)
&=&
\sum_{a=1}^{M}\sum_{\substack{b=1\\b\ne a}}^M
q_a^{\prime2}q'_b\,
\f{m_a^{\prime2}m_b^{\prime d-2}(p'_a.\mathbf n)^{d-5}}
{\left[(p'_a.p'_b)^2-p_a^{\prime2}p_b^{\prime2}\right]^{(d-1)/2}}
\left(p'_b.\mathbf n\,p_a^{\prime\mu}
-p'_a.\mathbf n\,p_b^{\prime\mu}\right).
\eea
For odd \(d\), the corresponding vectors are
\bea\label{eq:intro_odd_hard_tensors}
\mathcal C_{{\rm out},\,{\rm odd}}^\mu(\mathbf n)
&=&
\sum_{a=1}^{N}\sum_{\substack{b=1\\b\ne a}}^N
\f{q_a^2q_b\,(-p_a.\mathbf n)^{d-4}}{m_bm_a^{d-3}}\,
\mathcal F_{\rm out}^{(d)}
\left(\f{p_a.p_b}{m_am_b}\right)
\left(p_b^\mu-\f{p_b.\mathbf n}{p_a.\mathbf n}p_a^\mu\right),
\non\\
\mathcal C_{{\rm in},\,{\rm odd}}^\mu(\mathbf n)
&=&
\sum_{a=1}^{M}\sum_{\substack{b=1\\b\ne a}}^M
\f{q_a^{\prime2}q'_b\,(-p'_a.\mathbf n)^{d-4}}
{m'_bm_a^{\prime d-3}}\,
\mathcal F_{\rm in}^{(d)}
\left(\f{p'_a.p'_b}{m'_am'_b}\right)
\left(p_b^{\prime\mu}
-\f{p'_b.\mathbf n}{p'_a.\mathbf n}p_a^{\prime\mu}\right),
\non\\
\mathcal C_{{\rm mix},\,{\rm odd}}^\mu(\mathbf n)
&=&
\sum_{a=1}^{N}\sum_{b=1}^{M}
\f{q_a^2q'_b\,(-p_a.\mathbf n)^{d-4}}
{m'_bm_a^{d-3}}\,
\mathcal F_{\rm mix}^{(d)}
\left(-\f{p'_b.p_a}{m'_bm_a}\right)
\left(p_b^{\prime\mu}
-\f{p'_b.\mathbf n}{p_a.\mathbf n}p_a^\mu\right).
\eea
The three functions encoding the causal branch assignments in
\eqref{eq:intro_odd_hard_tensors} are
\be\label{eq:intro_odd_functions}
\begin{aligned}
\mathcal F_{\rm out}^{(d)}(z)
&=
\f{2\pi^{(d-2)/2}z}{z^2-1}
\left[
\f{\Gamma\left(\f12\right)}
{\Gamma\left(\f{d-1}{2}\right)}
+\sum_{j=1}^{(d-3)/2}
\f{\Gamma\left(\f12-j\right)}
{\Gamma\left(\f{d-1}{2}-j\right)}
\f{z^{2j-2}}{(z^2-1)^j}
\right],
\\
\mathcal F_{\rm in}^{(d)}(z)
&=
\f{2(-1)^{(d-3)/2}\pi^{d/2}}
{\Gamma\left(\f{d-2}{2}\right)(z^2-1)^{(d-1)/2}},
\\
\mathcal F_{\rm mix}^{(d)}(z)
&=
\f{2\pi^{d/2-1}}
{(d-3)\Gamma\left(\f{d-2}{2}\right)}
\left[(d-3)\mathcal H_d(z)+z\mathcal H'_d(z)\right].
\end{aligned}
\ee
Here \(\mathcal H_d\) is defined in \eqref{eq:intro_quantum_pair_data}.
Each vector in \eqref{eq:intro_even_hard_tensors} and
\eqref{eq:intro_odd_hard_tensors} is transverse to \(\mathbf n^\mu\), and its
denominators are nonzero under the generic-kinematics assumption.  The mixed term is
specific to odd dimensions: it represents the field of an incoming particle acting
inside the past light cone of an outgoing trajectory.

The frequency-space expansion
\eqref{eq:intro_classical_frequency_expansion} and the explicit results
\eqref{eq:intro_straight_frequency} and
\eqref{eq:intro_acceleration_frequency} hold in the ordered radiation-zone limit
\be\label{eq:intro_frequency_window}
r^{-1}\ll|\omega|\ll\mathcal R^{-1} .
\ee
The upper inequality in \eqref{eq:intro_frequency_window} is the classical
low-frequency condition, with \(\mathcal R^{-1}\) playing the role of the hard
scale, while the lower inequality places the observer in the radiation zone needed
for the saddle-point relation \eqref{eq:intro_radiative_relation}.  Within this
ordered limit, the results retain the leading radiative power
\(r^{-(d-2)/2}\).  The straight-line result
\eqref{eq:intro_straight_frequency} gives the leading term, while
\eqref{eq:intro_acceleration_frequency} gives the logarithmic term generated by the
leading long-range acceleration.  The analytic current contributions and the
non-logarithmic term at order \(\omega^{d-4}\) are therefore included in
\eqref{eq:intro_classical_frequency_expansion} only through their soft orders.
Additional logarithmic terms suppressed by powers of \(\omega\mathcal R\) are
included in the ellipsis in that equation.

We finally transform the universal frequency-space terms
\eqref{eq:intro_straight_frequency} and
\eqref{eq:intro_acceleration_frequency} using the Fourier convention in
\eqref{eq:intro_radiative_relation}.  The resulting retarded-time tails are valid
in the asymptotic radiation-zone window
\be\label{eq:intro_validity_window}
\mathcal R\ll |u|\ll r,
\ee
where \(\mathcal R\ll|u|\) suppresses corrections that resolve the hard region by
\(\mathcal R/|u|\), while \(|u|\ll r\) keeps the inverse transform within the
large-\(r\) expansion used to retain the leading radiative mode in
\eqref{eq:intro_radiative_relation}.  In the time-domain formulas below,
\(\simeq\) denotes equality at the leading universal order in the hierarchy
\eqref{eq:intro_validity_window}.  We write \(A_{\rm acc}^\mu\) for the
time-domain acceleration field, while displaying only the universal terms fixed by
\(\wt A_{\rm acc,\,log}^\mu\) in
\eqref{eq:intro_acceleration_frequency}.

For even \(d\geq6\), the inverse transform of
\eqref{eq:intro_straight_frequency} at leading radiative order is a derivative of a
delta function distributionally supported in retarded time within an interval of
order \(\mathcal R\), set by the hard-scattering region.  It therefore vanishes in
\eqref{eq:intro_validity_window}
\cite{Garfinkle:2017fre,Laddha:2019yaj}.  The leading universal early- and
late-time fields come from the acceleration term
\eqref{eq:intro_acceleration_frequency}:
\be\label{eq:intro_even_time_tails}
\begin{aligned}
A_{\rm acc}^\mu(x)\big|_{\substack{d=\mathrm{even}\\u\to+\infty}}
&\simeq
(-1)^{(d-2)/2}
\f{\Gamma\left(\f{3d-10}{2}\right)}
{(2\pi)^{(d-4)/2}(4\pi)^{d/2}
\Gamma\left(\f{d-2}{2}\right)}
\f{\mathcal C_{{\rm out},\,{\rm even}}^\mu(\mathbf n)}
{r^{(d-2)/2}u^{(3d-10)/2}},
\\
A_{\rm acc}^\mu(x)\big|_{\substack{d=\mathrm{even}\\u\to-\infty}}
&\simeq
-\f{\Gamma\left(\f{3d-10}{2}\right)}
{(2\pi)^{(d-4)/2}(4\pi)^{d/2}
\Gamma\left(\f{d-2}{2}\right)}
\f{\mathcal C_{{\rm in},\,{\rm even}}^\mu(\mathbf n)}
{r^{(d-2)/2}|u|^{(3d-10)/2}} .
\end{aligned}
\ee

For odd \(d\geq5\), the branch point of the retarded radiative kernel also converts
the analytic-current terms represented only by their orders in
\eqref{eq:intro_classical_frequency_expansion} into generally non-universal
late-time power laws.  The following formulas isolate the universal tails fixed by
the asymptotic trajectories.  In the window \eqref{eq:intro_validity_window}, the
straight-line frequency-space field gives
\be\label{eq:intro_odd_straight_time}
\begin{aligned}
A_{\rm st}^\mu(x)
&\simeq
\f{(-1)^{(d-5)/2}\Gamma\left(\f{d-4}{2}\right)}
{(2\pi)^{d/2}r^{(d-2)/2}}
\f{\Theta(u)}{u^{(d-4)/2}}
\Bigg[
\sum_{b=1}^M\f{q'_bv_b^{\prime\mu}}{\mathbf n.v'_b}
-\sum_{b=1}^N\f{q_bv_b^\mu}{\mathbf n.v_b}
\Bigg].
\end{aligned}
\ee
Equation \eqref{eq:intro_odd_straight_time} agrees with the odd-dimensional
straight-line tails obtained in
\cite{Satishchandran:2017pek,Laddha:2019yaj}.
Here \(\Theta\) denotes the Heaviside step function.
The straight-line field \eqref{eq:intro_odd_straight_time} vanishes as
\(u\to-\infty\) and falls as
\(r^{-(d-2)/2}u^{-(d-4)/2}\) as \(u\to+\infty\).  At the logarithmic acceleration
order in \eqref{eq:intro_classical_frequency_expansion}, transforming
\eqref{eq:intro_acceleration_frequency} gives the universal late- and early-time
contributions
\be\label{eq:intro_odd_acceleration_time}
\begin{aligned}
A_{\rm acc}^\mu(x)\big|_{\substack{d=\mathrm{odd}\\u\to+\infty}}
&\simeq
-\f{\Gamma\left(\f{3d-10}{2}\right)}
{4\pi^2(2\pi)^{3(d-2)/2}}
\f{\ln u}
{r^{(d-2)/2}u^{(3d-10)/2}}
\\[-1mm]
&\quad\times
\left[
\mathcal C_{{\rm out},\,{\rm odd}}^\mu(\mathbf n)
+\mathcal C_{{\rm mix},\,{\rm odd}}^\mu(\mathbf n)
+\mathcal C_{{\rm in},\,{\rm odd}}^\mu(\mathbf n)
\right]
\\[-1mm]
&\quad
+\mathcal O\left(
\f{1}{r^{(d-2)/2}u^{(3d-10)/2}}
\right),
\\
A_{\rm acc}^\mu(x)\big|_{\substack{d=\mathrm{odd}\\u\to-\infty}}
&\simeq
\f{(-1)^{(d-3)/2}\Gamma\left(\f{3d-10}{2}\right)}
{4\pi(2\pi)^{3(d-2)/2}}
\f{\mathcal C_{{\rm in},\,{\rm odd}}^\mu(\mathbf n)}
{r^{(d-2)/2}|u|^{(3d-10)/2}} .
\end{aligned}
\ee
The coefficients in \eqref{eq:intro_even_time_tails} are universal.  In
\eqref{eq:intro_odd_acceleration_time}, the coefficients of the displayed \(\ln u\)
and early-time tails are also universal, the latter being fixed by the
negative-frequency discontinuity between the two logarithmic boundary values in
\eqref{eq:intro_acceleration_frequency}.  Here \(\ln u\) denotes a logarithm
relative to a fixed time scale of order \(\mathcal R\).  Changing that scale shifts
only the undetermined non-logarithmic late-time term at the same power
\(u^{-(3d-10)/2}\), which is smaller than the displayed logarithm by
\(1/\ln u\).  Among the universal terms displayed in
\eqref{eq:intro_odd_acceleration_time}, the mixed vector in
\eqref{eq:intro_odd_hard_tensors} enters only the late-time logarithm.  The
nonvanishing fields in the window
\eqref{eq:intro_validity_window} constitute the electromagnetic tail memories at
leading radiative order.

The relation between the quantum soft theorem and the classical waveform at leading radiative order can now be
read directly from \eqref{eq:intro_radiative_relation}.  After contraction with the
photon polarization, \(i\varepsilon_\mu\widehat J^\mu(k)\) plays the role of the
classical soft factor.  Its leading \(1/\omega\) term is the classical realization
of \(\mathcal S_{\rm em}^{(-1)}\) in
\eqref{eq:intro_soft_factorization}; after the all-outgoing conversion
\eqref{eq:intro_all_outgoing_data}, it produces the straight-line waveform
\eqref{eq:intro_straight_frequency} at leading radiative order.  At logarithmic order, the radiative kernel
in \eqref{eq:intro_radiative_relation} maps the
\(\omega^{d-4}\ln\omega\) current term, whose positive-frequency coefficient is the
retarded classical part of \eqref{eq:intro_logarithmic_soft_theorem}, to the waveform order displayed in
\eqref{eq:intro_classical_frequency_expansion}, with its retarded coefficient given
in \eqref{eq:intro_acceleration_frequency}.

This correspondence does not equate the complete quantum logarithmic factor
\eqref{eq:intro_logarithmic_soft_theorem} with the classical current.  The quantum
amplitude uses Feynman boundary conditions, whereas the classical waveform at
leading radiative order is a retarded response.  Comparing the two complete integrals with the same contour
selects the retarded classical part; the additional Feynman photon-pole residue is
the intrinsically quantum remainder.

\pagebreak
Defining \(S_{\rm cl}(\varepsilon,k)=i\varepsilon_\mu\widehat J^\mu(k)\), its
retarded logarithmic term in the all-outgoing variables
\eqref{eq:intro_all_outgoing_data} is, for even \(d\),
\bea\label{eq:intro_classical_soft_factor_even}
\left.S_{\rm cl}(\varepsilon,k)
\right|_{\substack{d=\mathrm{even}\\\omega^{d-4}\ln}}
&\simeq&
-i\sum_{b=1}^{n}\sum_{\substack{a=1\\a\ne b}}^{n}
Q_aQ_b^2(k.P_b)^{d-4}\mathcal C_d
\f{\mathfrak m_a^{d-2}\mathfrak m_b^2}{\Delta_{ab}^{(d-1)/2}}
\left(\varepsilon.P_a-\f{k.P_a}{k.P_b}\varepsilon.P_b\right)
\non\\
&&\times\left[
\f{(1+\eta_a)(1+\eta_b)}{4}\ln(\omega+i\epsilon)
+\f{(1-\eta_a)(1-\eta_b)}{4}\ln(\omega-i\epsilon)
\right],
\eea
whereas for odd \(d\),
\bea\label{eq:intro_classical_soft_factor_odd}
\left.S_{\rm cl}(\varepsilon,k)
\right|_{\substack{d=\mathrm{odd}\\\omega^{d-4}\ln}}
&\simeq&
\f{1}{(2\pi)^{d-1}}
\sum_{b=1}^{n}\sum_{\substack{a=1\\a\ne b}}^{n}
\f{Q_aQ_b^2(k.P_b)^{d-4}}{\mathfrak m_a\mathfrak m_b^{d-3}}
\left(\varepsilon.P_a-\f{k.P_a}{k.P_b}\varepsilon.P_b\right)
\non\\[-1mm]
&&\times\Bigg\{
-\ln(\omega+i\epsilon)\Bigg[
\f{(1+\eta_a)(1+\eta_b)}{4}\,
\mathcal F_{\rm out}^{(d)}
\left(\f{P_a.P_b}{\mathfrak m_a\mathfrak m_b}\right)
\non\\[-1mm]
&&\hspace{3.8cm}
+\f{(1-\eta_a)(1+\eta_b)}{4}\,
\mathcal F_{\rm mix}^{(d)}
\left(\f{P_a.P_b}{\mathfrak m_a\mathfrak m_b}\right)
\Bigg]
\non\\[-1mm]
&&\hspace{1cm}
+\ln(\omega-i\epsilon)
\f{(1-\eta_a)(1-\eta_b)}{4}\,
\mathcal F_{\rm in}^{(d)}
\left(\f{P_a.P_b}{\mathfrak m_a\mathfrak m_b}\right)
\Bigg\}.
\eea
The projectors involving $\eta_a,\eta_b$ in \eqref{eq:intro_classical_soft_factor_even} select the two
same-branch assignments.  Those in \eqref{eq:intro_classical_soft_factor_odd}
select outgoing, incoming-to-outgoing and incoming pairs; the reverse mixed ordering
vanishes by retarded causality.  For \(\omega>0\), both boundary values reduce to
\(\ln\omega\), giving the classical part of
\eqref{eq:intro_logarithmic_soft_theorem}; keeping them distinct specifies the
retarded continuation and the corresponding time-domain tails.  In even dimensions,
\eqref{eq:intro_classical_soft_factor_even} contains only the first term in the
square brackets of \eqref{eq:intro_soft_factor_even_d}.  The finite rational series
and the inverse-hyperbolic-cosine term in its second part are absent and constitute
the intrinsic quantum contribution.  In odd dimensions the separation is not
termwise.  Equation \eqref{eq:intro_classical_soft_factor_odd} contains the
same-branch matter term of \eqref{eq:intro_soft_factor_odd_d} and
\((1+\eta_b)\) times its second, finite-binomial term, including the displayed
\(\eta_a\eta_b\) prefactor.  For \(\eta_b=+1\), twice this term reconstructs the
remaining outgoing contribution and the incoming-to-outgoing contribution; for
\(\eta_b=-1\), it is absent.  The intrinsic quantum remainder is therefore
\(-\eta_b\) times the finite-binomial term in
\eqref{eq:intro_soft_factor_odd_d}.  In particular, it contains the complete
reverse mixed sector selected by \((1+\eta_a)(1-\eta_b)/4\), which has no retarded
counterpart.

The remainder of the paper is organized as follows.  Section~\ref{S:classical_waveform}
constructs the retarded radiative current from the asymptotic particle
trajectories.  It first determines the straight-line contribution associated with
the leading soft factor and then evaluates the correction generated by the
long-range acceleration, obtaining the even- and odd-dimensional frequency-space
waveforms at leading radiative order and their early- and late-time tails.  Subsection~\ref{S:position_space_even_tail}
independently derives the even-dimensional tail from the retarded Green's function
on the detector's past light cone.  Section~\ref{S:quantum_soft} computes the
one-loop logarithmic soft-photon factor in scalar QED, isolates the scale-invariant
loop-momentum region that produces the logarithm, reduces the calculation to a
scalar Feynman master integral for each ordered pair of hard external lines, and
derives the explicit even- and
odd-dimensional soft factors.  It then separates the classical retarded
contribution from the additional quantum contribution by comparing the Feynman and
retarded prescriptions, gives the classical soft factor explicitly in both even and
odd dimensions, and verifies that these factors reproduce the frequency-space
waveforms at leading radiative order.  Section~\ref{S:outlook} summarizes the implications and
open questions.  Appendix~\ref{app:odd_fourier_transform} collects the branch-sensitive
odd-dimensional Fourier transforms used for the time-domain tails.
Appendix~\ref{app:acceleration_integral} evaluates the classical acceleration
integral for the different time-orientation assignments and gives an independent
proper-time check based on the asymptotic trajectory.  Finally,
appendix~\ref{app:feynman_integral} evaluates the quantum master integral by
separating its matter- and photon-pole contributions and relates them to the
corresponding terms in the classical calculation.

\section{Low-frequency expansion of the classical radiative waveform}
\label{S:classical_waveform}
\subsection{Classical scattering setup}\label{S:setup}
\begin{figure}[t]
\centering
\begin{tikzpicture}[scale=1.15,mid arrow/.style={postaction={decorate},decoration={markings,mark=at position 0.5 with {\arrow{Stealth}}}}]
\draw[mid arrow,line width=0.4mm,blue] (-0.5,0.5) to [out=120,in=-110] (0,3);
\draw[mid arrow,line width=0.4mm,blue] (0.5,0.5) to [out=60,in=-70] (0,3);
\draw[mid arrow,line width=0.4mm,blue] (0,-3) to [out=60,in=-70] (0.5,-0.5);
\draw[mid arrow,line width=0.4mm,blue] (0,-3) to [out=120,in=-110] (-0.5,-0.5);
\fill[blue] (0,1.5) circle [radius=0.05cm];
\fill[blue] (-0.3,1.5) circle [radius=0.05cm];
\fill[blue] (0.3,1.5) circle [radius=0.05cm];
\fill[blue] (0,-1.5) circle [radius=0.05cm];
\fill[blue] (-0.3,-1.5) circle [radius=0.05cm];
\fill[blue] (0.3,-1.5) circle [radius=0.05cm];
\node[above right] at (1.6,1.5) {$\mathscr{I}^+$};
\node[above left] at (-1.5,1.5) {$\mathscr{I}^+$};
\node[below right] at (1.5,-1.5) {$\mathscr{I}^-$};
\node[below left] at (-1.5,-1.5) {$\mathscr{I}^-$};
\node[right] at (3,0) {$i^0$};
\node[left] at (-3,0) {$i^0$};
\node[above] at (0,3) {$i^+$};
\node[below] at (0,-3) {$i^-$};
\draw[line width=0.3mm,decorate, decoration={snake, amplitude=0.8mm, segment length=2.8mm},color=red,line cap=round]  (0.2,1.8) -- (0.7,2.3);
\draw[line width=0.3mm,decorate, decoration={snake, amplitude=0.8mm, segment length=2.8mm},color=red,line cap=round]  (0.1,2.0) -- (0.55,2.45);
\draw[line width=0.3mm,decorate, decoration={snake, amplitude=0.8mm, segment length=2.8mm},color=red,line cap=round]  (-0.2,1.8) -- (-0.7,2.3);
\draw[line width=0.3mm,decorate, decoration={snake, amplitude=0.8mm, segment length=2.8mm},color=red,line cap=round]  (-0.1,2.0) -- (-0.55,2.45);
\draw[line width=0.3mm,decorate, decoration={snake, amplitude=0.8mm, segment length=2.8mm},color=red,line cap=round]  (0.9,-0.8) -- (2.35,0.65);
\draw[line width=0.3mm,decorate, decoration={snake, amplitude=0.8mm, segment length=2.8mm},color=red,line cap=round]  (1.2,-1.2) -- (2.7,0.3);
\draw[line width=0.3mm,decorate, decoration={snake, amplitude=0.8mm, segment length=2.8mm},color=red,line cap=round]  (-0.9,-0.8) -- (-2.35,0.65);
\draw[line width=0.3mm,decorate, decoration={snake, amplitude=0.8mm, segment length=2.8mm},color=red,line cap=round]  (-1.2,-1.2) -- (-2.7,0.3);
\draw[line width=0.4mm] (0,3) -- (-3,0) -- (0,-3) -- (3,0) -- (0,3) -- (-3,0);
\node[right] at (0.4,2.8) {$\mathscr{I}^+_+$};
\node[left] at (-0.4,2.8) {$\mathscr{I}^+_+$};
\node[right] at (2.7,0.5) {$\mathscr{I}^+_-$};
\node[left] at (-2.7,0.5) {$\mathscr{I}^+_-$};
\draw[pattern=north east lines,line width=0.4mm] (0,0) circle [radius=0.7cm];
\end{tikzpicture}
\caption{Penrose diagram for charged-particle scattering.  The hatched central disk
denotes the finite interaction region, blue solid curves with arrows denote charged-particle
trajectories, and red wavy lines denote early- and late-time electromagnetic radiation.}
\label{f:scattering_penrose}
\end{figure}
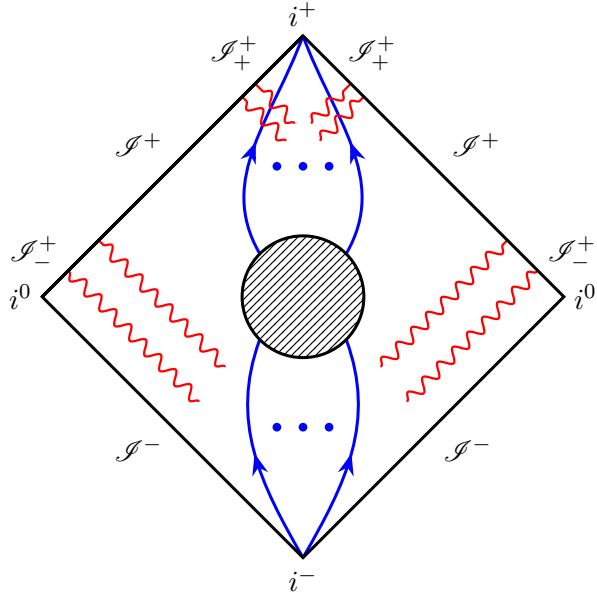

Consider a classical scattering process involving \(M\) incoming and \(N\) outgoing
charged particles in \(d\)-dimensional Minkowski spacetime with mostly-plus metric.  We
assume that the non-asymptotic dynamics is localized within a finite spacetime region of
characteristic length \(\mathcal R\).  Arbitrary short-range interactions, including hard
scattering, splitting and fusion, may occur within this region, whereas outside it we
retain only the long-range electromagnetic interaction.  The particle trajectories and
the emitted electromagnetic radiation are illustrated in the Penrose diagram in
Figure~\ref{f:scattering_penrose}.

We denote the masses, asymptotic velocities and charges of the incoming particles by
\(\lbrace m'_a,v'_a,q'_a\rbrace\) and those of the outgoing particles by
\(\lbrace m_a,v_a,q_a\rbrace\).  Outside the finite interaction region, we parametrize
their trajectories as
\bea\label{eq:asymptotic_trajectory}
\begin{gathered}
X_a^\mu(\tau_a)=r_a^\mu +v_a^\mu\tau_a +Y_a^\mu(\tau_a), \qquad \tau_a\in [0,\infty)\quad \text{for $a=1,2,\cdots,N$,} \\
X_a^{\prime\mu}(\tau'_a)=r_a^{\prime\mu} +v_a^{\prime\mu}\tau'_a+Y_a^{\prime\mu}(\tau'_a), \qquad \tau'_a\in (-\infty,0]\quad \text{for $a=1,2,\cdots,M$}.
\end{gathered}
\eea
The corresponding physical momenta are
\be\label{eq:hard_momenta}
p_a^\mu=m_av_a^\mu,\qquad p_a^{\prime\mu}=m'_av_a^{\prime\mu},
\qquad p_a^2=-m_a^2,\qquad p_a^{\prime 2}=-m_a^{\prime 2}.
\ee
Both the incoming and outgoing momenta in \eqref{eq:hard_momenta} are future-directed.
We assume generic asymptotic kinematics: no two distinct particles on the same
asymptotic branch have identical four-velocity.  Equivalently, for every same-branch
pair that occurs below,
\(
(v_a.v_b)^2-v_a^2v_b^2>0
\)
(and likewise for the primed velocities). We choose the spacetime origin within the interaction region and set each trajectory
parameter to zero where the corresponding asymptotic trajectory meets its boundary.  The
corrections to the straight-line trajectories obey
\bea\label{eq:Y_bdy_condition}
Y_a^\mu(0)=0,\quad \lim_{\tau_a\rightarrow\infty}\f{d Y_a^\mu(\tau_a)}{d\tau_a}=0\, , \qquad Y_a^{\prime \mu}(0)=0,\quad\lim_{\tau'_a\rightarrow -\infty}\f{d Y_a^{\prime\mu}(\tau'_a)}{d\tau'_a}=0\, .
\eea
Thus \(r'_a\) is the endpoint of the \(a\)-th incoming asymptotic trajectory and \(r_a\)
is the starting point of the \(a\)-th outgoing asymptotic trajectory.  Both lie on the
boundary of the finite interaction region, whose characteristic scale is \(\mathcal R\), while
\(Y'_a\) and \(Y_a\) encode the long-range deviations from straight-line motion outside
this region.

For the trajectories in \eqref{eq:asymptotic_trajectory}, the asymptotic current
density is given by
\bea\label{eq:current_full}
J^\mu(y)&=&\sum_{b=1}^{M}q'_b\int_{-\infty}^0 d\tau'_b\,   \delta^{(d)}\left(y-r'_b-v'_b\tau'_b-Y_b^{\prime}(\tau'_b)\right)\left(v_b^{\prime\mu}+\f{d Y_b^{\prime\mu}(\tau'_b)}{d\tau'_b}\right)\non\\
&+&\sum_{b=1}^{N}q_b\int_0^\infty d\tau_b\,   \delta^{(d)}\left(y-r_b-v_b\tau_b-Y_b(\tau_b)\right)\left(v_b^\mu+\f{dY_b^\mu(\tau_b)}{d\tau_b}\right).
\eea
Equation \eqref{eq:current_full} is the part of the current supported outside the
finite interaction region.  The complete conserved current also contains a
contribution supported inside that region.  Its Fourier transform is analytic for
\(|\omega|\mathcal R\ll1\), so it can affect the process-dependent analytic terms
but not the leading \(1/\omega\) term or the logarithmic coefficient derived below.
We therefore suppress this compactly supported contribution when extracting the
universal terms.
We would like to find the gauge field $A^\mu(x)$ near future null infinity
($\mathscr{I}^+$) by solving Maxwell's equation in harmonic gauge $\p_\mu A^\mu(x)=0$,
given by
\be
\square A^\mu(x)=-J^\mu(x).\label{eq:Maxwell}
\ee
The retarded gauge field solution from \eqref{eq:Maxwell} reads
\bea
A^{\mu}(x)=-\int d^dy
G_R(x-y) J^\mu(y)\, ,\label{eq:A_mu_gen}
\eea
with the retarded Green's function satisfying
\bea
\square_x G_R(x-y)=\delta^{(d)}(x-y),\qquad G_R(x-y)=0\quad \text{for}\quad x^0<y^0.\label{eq:box_G}
\eea
The retarded Green's function has the momentum-space representation
\be\label{eq:GR_mom_space}
G_R(x-y)=\int\f{d^d\ell}{(2\pi)^d}
\f{e^{i\ell\cdot(x-y)}}{(\ell^0+i\epsilon)^2-\vec{\ell}^{\,2}}.
\ee

We now extract the radiative part of the retarded solution \eqref{eq:A_mu_gen}.  The
calculation proceeds in two steps.  We first derive the general relation between the
frequency-space field and the Fourier-transformed current and apply it to the straight-line
current.  We then return to the long-range corrections in
\eqref{eq:asymptotic_trajectory} and isolate their leading non-analytic contribution.  The
derivation of the radiative kernel follows
\cite{Laddha:2018rle,Laddha:2019yaj}. 

Substituting the retarded Green's function \eqref{eq:GR_mom_space} into
\eqref{eq:A_mu_gen} and Fourier transforming in time gives
\bea\label{eq:retarded_field_momentum_space}
\wt{A}^{\mu}(\omega,\vec{x})\equiv \int_{-\infty}^\infty dx^0 e^{i\omega x^0}A^\mu(x)=-\int d^dy\,  e^{i\omega y^0} J^\mu(y)\int\f{d^{d-1}\vec\ell}{(2\pi)^{d-1}}
\f{e^{i\vec{\ell}.(\vec{x}-\vec{y})}}{(\omega+i\epsilon)^2-\vec{\ell}^2}.
\eea
We decompose $\vec{\ell}$ into the component $\ell_{\parallel}$ parallel to
$\vec{x}-\vec{y}$ and the $(d-2)$-dimensional transverse component
$\vec{\ell}_{\perp}$.  Closing the $\ell_{\parallel}$ contour in the upper half-plane
selects the outgoing pole
$\ell_{\parallel}=\sqrt{(\omega+i\epsilon)^2-\vec{\ell}_{\perp}^{\,2}}$ and yields
\bea\label{eq:retarded_field_after_longitudinal_contour}
\wt{A}^{\mu}(\omega,\vec{x})=\f{i}{2}\int d^dy\,  e^{i\omega y^0} J^\mu(y)\int\f{d^{d-2}\vec\ell_\perp}{(2\pi)^{d-2}} \quad \f{e^{i|\vec x-\vec y|\sqrt{(\omega+i\epsilon)^2-\vec\ell_\perp^2}}}{\sqrt{(\omega+i\epsilon)^2-\vec\ell_\perp^2}}.
\eea
We now impose the radiation-zone condition $|\vec x|\gg|\vec y|$.  In this
regime,
\be\label{eq:radiation_zone_distance_expansion}
|\vec x-\vec y|
=|\vec x|-\f{\vec x.\vec y}{|\vec x|}
+\mathcal O\left(\f{|\vec y|^2}{|\vec x|}\right).
\ee
Equation~\eqref{eq:radiation_zone_distance_expansion} also shows that
$|\vec x-\vec y|$ is large.  The transverse integral in
\eqref{eq:retarded_field_after_longitudinal_contour} is therefore controlled by
the saddle at $|\vec\ell_\perp|=0$.
Expanding around this saddle gives
\bea\label{eq:retarded_field_saddle_expansion}
\wt{A}^{\mu}(\omega,\vec{x})&=&\f{i}{2}\f{1}{2^{d-3}\pi^\f{d-2}{2}\Gamma\left(\f{d-2}{2}\right)}\int d^dy\,  e^{i\omega y^0} J^\mu(y)\int_0^\infty d|\vec\ell_\perp|\, |\vec\ell_\perp|^{d-3} \non\\
&\times &\exp\left\lbrace i|\vec x-\vec y|\left((\omega+i\epsilon)-\f{1}{2}\f{\vec\ell_\perp^2}{\omega+i\epsilon}+\cdots\right)\right\rbrace\times \left(\f{1}{\omega+i\epsilon}+\cdots\right),
\eea
where the omitted terms in \eqref{eq:retarded_field_saddle_expansion} are
subleading in the large-distance expansion and will be discarded under the sign
$\simeq$.  Using \eqref{eq:radiation_zone_distance_expansion} in
\eqref{eq:retarded_field_saddle_expansion}, the Gaussian integral gives
\bea
\wt{A}^{\mu}(\omega,\vec{x})&\simeq &i\f{e^{i\omega |\vec x|}}{2^{d-2}\pi^\f{d-2}{2}\Gamma\left(\f{d-2}{2}\right)}\f{1}{\omega+i\epsilon}\, \int_{|\vec{y}|\ll|\vec x|} d^dy\,  e^{i\omega y^0-i\omega \f{\vec x.\vec y}{|\vec x|}} J^\mu(y)\non\\
&&\times \int_0^\infty d|\vec\ell_\perp|\, |\vec\ell_\perp|^{d-3}
\exp\left\lbrace -\f{i}{2(\omega+i\epsilon)}|\vec x|\times |\vec\ell_\perp|^2\right\rbrace\non\\
&=& \f{1}{4\pi\, |\vec x|^\f{d-2}{2}}\, \left(\f{\omega+i\epsilon}{2\pi i}\right)^\f{d-4}{2}e^{i\omega|\vec x|}\times \widehat{J}^\mu(k),\label{eq:AJ_relation}
\eea
where the expression for $\widehat{J}^\mu(k)$ is given by
\be\label{eq:Jhat_def}
\widehat{J}^\mu(k)\equiv \int d^dy\,  e^{-ik.y} J^\mu(y)\, ,\quad k=\omega\left(1,\f{\vec x}{|\vec x|}\right)\equiv \omega\mathbf{n}\,.
\ee
The source integral in \eqref{eq:Jhat_def} is restricted to
$|\vec y|\ll|\vec x|$, and the proper-time integrals below are defined with the
corresponding convergence prescriptions.  Equation \eqref{eq:AJ_relation} is the standard
relation between the radiative field and the Fourier-transformed current
\cite{Goldberger:2016iau,Laddha:2018rle,Laddha:2019yaj}.

\subsection{Leading soft waveform from straight-line trajectories}\label{S:rad_A_straight}
We first isolate the leading low-frequency radiative waveform produced by the straight-line part of the
asymptotic trajectories. In this subsection only, we set $Y'_a=Y_a=0$ in
\eqref{eq:asymptotic_trajectory}, so that
\bea\label{eq:straight_trajectory}
X_a^\mu(\tau_a)&=&r_a^\mu +v_a^\mu\tau_a, \qquad \tau_a\in [0,\infty)\quad \text{for $a=1,2,\cdots,N$,} \\
X_a^{\prime\mu}(\tau'_a)&=&r_a^{\prime\mu} +v_a^{\prime\mu}\tau'_a, \qquad \tau'_a\in (-\infty,0]\quad \text{for $a=1,2,\cdots,M$}.
\eea
The corresponding current is denoted by $J_{\rm st}^\mu$, where the subscript ``st''
refers to the straight-line approximation:
\be\label{eq:current_st}
J_{\rm st}^\mu(y)=\sum_{b=1}^{M}q'_bv_b^{\prime\mu}\int_{-\infty}^0 d\tau'_b\,  \delta^{(d)}\left(y-r'_b-v'_b\tau'_b\right)+\sum_{b=1}^{N}q_bv_b^\mu\int_0^\infty d\tau_b\,  \delta^{(d)}\left(y-r_b-v_b\tau_b\right).
\ee
Substituting \eqref{eq:current_st} in \eqref{eq:Jhat_def}, we get
\bea\label{eq:Jhat_st_tau}
\widehat{J}_{\rm st}^\mu(k)=\sum_{b=1}^{M}q'_bv_b^{\prime\mu}\int_{-\infty}^0 d\tau'_b \, e^{-ik.(r'_b+v'_b\tau'_b)} +\sum_{b=1}^{N}q_bv_b^\mu\int_0^\infty d\tau_b  \, e^{-ik.(r_b+v_b\tau_b)}.
\eea
Performing the $\tau'_b$ and $\tau_b$ integrals with the appropriate $i\epsilon$
prescription at $\pm\infty$ gives
\bea\label{eq:Jhat_st}
\widehat{J}_{\rm st}^\mu(k)=\sum_{b=1}^{M}q'_bv_b^{\prime\mu}\, e^{-ik.r'_b}\f{i}{k.v'_b+i\epsilon} -\sum_{b=1}^{N}q_bv_b^\mu\, e^{-ik.r_b}\f{i}{k.v_b-i\epsilon}.
\eea
Substituting \eqref{eq:Jhat_st} in \eqref{eq:AJ_relation}, we get
\bea\label{eq:A_st_frequency}
\wt{A}_{\rm st}^\mu(\omega,\vec x)
&\simeq&
i\f{e^{i\omega|\vec x|}}{4\pi |\vec x|^\f{d-2}{2}}
\left(\f{\omega+i\epsilon}{2\pi i}\right)^\f{d-4}{2}
\Bigg[
\f{1}{\omega-i\epsilon}\sum_{b=1}^M
\f{q'_bv_b^{\prime\mu}}{\mathbf{n}.v'_b}e^{-i\omega\mathbf{n}.r'_b}
\non\\
&&\hspace{3.0cm}
-\f{1}{\omega+i\epsilon}\sum_{b=1}^N
\f{q_b v_b^\mu}{\mathbf{n}.v_b}e^{-i\omega\mathbf{n}.r_b}
\Bigg].
\eea
For all \(d>4\), the difference
\be\label{eq:straight_wave_difference_identity}
\f{1}{\omega-i\epsilon}-\f{1}{\omega+i\epsilon}=2i\pi\delta(\omega)
\ee
is multiplied in \eqref{eq:A_st_frequency} by
\((\omega+i\epsilon)^{(d-4)/2}\), whose boundary value vanishes at \(\omega=0\).
With the same regulator used in both factors, their product tends to zero as a
distribution for every \(d>4\).  Therefore the outgoing denominator
\(1/(\omega-i\epsilon)\) in \eqref{eq:A_st_frequency} will be replaced by
\(1/(\omega+i\epsilon)\) while evaluating the position-space waveform below.  The only exception is \(d=4\), where the prefactor becomes
unity and the pole at \(\omega=0\) gives the usual memory contribution.

We now transform back to the position-space waveform, introducing the retarded time
\(u=t-r\) and radial distance \(r=|\vec x|\):
\be \label{AAtilde}
A^\mu( x)=\int_{-\infty}^\infty \f{d\omega}{2\pi} \, e^{-i\omega t} \wt{A}^\mu(\omega,\vec x),\qquad x^0=t=u+r.
\ee
In even spacetime dimensions the integrand has a pole at \(\omega=0\) only for \(d=4\);
for even \(d\geq6\), it is analytic in a neighbourhood of the origin.  For odd \(d\),
it instead has a branch point at \(\omega=0\).  We therefore analyze the even- and
odd-dimensional Fourier transforms separately, since they lead to qualitatively
different behavior as \(u\to\pm\infty\).

Here and below, the limits \(u\to\pm\infty\) are understood within the asymptotic
hierarchy
\be\label{eq:asymptotic_waveform_window}
\mathcal R\ll |u|\ll r,\qquad
r^{-1}\ll |\omega|\sim |u|^{-1}\ll \mathcal R^{-1}.
\ee
The inverse Fourier transform at large \(|u|\) is controlled by frequencies
\(|\omega|\sim |u|^{-1}\).  The two inequalities in the second part of
\eqref{eq:asymptotic_waveform_window} therefore follow from the first hierarchy.
The upper bound \(|\omega|\ll\mathcal R^{-1}\), equivalently
\(\mathcal R\ll|u|\), places the transform in the low-frequency regime in which
the asymptotic trajectories outside the finite interaction region determine the
universal waveform.  The lower bound \(r^{-1}\ll|\omega|\), equivalently
\(|u|\ll r\), ensures \(|\omega|r\gg1\) and keeps the calculation within the
large-distance expansion used to retain the leading radiative mode in
\eqref{eq:AJ_relation}.  

\paragraph{Even dimensional waveform:}
For even $d\geq6$, substituting \eqref{eq:A_st_frequency} into \eqref{AAtilde} and using
the Fourier representation of the derivatives of a delta function gives
\be\label{eq:A_st_even}
\boxed{\begin{aligned}
A_{\rm st}^\mu(x)\big|_{d=\text{even}}
&\simeq
\f{1}{r^\f{d-2}{2}}\f{1}{2(2\pi)^\f{d-2}{2}}
\Bigg[
\sum_{b=1}^M\f{q'_bv_b^{\prime\mu}}{\mathbf{n}.v'_b}
\delta^{\left(\f{d-6}{2}\right)}\left(u+\mathbf{n}.r'_b\right)
\\[-1mm]
&\hspace{3.0cm}
-\sum_{b=1}^N\f{q_b v_b^\mu}{\mathbf{n}.v_b}
\delta^{\left(\f{d-6}{2}\right)}\left(u+\mathbf{n}.r_b\right)
\Bigg].
\end{aligned}}
\ee
The distributions in
\eqref{eq:A_st_even} are supported at \(u=-\mathbf n.r_b\) and
\(u=-\mathbf n.r'_b\).  Since these offsets are set by the finite scale \(\mathcal R\), the
radiative field vanishes in the asymptotic regime
\eqref{eq:asymptotic_waveform_window}.  The straight-line radiative mode therefore carries no
electromagnetic memory for even $d\geq6$, in agreement with
\cite{Garfinkle:2017fre,Laddha:2019yaj}. 

\paragraph{Odd dimensional waveform:}
For odd spacetime dimensions $d\geq 5$, replacing
\(1/(\omega-i\epsilon)\) by \(1/(\omega+i\epsilon)\) in
\eqref{eq:A_st_frequency}, as justified above, and then substituting into
\eqref{AAtilde}, we get
\bea
A_{\rm st}^\mu(x)&\simeq& -\f{1}{r^\f{d-2}{2}}\f{1}{2(2\pi i)^\f{d-2}{2}}
\non\\
&&\times
\Bigg[
\sum_{b=1}^M\f{q'_bv_b^{\prime\mu}}{\mathbf{n}.v'_b}
{\cal I}_{\f{d-6}{2}}\left(u+\mathbf{n}.r'_b\right)
-\sum_{b=1}^N\f{q_b v_b^\mu}{\mathbf{n}.v_b}
{\cal I}_{\f{d-6}{2}}\left(u+\mathbf{n}.r_b\right)
\Bigg],
\eea
where the integral
\be\label{eq:odd_tail_integral_def}
{\cal I}_\alpha(s)\equiv
\int_{-\infty}^\infty \f{d\omega}{2\pi}\,
e^{-i\omega s}(\omega+i\epsilon)^\alpha,\qquad \epsilon\rightarrow 0^+
\ee
is evaluated for general non-integer \(\alpha\) in
appendix~\ref{app:odd_fourier_transform}; see
\eqref{eq:odd_power_transform}.  In the specialized form needed here,
\be\label{eq:odd_tail_integral_specialized}
{\cal I}_{\f{d-6}{2}}(s)
=
\f{e^{\f{i\pi}{4}(d-6)}}{\Gamma\left(\f{6-d}{2}\right)}
\f{\Theta(s)}{s^\f{d-4}{2}}.
\ee
Substituting \eqref{eq:odd_tail_integral_specialized} into the preceding line, and then
using the following identities for odd values of $d$
\be\label{eq:odd_gamma_identities}
\begin{aligned}
\Gamma\left(\f{6-d}{2}\right)\Gamma\left(\f{d-4}{2}\right)
&=\f{\pi}{\sin\left[\pi(d-4)/2\right]},\\
\sin\left[\pi(d-4)/2\right]&=(-1)^\f{d-5}{2},
\end{aligned}
\ee
we obtain
\be\label{eq:A_st_odd}
\boxed{\begin{aligned}
A_{\rm st}^\mu(x)\big|_{d=\text{odd}}
&\simeq
\f{(-1)^\f{d-5}{2}\Gamma\left(\f{d-4}{2}\right)}
{(2\pi )^\f{d}{2}r^\f{d-2}{2}}
\Bigg[
\sum_{b=1}^M\f{q'_bv_b^{\prime\mu}}{\mathbf{n}.v'_b}
\f{\Theta(u+\mathbf{n}.r'_b)}{(u+\mathbf{n}.r'_b)^\f{d-4}{2}}
\\[-1mm]
&\hspace{2.2cm}
-\sum_{b=1}^N\f{q_b v_b^\mu}{\mathbf{n}.v_b}
\f{\Theta(u+\mathbf{n}.r_b)}{(u+\mathbf{n}.r_b)^\f{d-4}{2}}
\Bigg].
\end{aligned}}
\ee
Thus, at late positive retarded time in the regime
\eqref{eq:asymptotic_waveform_window}, the radiative gauge field falls off as
\(u^{-(d-4)/2}\).  Hence there is no static electromagnetic memory at radiative order, but a power-law
\emph{electromagnetic tail memory} remains.  Here ``tail memory'' refers to the
decaying radiative field that retains universal scattering information at large retarded
time.
Equation \eqref{eq:A_st_odd} is consistent
with the results of \cite{Satishchandran:2017pek,Laddha:2019yaj}. 

\subsection{Leading logarithmic correction from long-range trajectories}\label{S:log_correction}
We now isolate the part of the low-frequency expansion that can produce a waveform at
large retarded time.  In even dimensions the propagation factor in
\eqref{eq:AJ_relation} is an integer power of \(\omega\).  Consequently, a term that
is analytic at \(\omega=0\) can be expanded in non-negative integer powers of
\(\omega\), whose inverse Fourier transforms are
\be\label{eq:analytic_frequency_contact}
\int_{-\infty}^{\infty}\f{d\omega}{2\pi}\,
e^{-i\omega u}\omega^s
=i^s\delta^{(s)}(u),\qquad s=0,1,2,\ldots .
\ee
The finite phases \(e^{-i\omega\mathbf n.r_a}\) only shift this contact support by an
amount of order \(\mathcal R\).  Analytic terms in even dimensions therefore give
retarded-time distributions supported within an interval of order \(\mathcal R\),
set by the hard-scattering region, and vanish in the regime
\eqref{eq:asymptotic_waveform_window}.  The tail memory of interest here is a
power-law contribution at late times and must instead come from a non-analytic
dependence at \(\omega=0\).\footnote{\label{fn:nonanalytic_definition}Here a term is called non-analytic at
\(\omega=0\) if it does not admit a Taylor expansion in a neighbourhood of
\(\omega=0\), with its boundary value understood using the stated \(i\epsilon\)
prescription.  Such a term may nevertheless vanish in the soft limit.  For example,
when \(d>4\), \(\omega^{d-4}\ln(\omega\mathbin{\pm}i\epsilon)\to0\), but its
\((d-4)\)-th derivative behaves as
\((d-4)!\ln(\omega\mathbin{\pm}i\epsilon)+\mathcal O(1)\) and diverges
logarithmically; derivatives of higher order are power-law singular.}
The long-range trajectory corrections \(Y'_a\) and \(Y_a\) in
\eqref{eq:asymptotic_trajectory} generate terms of this kind in the current
\eqref{eq:Jhat_def}.  In particular,
\(\omega^s\ln(\omega\mathbin{\pm}i\epsilon)\) does not transform only to contact terms
at \(u=0\); away from \(u=0\), its Fourier transform is a power-law distribution
supported on the half-line selected by the \(i\epsilon\) prescription.  In even
dimensions the common acceleration integral \eqref{eq:Ib_log_integral} evaluates to
\eqref{eq:Ib_log_result} and produces the late-time waveform
\eqref{eq:A_acc_log_late}, which is absent from the straight-line result
\eqref{eq:A_st_even}.  In odd dimensions the radiative kernel already has a branch
point and produces the straight-line tail \eqref{eq:A_st_odd}.  The logarithmic
current separates into outgoing, incoming-to-outgoing, and incoming contributions in
\eqref{eq:Ib_odd_log_result}; the mixed term is present because the retarded Green's
function has support inside the past light cone.  Their combination gives the
logarithmically enhanced late-time subleading correction and the associated early-time
power-law tail in \eqref{eq:A_acc_odd_late_early}.  These results
generalize the four-dimensional classical logarithmic soft-photon theorem
\cite{Laddha:2018myi,Sahoo:2018lxl,Saha:2019tub,Sahoo:2020ryf}.

Throughout this subsection, all low-frequency expansions and asymptotic waveform
limits are understood in the hierarchy
\eqref{eq:asymptotic_waveform_window} and with the order of limits described there.

\subsubsection{Acceleration-induced current}\label{S:acceleration_current}

The Fourier transform of the asymptotic current \eqref{eq:current_full} is
\bea\label{eq:Jhat_full}
\widehat{J}^\mu(k)&=&\sum_{b=1}^{M}q'_b\int_{-\infty}^0 d\tau'_b \, e^{-ik.(r'_b+v'_b\tau'_b+Y'_b(\tau'_b))} \left(v_b^{\prime\mu}+\f{d Y_b^{\prime\mu}(\tau'_b)}{d\tau'_b}\right)\non\\
&+&\sum_{b=1}^{N}q_b\int_0^\infty d\tau_b  \, e^{-ik.(r_b+v_b\tau_b+Y_b(\tau_b))}\left(v_b^\mu+\f{dY_b^\mu(\tau_b)}{d\tau_b}\right).
\eea
Expanding \eqref{eq:Jhat_full} around the straight-line part and keeping terms up to linear
order in $Y_a^{\prime\mu}(\tau'_a)$ and $Y_a^\mu(\tau_a)$ gives the decomposition
\bea\label{eq:Jhat_decomposition}
\widehat{J}^\mu(k)&=&\widehat{J}_{\rm st}^\mu(k) +\widehat{J}_{\rm acc}^\mu(k) +\mathcal{O}\left(Y^2,Y^{\prime 2}\right).
\eea
Here $\widehat{J}_{\rm st}^\mu(k)$ is the straight-line result in
\eqref{eq:Jhat_st}, while $\widehat{J}_{\rm acc}^\mu(k)$ is the leading order acceleration-induced
piece
\bea\label{eq:Jacc_linear}
\widehat{J}_{\rm acc}^\mu(k)&=& \sum_{b=1}^{M}q'_b e^{-ik.r'_b}\int_{-\infty}^0 d\tau'_b \, e^{-i(k.v'_b+i\epsilon)\tau'_b} \left(-ik.Y'_b(\tau'_b)\, v_b^{\prime\mu}+\f{d Y_b^{\prime\mu}(\tau'_b)}{d\tau'_b}\right)\non\\
&+&\sum_{b=1}^{N}q_be^{-ik.r_b}\int_0^\infty d\tau_b  \, e^{-i(k.v_b-i\epsilon)\tau_b}\left(-ik.Y_b(\tau_b)\, v_b^\mu+\f{dY_b^\mu(\tau_b)}{d\tau_b}\right).
\eea
The $i\epsilon$ prescriptions in \eqref{eq:Jacc_linear} ensure convergence at the two ends of the
asymptotic trajectories along with the boundary conditions \eqref{eq:Y_bdy_condition}.  Integrating the terms proportional to $k.Y'_b$ and $k.Y_b$ by
parts, and using the boundary conditions in \eqref{eq:Y_bdy_condition}, we get
\bea\label{eq:Jacc_int}
\widehat{J}_{\rm acc}^\mu(k)
&=&
\sum_{b=1}^{M}q'_b e^{-ik.r'_b}
\left(\delta^\mu_\rho-\f{\mathbf n_\rho v_b^{\prime\mu}}
{\mathbf n.v'_b}\right)I_b^{\prime\rho}(k)
\non\\
&&+
\sum_{b=1}^{N}q_b e^{-ik.r_b}
\left(\delta^\mu_\rho-\f{\mathbf n_\rho v_b^\mu}
{\mathbf n.v_b}\right)I_b^\rho(k),
\eea
where
\bea\label{eq:Ib_def}
I_b^\rho(k)
&\equiv&
\int_0^\infty d\tau_b\,
e^{-i(k.v_b-i\epsilon)\tau_b}
\f{dY_b^\rho(\tau_b)}{d\tau_b},
\non\\
I_b^{\prime\rho}(k)
&\equiv&
\int_{-\infty}^0 d\tau'_b\,
e^{-i(k.v'_b+i\epsilon)\tau'_b}
\f{dY_b^{\prime\rho}(\tau'_b)}{d\tau'_b}.
\eea

Equation \eqref{eq:Jacc_int} shows that all non-analytic low-frequency dependence is
contained in the two integrals in \eqref{eq:Ib_def}; the tensors multiplying them are
transverse to \(\mathbf n^\mu\).  The treatment of \(I_b^\rho(k)\) depends on the causal
support of the retarded Green's function, since the trajectory correction
\(Y_b^\rho(\tau_b)\) is determined by the electromagnetic field of the other
particles.  The boundary conditions in \eqref{eq:Y_bdy_condition} imply
\(Y_b(\tau_b)=o(\tau_b)\) and \(Y'_a(\tau'_a)=o(-\tau'_a)\) at the corresponding
asymptotic ends.\footnote{Here \(f^\mu(\lambda)=o(g(\lambda))\) means that
\(f^\mu(\lambda)/g(\lambda)\to0\) componentwise in any fixed inertial frame in the
stated limit; for scalar quantities the same notation has its usual ratio meaning.}
Hence, for \(\tau_b>0\) and \(\tau'_a<0\),
\be\label{eq:incoming_outgoing_separation}
X_b(\tau_b)-X'_a(\tau'_a)
=v_b\tau_b+v'_a(-\tau'_a)+o(\tau_b-\tau'_a),
\ee
where the fixed displacement \(r_b-r'_a\) is included in the last term.  Since
\(v_b^2=v_a^{\prime 2}=-1\) and \(v_b\mathbin{.}v'_a\leq-1\),
\be\label{eq:incoming_outgoing_timelike}
\begin{aligned}
\left[X_b(\tau_b)-X'_a(\tau'_a)\right]^2
&=-\tau_b^2-(-\tau'_a)^2
+2\tau_b(-\tau'_a)v_b\mathbin{.}v'_a
+o\!\left((\tau_b-\tau'_a)^2\right) \\
&\leq-(\tau_b-\tau'_a)^2
+o\!\left((\tau_b-\tau'_a)^2\right)<0
\end{aligned}
\ee
for sufficiently late \(\tau_b\), uniformly for \(\tau'_a\leq0\).  Moreover,
\(X_b^0(\tau_b)-X_a^{\prime0}(\tau'_a)>0\) in the same limit.
Thus the incoming point lies strictly inside, rather than on the past light cone of the
late outgoing point with generic velocity configuration. 

In view of \eqref{eq:incoming_outgoing_timelike}, whether the incoming trajectory
contributes is determined by the support of the retarded Green's function.  In even
spacetime dimensions it is supported only on null separation, so
an incoming asymptotic trajectory does not contribute to the field acting on a sufficiently
late outgoing trajectory.  In odd spacetime dimensions the retarded Green's function is
also supported at timelike separation, so the preceding causal argument does not exclude
this contribution.  Accordingly, we first isolate the contribution to \(I_b^\rho(k)\) sourced
by the other outgoing particles.  This is the complete asymptotic-source contribution in
even dimensions.  In odd dimensions, however, the \(N-1\) outgoing-source terms must be
supplemented by contributions from all \(M\) incoming trajectories.  To keep the
intermediate expressions concise, we display only the outgoing-source force in
\eqref{eq:trajectory}--\eqref{eq:Ib_log_integral}.  The additional incoming-to-outgoing
contribution involves primed velocities and a different pole prescription; it is
evaluated separately in
\eqref{eq:Jab_mix_def}--\eqref{eq:odd_I_mix_covariant} and included in the full result
\eqref{eq:Ib_odd_log_result}.  The same-branch incoming integral
\(I_b^{\prime\rho}(k)\) is given for even dimensions in \eqref{eq:Ib_log_result}; its
odd-dimensional evaluation is carried out in
\eqref{eq:Jab_in_def}--\eqref{eq:odd_I_in_covariant} and is likewise included in
\eqref{eq:Ib_odd_log_result}.

For this outgoing-source contribution, the trajectory equation is
\bea\label{eq:trajectory}
m_b\f{d^2 X_b^\rho(\tau_b)}{d\tau_b^2}&=& q_b \f{d X_{b\sigma}(\tau_b)}{d\tau_b}
\sum_{\substack{a=1\\a\ne b}}^N F_{(a)}^{\rho\sigma}(X_b(\tau_b)),
\eea
where \(F_{(a)}^{\rho\sigma}(x)\) denotes the electromagnetic field strength produced at
\(x\) by outgoing particle \(a\).  We work at the leading nontrivial order in the
long-range electromagnetic interaction.  At this order, radiation reaction,
self-force effects and corrections to the source trajectories can be neglected.
To the first order in the long-range correction to the
straight-line motion, the right hand side of \eqref{eq:trajectory} may be evaluated on the
straight-line trajectory.  This gives
\bea\label{eq:Y_acceleration}
\f{d^2 Y_b^\rho(\tau_b)}{d\tau_b^2}&=& \f{q_b v_{b\sigma}}{m_b}
\sum_{\substack{a=1\\a\ne b}}^N F_{(a)}^{\rho\sigma}(r_b+v_b\tau_b), \quad\text{with}\qquad \lim_{\tau_b\rightarrow\infty} \f{dY_b^\rho(\tau_b)}{d\tau_b}=0.
\eea
The boundary condition in \eqref{eq:Y_acceleration} gives
\be\label{eq:dY}
\f{d Y_b^\rho(\tau_b)}{d\tau_b}= -\f{q_b v_{b\sigma}}{m_b}
\sum_{\substack{a=1\\a\ne b}}^N\int_{\tau_b}^\infty ds_b F_{(a)}^{\rho\sigma}(r_b+v_bs_b),
\ee
where, at this order, the field strength produced by outgoing particle \(a\) is evaluated
on its straight-line trajectory:
\be\label{eq:F_straight}
F_{(a)}^{\rho\sigma}(x)
=iq_a \int\f{d^d\ell}{(2\pi)^d}
\f{v_a^\rho\ell^\sigma-v_a^\sigma\ell^\rho}
{(\ell^0+i\epsilon)^2-\vec\ell^2}
\int_0^\infty d\tau_a\, e^{i\ell.(x-r_a-v_a\tau_a)}.
\ee
Substituting \eqref{eq:F_straight} into \eqref{eq:dY} and performing the $\tau_a$ and
$s_b$ integrals gives
\bea\label{eq:dY_momentum}
\f{d Y_b^\rho(\tau_b)}{d\tau_b}
&=&-i\f{q_b}{m_b}
\sum_{\substack{a=1\\a\ne b}}^N q_a
\int\f{d^d\ell}{(2\pi)^d}
\f{e^{i\ell.(r_b-r_a)}e^{i(\ell.v_b+i\epsilon)\tau_b}}
{(\ell^0+i\epsilon)^2-\vec\ell^2}
\non\\
&&\times
\f{v_a^\rho v_b.\ell-v_a.v_b\ell^\rho}
{(\ell.v_a-i\epsilon)(\ell.v_b+i\epsilon)}.
\eea
Substituting \eqref{eq:dY_momentum} into \eqref{eq:Ib_def} and performing the $\tau_b$
integral gives
\bea\label{eq:Ib_momentum_integral}
I_b^\rho(k)
&=&\f{q_b}{m_b}
\sum_{\substack{a=1\\a\ne b}}^N q_a
\int\f{d^d\ell}{(2\pi)^d}
\f{e^{i\ell.(r_b-r_a)}}{(\ell^0+i\epsilon)^2-\vec\ell^2}
\non\\
&&\times
\f{v_a^\rho v_b.\ell-v_a.v_b\ell^\rho}
{(\ell.v_a-i\epsilon)(\ell.v_b+i\epsilon)}
\f{1}{(\ell-k).v_b+i\epsilon}.
\eea
Our aim is not to determine every term in the small-\(k\) expansion of
\eqref{eq:Ib_momentum_integral}, but only its leading non-analytic term in the sense of
footnote~\ref{fn:nonanalytic_definition}: a leading power in \(\omega\) multiplying
\(\ln(\omega\mathbin{\pm}i\epsilon)\).  Here \(\ell\) is the momentum carried by
the retarded electromagnetic propagator.  We first exclude the ranges of this
momentum that cannot produce the logarithm.  In the region
\(|\ell|\ll|\omega|\), the last denominator in
\eqref{eq:Ib_momentum_integral} may be expanded in \(\ell/k\).  For any fixed
\(0<c<1\), the term of order \(n\) has, after the angular integrations, the radial
scaling
\be\label{eq:classical_small_momentum_scaling}
 \f{1}{|\omega|^{n+1}}\int_0^{c|\omega|}d|\ell|\,
 |\ell|^{d-4+n}\ \propto\ |\omega|^{d-4}.
\ee
For \(d>4\), the radial exponent in
\eqref{eq:classical_small_momentum_scaling} is never \(-1\), so this region cannot
generate \(\ln\omega\).  In the transition region \(|\ell|\sim|\omega|\), where the
expansion in \(\ell/k\) is not valid, the rescaling \(\ell=|\omega|L\) of the full
degree-\(-4\) integrand again gives \(|\omega|^{d-4}\) times an integral over a fixed
range of \(L\).  Thus the entire region \(|\ell|\lesssim|\omega|\) can affect a
non-logarithmic term at order \(\omega^{d-4}\), but not the logarithmic coefficient.

At the other end, let \(r_{ba}\equiv r_b-r_a=\mathcal O(\mathcal R)\).  For
\(|\ell|\gtrsim\mathcal R^{-1}\), one has \(|k|/|\ell|\ll1\), and the last
denominator in \eqref{eq:Ib_momentum_integral} has a Taylor expansion in \(k/\ell\).
The transition region \(|\ell|\sim\mathcal R^{-1}\) has \(\omega\)-independent
extent and therefore gives Taylor coefficients in \(k\).  For
\(|\ell|\gg\mathcal R^{-1}\), the phase \(e^{i\ell.r_{ba}}\) is rapidly oscillatory;
successive integrations by parts suppress the large-\(|\ell|\) contribution while
leaving the same Taylor expansion in \(k\).  Hence this region also cannot produce
\(\ln\omega\).  The logarithmic contribution can therefore arise only from the
parametrically separated intermediate region
\be\label{eq:logarithmic_overlap_region}
 |\omega|\ll|\ell|\ll \mathcal R^{-1}.
\ee
This intermediate region is the only one not excluded by the two endpoint
analyses.  Within \eqref{eq:logarithmic_overlap_region}, the denominator and source
phase admit the two simultaneous expansions
\be\label{eq:soft_denominator_expansion}
\f{1}{(\ell-k).v_b+i\epsilon}
=\sum_{n=0}^{\infty}\f{(k.v_b)^n}{(\ell.v_b+i\epsilon)^{n+1}} .
\ee
\be\label{eq:source_phase_expansion}
 e^{i\ell.r_{ba}}
 =\sum_{m=0}^{\infty}\f{(i\ell.r_{ba})^m}{m!}.
\ee
For the term \((m,n)\), the retarded electromagnetic propagator, numerator, and
\((\ell.v_a-i\epsilon)^{-1}\) have degrees \(-2,+1\), and \(-1\), respectively.
The two \(v_b\)-denominators and the phase have degrees \(-(n+2)\) and \(+m\).
The integrand therefore has degree \(-n-4+m\), and its radial dependence is
\be\label{eq:mn_radial_scaling}
 (k.v_b)^n(r_{ba})^m
 \int_{|\omega|}^{\mathcal R^{-1}}d|\ell|\,
 |\ell|^{d-5-n+m}.
\ee
The radial integral is
\be\label{eq:radial_endpoint_evaluation}
\int_{|\omega|}^{\mathcal R^{-1}}d|\ell|\,|\ell|^{d-5-n+m}
=
\begin{cases}
\displaystyle
\f{(\mathcal R^{-1})^{d-4-n+m}-|\omega|^{d-4-n+m}}{d-4-n+m},
&n\ne d-4+m,\\[3mm]
\displaystyle
\ln\f{\mathcal R^{-1}}{|\omega|},
&n=d-4+m.
\end{cases}
\ee
Equation~\eqref{eq:radial_endpoint_evaluation} makes the three cases immediate:
\begin{enumerate}
\item If \(n<d-4+m\), the upper endpoint represents matching to the fixed region
\(|\ell|\sim\mathcal R^{-1}\).  Its coefficient is independent of \(\omega\), so
multiplication by \((k.v_b)^n\propto\omega^n\) gives an analytic term.  The lower
endpoint belongs to \(|\ell|\sim|\omega|\) and, by
\eqref{eq:classical_small_momentum_scaling}, cannot contribute to the logarithmic
coefficient.

\item If \(n=d-4+m\), the radial power is \(-1\), and
\eqref{eq:radial_endpoint_evaluation} directly gives
\(\ln(\mathcal R^{-1}/|\omega|)\).  For fixed \(m\), this is the unique logarithmic,
and hence non-analytic, contribution.

\item If \(n>d-4+m\), the lower endpoint dominates, but there
\(|\omega|/|\ell|=\mathcal O(1)\) and
\eqref{eq:soft_denominator_expansion} must be resummed.  This is precisely the
\(|\ell|\sim|\omega|\) region already covered by
\eqref{eq:classical_small_momentum_scaling}; it may change a non-logarithmic term but
cannot alter the logarithmic coefficient.
\end{enumerate}

Consequently, only the marginality condition \(n=d-4+m\) gives
\be\label{eq:higher_phase_log_order}
 \mathcal O\!\left(
 \omega^{d-4+m}\mathcal R^m\ln\omega
 \right).
\ee
The leading choice is \(m=0\), \(n=d-4\), yielding
\(\omega^{d-4}\ln\omega\).  Terms with \(m\geq1\) are suppressed by
\((\omega\mathcal R)^m\), or by \((\mathcal R/|u|)^m\) after Fourier transformation, so the phase may
be set to one at leading order.\footnote{The \(m=1\) contribution at order
\(\omega\ln\omega\) to the four-dimensional gravitational waveform was analyzed in
\cite{Ghosh:2021bam}.}  Dropping the \(\omega\)-independent term \(\ln \mathcal R^{-1}\) and
restoring the \(i\epsilon\) prescription fixed by the proper-time range gives
\bea\label{eq:Ib_log_integral}
I_b^\rho(k)\big|_{\mathrm{out},\,\ln}
&\simeq&
\f{q_b}{m_b} (v_b.k)^{d-4}
\sum_{\substack{a=1\\a\ne b}}^N q_a
\int_\omega\f{d^d\ell}{(2\pi)^d}
\f{1}{(\ell^0+i\epsilon)^2-\vec\ell^2}
\f{v_a^\rho v_b.\ell-v_a.v_b\,\ell^\rho}
{(\ell.v_a-i\epsilon)(\ell.v_b+i\epsilon)^{d-2}} .
\eea
The subscript on \(\int_\omega\) denotes a radial lower scale of order
\(|\omega|\), as in \eqref{eq:logarithmic_overlap_region}, rather than a signed
radial endpoint.  Here and below \(\simeq\) means that analytic and other non-logarithmic terms,
and logarithmic terms at higher powers of $\omega$ are omitted, so only the leading
\(\omega^{d-4}\ln\omega\) contribution is retained.  The momentum integral in
\eqref{eq:Ib_log_integral} is evaluated in
appendix~\ref{app:acceleration_integral}.  We now use the momentum-space evaluation
separately in even and odd spacetime dimensions.

\subsubsection{Even spacetime dimensions}
For even \(d\), \((d-4)/2\) is an integer.  The evaluation in
appendix~\ref{app:acceleration_integral}, summarized in
\eqref{eq:Ib_log_result_app}, gives
\bea\label{eq:Ib_log_result}
I_b^\rho(k)\big|_{d=\mathrm{even},\,\ln}
&\simeq&
\omega^{d-4}\ln(\omega+i\epsilon)\,
\f{q_b}{m_b}(v_b.\mathbf{n})^{d-4}\,
\f{(-1)^\f{d-4}{2}}
{(4\pi)^\f{d-2}{2}\Gamma\left(\f{d-2}{2}\right)}
\non\\
&&\times
\sum_{\substack{a=1\\a\ne b}}^N q_a
\f{v_a^\rho v_b^2-v_a.v_b v_b^\rho}
{\left[(v_a.v_b)^2-v_a^2 v_b^2\right]^\f{d-1}{2}},
\non\\
I_b^{\prime\rho}(k)\big|_{d=\mathrm{even},\,\ln}
&\simeq&
\omega^{d-4}\ln(\omega-i\epsilon)\,
\f{q'_b}{m'_b}(v'_b.\mathbf{n})^{d-4}\,
\f{(-1)^\f{d-4}{2}}
{(4\pi)^\f{d-2}{2}\Gamma\left(\f{d-2}{2}\right)}
\non\\
&&\times
\sum_{\substack{a=1\\a\ne b}}^M q'_a
\f{v_a^{\prime\rho}v_b^{\prime2}-v'_a.v'_b\,v_b^{\prime\rho}}
{\left[(v'_a.v'_b)^2-v_a^{\prime2}v_b^{\prime2}\right]^\f{d-1}{2}}.
\eea
For even \(d\geq6\), subsection~\ref{S:position_space_even_tail} obtains the
large-proper-time trajectory from \eqref{eq:Y_acceleration} and derives the
waveform directly from the retarded Green's function.  The proper-time
calculation in appendix~\ref{app:position_space_integral_check} then inserts
that trajectory into \eqref{eq:Ib_def} and reproduces both the normalization
and the \(\ln(\omega\mathbin{\pm}i\epsilon)\) prescriptions in
\eqref{eq:Ib_log_result}.  Together these calculations independently verify
the marginal momentum-space term and show why the discarded source-phase,
finite-time and subleading trajectory terms cannot affect its logarithmic
coefficient. The two logarithms
carry the $i\epsilon$ prescriptions inherited from the corresponding proper-time ranges in
\eqref{eq:Ib_def}.

\paragraph{Frequency-space waveform:}
At the order retained in \eqref{eq:Ib_log_result}, the factors $e^{-ik.r_b}$ and
$e^{-ik.r'_b}$ in \eqref{eq:Jacc_int} may be set to one: their first corrections carry an
extra power of $\omega$ and are subleading.  The contraction of the transverse projector
in \eqref{eq:Jacc_int} with the numerator in \eqref{eq:Ib_log_result} is
\bea\label{eq:projected_tensor}
\left(\delta^\mu_\rho-\f{\mathbf n_\rho v_b^\mu}{\mathbf n.v_b}\right)
\left(v_a^\rho v_b^2-v_a.v_b\,v_b^\rho\right)
&=&
\f{v_b^2}{\mathbf n.v_b}
\left(\mathbf n.v_b\,v_a^\mu-\mathbf n.v_a\,v_b^\mu\right).
\eea
Using $v_b^2=-1$, equation \eqref{eq:projected_tensor} combines with the factor
$(v_b.\mathbf n)^{d-4}$ in \eqref{eq:Ib_log_result}.  Relabeling the two dummy particle
indices so that \(a\) denotes the accelerated particle, and using
\be\label{eq:velocity_momentum_invariant}
(v_a.v_b)^2-v_a^2v_b^2
=\f{(p_a.p_b)^2-p_a^2p_b^2}{m_a^2m_b^2},
\ee
we obtain from \eqref{eq:velocity_momentum_invariant} the transverse hard-particle
vectors
\bea\label{eq:hard_tensors}
\mathcal C_{{\rm out},\,{\rm even}}^\mu(\mathbf n)
&\equiv&
\sum_{a=1}^{N}\sum_{\substack{b=1\\b\ne a}}^N
q_a^2q_b\,
\f{m_a^2m_b^{d-2}(p_a.\mathbf n)^{d-5}}
{\left[(p_a.p_b)^2-p_a^2p_b^2\right]^\f{d-1}{2}}
\left(p_b.\mathbf n\,p_a^\mu-p_a.\mathbf n\,p_b^\mu\right),
\non\\
\mathcal C_{{\rm in},\,{\rm even}}^\mu(\mathbf n)
&\equiv&
\sum_{a=1}^{M}\sum_{\substack{b=1\\b\ne a}}^M
q_a^{\prime 2}q'_b\,
\f{m_a^{\prime 2}m_b^{\prime d-2}(p'_a.\mathbf n)^{d-5}}
{\left[(p'_a.p'_b)^2-p_a^{\prime2}p_b^{\prime2}\right]^\f{d-1}{2}}
\left(p'_b.\mathbf n\,p_a^{\prime\mu}
-p'_a.\mathbf n\,p_b^{\prime\mu}\right).
\eea
Substituting \eqref{eq:Ib_log_result} into \eqref{eq:Jacc_int} and using
\eqref{eq:projected_tensor} and \eqref{eq:hard_tensors} gives the even-dimensional
current
\bea\label{eq:Jacc_even_log}
\widehat J_{\rm acc}^\mu(k)\big|_{d=\mathrm{even},\,\ln}
&\simeq&
\f{(-1)^\f{d-4}{2}\omega^{d-4}}
{(4\pi)^\f{d-2}{2}\Gamma\left(\f{d-2}{2}\right)}
\left[
\ln(\omega+i\epsilon)\mathcal C_{{\rm out},\,{\rm even}}^\mu(\mathbf n)
\right.
\non\\
&&\left.\hspace{4.5cm}
+\ln(\omega-i\epsilon)\mathcal C_{{\rm in},\,{\rm even}}^\mu(\mathbf n)
\right].
\eea
Applying the radiative kernel \eqref{eq:AJ_relation} to
\eqref{eq:Jacc_even_log} gives
\be\label{eq:A_acc_log_frequency}
\boxed{\begin{aligned}
\wt{A}_{\rm acc,\, log}^\mu(\omega,\vec x)
&\overset{d=\text{even}}{\simeq}
\f{e^{i\omega r}}{r^\f{d-2}{2}}\,
\f{i^\f{d-4}{2}\omega^\f{3(d-4)}{2}}
{(2\pi)^\f{d-4}{2}(4\pi)^\f{d}{2}
\Gamma\left(\f{d-2}{2}\right)}
\\[-1mm]
&\quad\times
\left[
\ln(\omega+i\epsilon)\mathcal C_{{\rm out},\,{\rm even}}^\mu(\mathbf n)
+\ln(\omega-i\epsilon)\mathcal C_{{\rm in},\,{\rm even}}^\mu(\mathbf n)
\right].
\end{aligned}}
\ee

\paragraph{Retarded-time waveform:}
The logarithms carry the retarded prescriptions inherited from the outgoing and incoming
proper-time integrals.  The transforms below are understood in the hierarchy
\eqref{eq:asymptotic_waveform_window}.  With the Fourier
convention \eqref{AAtilde}, and away from the contact terms at $u=0$,
\bea\label{eq:log_basic_transforms}
\int_{-\infty}^{\infty}\f{d\omega}{2\pi}e^{-i\omega u}\ln(\omega+i\epsilon)
=-\f{\Theta(u)}{u},\qquad
\int_{-\infty}^{\infty}\f{d\omega}{2\pi}e^{-i\omega u}\ln(\omega-i\epsilon)
=\f{\Theta(-u)}{u}.
\eea
These are the same regulated large-$|u|$ transforms used in
\cite{Saha:2019tub,Sahoo:2020ryf}.  Applying \(3(d-4)/2\) derivatives to
\eqref{eq:log_basic_transforms} gives the following away from $u=0$
\bea\label{eq:log_power_transforms}
\int_{-\infty}^{\infty}\f{d\omega}{2\pi}e^{-i\omega u}
\omega^\f{3(d-4)}{2}\ln(\omega+i\epsilon)
&\simeq&
(-1)^{\f{3(d-4)}{2}+1}i^\f{3(d-4)}{2}
\Gamma\left(\f{3d-10}{2}\right)
\f{\Theta(u)}{u^\f{3d-10}{2}},\non\\
\int_{-\infty}^{\infty}\f{d\omega}{2\pi}e^{-i\omega u}
\omega^\f{3(d-4)}{2}\ln(\omega-i\epsilon)
&\simeq&
\,-i^\f{3(d-4)}{2}
\Gamma\left(\f{3d-10}{2}\right)
\f{\Theta(-u)}{|u|^\f{3d-10}{2}}.
\eea
Here \(\simeq\) denotes equality modulo the contact terms at \(u=0\) generated
when the derivatives act on \(\Theta(\pm u)\).
Equations \eqref{eq:A_acc_log_frequency}--\eqref{eq:log_power_transforms} therefore give
the late-time waveform
\be\label{eq:A_acc_log_late}
\boxed{
 A_{\rm acc,\, log}^\mu(x)\big|_{u\to+\infty}^{d=\text{even}}
\simeq
(-1)^\f{d-2}{2}
\f{\Gamma\left(\f{3d-10}{2}\right)}
{(2\pi)^\f{d-4}{2}(4\pi)^\f d2
\Gamma\left(\f{d-2}{2}\right)}
\f{\mathcal C_{{\rm out},\,{\rm even}}^\mu(\mathbf n)}
{r^\f{d-2}{2}u^\f{3d-10}{2}},
}
\ee
and early-time waveform
\be\label{eq:A_acc_log_early}
\boxed{
 A_{\rm acc,\, log}^\mu(x)\big|_{u\to-\infty}^{d=\text{even}}
\simeq
-
\f{\Gamma\left(\f{3d-10}{2}\right)}
{(2\pi)^\f{d-4}{2}(4\pi)^\f d2
\Gamma\left(\f{d-2}{2}\right)}
\f{\mathcal C_{{\rm in},\,{\rm even}}^\mu(\mathbf n)}
{r^\f{d-2}{2}|u|^\f{3d-10}{2}},
}
\ee
At \(d=4\), these expressions reduce to the electromagnetic waveforms of
\cite{Sahoo:2018lxl,Saha:2019tub,Sahoo:2020ryf}.

\subsubsection{Odd spacetime dimensions}
The momentum-region analysis leading to \(\omega^{d-4}\ln\omega\) is unchanged in
odd dimensions; the new ingredient is the causal support of the retarded Green's
function and the corresponding pole content of \eqref{eq:Ib_log_integral}.  As shown
by the causal argument following
\eqref{eq:incoming_outgoing_timelike}, the incoming-to-outgoing contribution is absent
in even dimensions but is allowed in odd dimensions, while the reverse mixed ordering
vanishes for the retarded field.  Thus, in odd dimensions, the outgoing proper-time
integral contains same-branch and mixed contributions, whereas the incoming integral
contains only the same-branch contribution.  The corresponding contour derivation and
pole interpretation are given in appendix~\ref{app:acceleration_integral} and
summarized around \eqref{eq:odd_d5_out_in_comparison}.

All three contributions have the same soft scaling, while the same-branch terms retain
the even-dimensional covariant tensor structure with different scalar coefficients.
The evaluations
\eqref{eq:odd_I_out_covariant}, \eqref{eq:odd_I_mix_covariant} and
\eqref{eq:odd_I_in_covariant} give
\bea\label{eq:Ib_odd_log_result}
I_b^\rho(k)\big|_{d=\mathrm{odd},\,\ln}
&\simeq&
\left.I_{b,\,{\rm out}}^\rho(k)\right|_{\ln}
+\left.I_{b,\,{\rm mix}}^\rho(k)\right|_{\ln},
\non\\
\left.I_{b,\,{\rm out}}^\rho(k)\right|_{\ln}
&\simeq&
\f{i\,q_b}{m_b}
\f{\omega^{d-4}\ln(\omega+i\epsilon)}{(2\pi)^{d-1}}
\sum_{\substack{a=1\\a\ne b}}^N q_a
(-\mathbf n.v_b)^{d-4}
\non\\
&&\times
\mathcal F_{\rm out}^{(d)}(v_a.v_b)
\left(v_a^\rho v_b^2-v_a.v_b\,v_b^\rho\right),
\non\\
\left.I_{b,\,{\rm mix}}^\rho(k)\right|_{\ln}
&\simeq&
\f{i\,q_b}{m_b}
\f{\omega^{d-4}\ln(\omega+i\epsilon)}{(2\pi)^{d-1}}
\sum_{a=1}^{M} q'_a
(-\mathbf n.v_b)^{d-4}
\non\\
&&\times
\mathcal F_{\rm mix}^{(d)}(-v'_a.v_b)
\left(v_a^{\prime\rho}v_b^2-v'_a.v_b\,v_b^\rho\right),
\non\\
I_b^{\prime\rho}(k)\big|_{d=\mathrm{odd},\,\ln}
&\simeq&
\f{i\,q'_b}{m'_b}
\f{\omega^{d-4}\ln(\omega-i\epsilon)}{(2\pi)^{d-1}}
\sum_{\substack{a=1\\a\ne b}}^M q'_a
(-\mathbf n.v'_b)^{d-4}
\non\\
&&\times
\mathcal F_{\rm in}^{(d)}(v'_a.v'_b)
\left(v_a^{\prime\rho}v_b^{\prime2}
-v'_a.v'_b\,v_b^{\prime\rho}\right).
\eea
The invariant functions appearing in \eqref{eq:Ib_odd_log_result} are
\be\label{eq:odd_F_main}
\begin{aligned}
\mathcal F_{\rm out}^{(d)}(z)
&=
\f{2\pi^\f{d-2}{2}z}{z^2-1}
\left[
\f{\Gamma\left(\f12\right)}
{\Gamma\left(\f{d-1}{2}\right)}
+\sum_{j=1}^{\f{d-3}{2}}
\f{\Gamma\left(\f12-j\right)}
{\Gamma\left(\f{d-1}{2}-j\right)}
\f{z^{2j-2}}{(z^2-1)^j}
\right],
\\
\mathcal F_{\rm in}^{(d)}(z)
&=
\f{2(-1)^\f{d-3}{2}\pi^\f d2}
{\Gamma\left(\f{d-2}{2}\right)
(z^2-1)^\f{d-1}{2}},
\\
\mathcal F_{\rm mix}^{(d)}(z)
&=
\f{2\pi^{\frac d2-1}}
{(d-3)\Gamma\left(\f{d-2}{2}\right)}
\left[
(d-3)\mathcal H_d(z)+z\,\mathcal H_d'(z)
\right],
\\
\mathcal H_d(z)
&=
\f{1}{z}\,
\mathrm B\left(\f12,\f{d-2}{2}\right)
{}_2F_1\left(
1,\f12;\f{d-1}{2};1-\f{1}{z^2}
\right)
\\
&=
\f{\pi}{2^{d-3}}
\left[
\binom{d-3}{\frac{d-3}{2}}
+2\sum_{j=1}^{\frac{d-3}{2}}(-1)^j
\binom{d-3}{\frac{d-3}{2}-j}
\left(\f{z-1}{z+1}\right)^j
\right],\qquad d\ \mathrm{odd}.
\end{aligned}
\ee
Here \(\mathrm B\) denotes the Euler beta function.  The second representation of
\(\mathcal H_d\) in \eqref{eq:odd_F_main} is a finite
series in \((z-1)/(z+1)\).
Appendix~\ref{app:acceleration_integral} derives the three expressions in
\eqref{eq:odd_F_main}; see \eqref{eq:odd_F_out}, \eqref{eq:odd_F_mix} and
\eqref{eq:odd_F_in}.
For normalized velocities, each factor \(z^2-1\) in \eqref{eq:odd_F_main} is the
corresponding relative-velocity invariant: \((v_a.v_b)^2-v_a^2v_b^2\) for an outgoing
pair, its primed counterpart for an incoming pair, and
\((v'_a.v_b)^2-v_a^{\prime2}v_b^2\) for a mixed pair.  Their collinear
representations are displayed in \eqref{eq:relative_velocity_invariant},
\eqref{eq:odd_incoming_collinear_invariant} and
\eqref{eq:odd_mix_invariant}, respectively.  Thus the even- and odd-dimensional results use
the same invariant and tensor structures for each same-branch pair, while their scalar
coefficient functions differ.  In particular, the propagator-pole contribution to
\(\mathcal F_{\rm mix}^{(d)}\) is the momentum-space manifestation of the
odd-dimensional tail.  The relation among the three pole contributions and their
causal interpretation is summarized around
\eqref{eq:odd_d5_out_in_comparison}.

\paragraph{Frequency-space waveform:}
The tensor numerators in \eqref{eq:Ib_odd_log_result} are the same as in
\eqref{eq:Ib_log_result}, so their contraction with the transverse projector is given by
\eqref{eq:projected_tensor} and by the same identity with \(v_a^\rho\) replaced by
\(v_a^{\prime\rho}\).  Using these identities and relabeling \(a\) as the accelerated
particle (outgoing in the outgoing and mixed terms, and incoming in the incoming
term), we define
\bea\label{eq:odd_hard_tensors}
\mathcal C_{{\rm out},\,{\rm odd}}^\mu(\mathbf n)
&\equiv&
\sum_{a=1}^{N}\sum_{\substack{b=1\\b\ne a}}^N
\f{q_a^2q_b\,(-p_a.\mathbf n)^{d-4}}{m_bm_a^{d-3}}\,
\mathcal F_{\rm out}^{(d)}
\left(\f{p_a.p_b}{m_am_b}\right)
\left(p_b^\mu-\f{p_b.\mathbf n}{p_a.\mathbf n}p_a^\mu\right),
\non\\
\mathcal C_{{\rm in},\,{\rm odd}}^\mu(\mathbf n)
&\equiv&
\sum_{a=1}^{M}\sum_{\substack{b=1\\b\ne a}}^M
\f{q_a^{\prime 2}q'_b\,(-p'_a.\mathbf n)^{d-4}}
{m'_bm_a^{\prime d-3}}\,
\mathcal F_{\rm in}^{(d)}
\left(\f{p'_a.p'_b}{m'_am'_b}\right)
\left(p_b^{\prime\mu}
-\f{p'_b.\mathbf n}{p'_a.\mathbf n}p_a^{\prime\mu}\right),
\non\\
\mathcal C_{{\rm mix},\,{\rm odd}}^\mu(\mathbf n)
&\equiv&
\sum_{a=1}^{N}\sum_{b=1}^{M}
\f{q_a^2q'_b\,(-p_a.\mathbf n)^{d-4}}
{m'_bm_a^{d-3}}\,
\mathcal F_{\rm mix}^{(d)}
\left(
-\f{p'_b.p_a}{m'_bm_a}
\right)
\left(
p_b^{\prime\mu}
-\f{p'_b.\mathbf n}{p_a.\mathbf n}p_a^\mu
\right).
\eea
Substituting \eqref{eq:Ib_odd_log_result} into \eqref{eq:Jacc_int} and using
\eqref{eq:projected_tensor} and \eqref{eq:odd_hard_tensors} gives
\bea\label{eq:Jacc_odd_log}
\widehat J_{\rm acc}^\mu(k)\big|_{d=\mathrm{odd},\,\ln}
&\simeq&
-\f{i\,\omega^{d-4}}{(2\pi)^{d-1}}
\Bigg[
\ln(\omega+i\epsilon)
\left(
\mathcal C_{{\rm out},\,{\rm odd}}^\mu(\mathbf n)
+\mathcal C_{{\rm mix},\,{\rm odd}}^\mu(\mathbf n)
\right)
\non\\
&&\hspace{1.5cm}
+\ln(\omega-i\epsilon)\mathcal C_{{\rm in},\,{\rm odd}}^\mu(\mathbf n)
\Bigg].
\eea
Substituting \eqref{eq:Jacc_odd_log} into the radiative relation
\eqref{eq:AJ_relation} gives the odd-dimensional frequency-space waveform
\be\label{eq:A_acc_odd_frequency}
\boxed{\begin{aligned}
\wt A_{\rm acc,\,log}^\mu(\omega,\vec x)\big|_{d=\mathrm{odd}}
&\simeq
-\f{i\,e^{i\omega r}}{4\pi r^\f{d-2}{2}}
\f{(\omega+i\epsilon)^\f{d-4}{2}\omega^{d-4}}
{(2\pi i)^\f{d-4}{2}(2\pi)^{d-1}}
\\[-1mm]
&\quad\times
\left[
\ln(\omega+i\epsilon)
\left(
\mathcal C_{{\rm out},\,{\rm odd}}^\mu(\mathbf n)
+\mathcal C_{{\rm mix},\,{\rm odd}}^\mu(\mathbf n)
\right)
+\ln(\omega-i\epsilon)\mathcal C_{{\rm in},\,{\rm odd}}^\mu(\mathbf n)
\right].
\end{aligned}}
\ee

\paragraph{Logarithmically enhanced retarded-time waveform:}
Using the Fourier convention \eqref{AAtilde}, we next transform
\eqref{eq:A_acc_odd_frequency} to retarded time.  The required
branch-sensitive Fourier transforms are derived in
appendix~\ref{app:odd_fourier_transform}; see
\eqref{eq:odd_power_transform}, \eqref{eq:odd_log_power_transform} and
\eqref{eq:odd_negative_frequency_transform}.
Applying \eqref{eq:odd_log_power_transform} to the logarithmic integral appearing in
\eqref{eq:A_acc_odd_frequency} gives
\be\label{eq:odd_log_transform_base}
\int_{-\infty}^{\infty}\f{d\omega}{2\pi}e^{-i\omega u}
(\omega+i\epsilon)^\f{d-4}{2}\ln(\omega+i\epsilon)
=
\f{e^{\f{i\pi}{4}(d-4)}}{\Gamma\left(\f{4-d}{2}\right)}
\f{\Theta(u)}{u^\f{d-2}{2}}
\left[
\f{\Gamma'\left(\f{4-d}{2}\right)}{\Gamma\left(\f{4-d}{2}\right)}
+\f{i\pi}{2}-\ln u
\right].
\ee
Multiplication by \(\omega^{d-4}\) is equivalent to applying
\(i^{d-4}\partial_u^{d-4}\).  At late positive retarded time, the term proportional
to \(\ln u/u^{(3d-10)/2}\) dominates the accompanying pure power-law term
\(u^{-(3d-10)/2}\).  To extract this leading late-time behavior, we therefore retain
only the coefficient of \(\ln u\) and find
\bea\label{eq:odd_log_transform_coefficient}
\int_{-\infty}^{\infty}\f{d\omega}{2\pi}e^{-i\omega u}
(\omega+i\epsilon)^\f{d-4}{2}\omega^{d-4}
\ln(\omega+i\epsilon)\bigg|_{\ln u}
&=&
-\f{\Gamma\left(\f{3d-10}{2}\right)}{\pi}
e^{-\f{i\pi}{2}\left(\f{3d-10}{2}\right)}
\f{\Theta(u)\ln u}{u^\f{3d-10}{2}}.
\eea
The two logarithmic prescriptions in \eqref{eq:A_acc_odd_frequency} are related by
\be\label{eq:odd_log_prescription_relation}
\ln(\omega-i\epsilon)
=\ln(\omega+i\epsilon)-2\pi i\Theta(-\omega).
\ee
The transform of the second term in \eqref{eq:odd_log_prescription_relation} follows
from the piecewise evaluation \eqref{eq:odd_negative_frequency_piecewise}, or
equivalently \eqref{eq:odd_negative_frequency_transform}: setting
\(\alpha=(d-4)/2\) there and then applying \(i^{d-4}\partial_u^{d-4}\) separately to
the \(u>0\) and \(u<0\) branches gives
\bea\label{eq:odd_prescription_difference_transform}
&&\int_{-\infty}^{\infty}\f{d\omega}{2\pi}e^{-i\omega u}
(\omega+i\epsilon)^\f{d-4}{2}\omega^{d-4}\Theta(-\omega)
\non\\
&&\qquad=
-\f{\Gamma\left(\f{3d-10}{2}\right)}{2\pi}
e^{i\pi(d-4)/2}
\left[
e^{\f{i\pi}{2}\left(\f{3d-10}{2}\right)}
\f{\Theta(u)}{u^\f{3d-10}{2}}
+e^{-\f{i\pi}{2}\left(\f{3d-10}{2}\right)}
\f{\Theta(-u)}{|u|^\f{3d-10}{2}}
\right].
\eea
Equation \eqref{eq:odd_prescription_difference_transform} is a pure power-law tail and
contains no \(\ln|u|\).  Consequently, both logarithms in
\eqref{eq:A_acc_odd_frequency} have the same logarithmically enhanced transform
\eqref{eq:odd_log_transform_coefficient}, although their non-logarithmic power-law
parts differ.
Combining \eqref{eq:A_acc_odd_frequency} and
\eqref{eq:odd_log_transform_coefficient}, including all phases, gives
\bea\label{eq:A_acc_odd_time}
&&A_{\rm acc,\,log}^\mu(x)\big|_{d=\mathrm{odd},\,\ln|u|}\non\\
&\simeq&
-\f{\Gamma\left(\f{3d-10}{2}\right)}
{4\pi^2(2\pi)^\f{3(d-2)}{2}}
\f{\Theta(u)\ln u}
{r^\f{d-2}{2}u^\f{3d-10}{2}}
\left[
\mathcal C_{{\rm out},\,{\rm odd}}^\mu(\mathbf n)
+\mathcal C_{{\rm mix},\,{\rm odd}}^\mu(\mathbf n)
+\mathcal C_{{\rm in},\,{\rm odd}}^\mu(\mathbf n)
\right].
\eea
Equation \eqref{eq:A_acc_odd_time} fixes the logarithmically enhanced term at late
positive retarded time.  Non-logarithmic terms at the same order in the asymptotic
current are analytic at \(\omega=0\).  In odd dimensions the retarded radiative
kernel converts them into power-law tails supported at \(u>0\), as follows from
\eqref{eq:odd_power_transform}.  Their coefficient is not determined by the
logarithmic calculation, but they are subleading to the logarithm by a factor of
\(1/\ln u\).  At early times \eqref{eq:A_acc_odd_time} vanishes.  The
\(\Theta(-\omega)\) term in \eqref{eq:odd_log_prescription_relation}, however, has
support for \(u<0\) through \eqref{eq:odd_prescription_difference_transform} and
fixes the early-time power-law coefficient.  The two asymptotic limits are therefore
\be\label{eq:A_acc_odd_late_early}
\boxed{\begin{aligned}
A_{\rm acc}^\mu(x)\big|_{\substack{d=\mathrm{odd}\\u\to+\infty}}
&\simeq
-\f{\Gamma\left(\f{3d-10}{2}\right)}
{4\pi^2(2\pi)^\f{3(d-2)}{2}}
\f{\ln u}
{r^\f{d-2}{2}u^\f{3d-10}{2}}
\\[-1mm]
&\quad\times
\left[
\mathcal C_{{\rm out},\,{\rm odd}}^\mu(\mathbf n)
+\mathcal C_{{\rm mix},\,{\rm odd}}^\mu(\mathbf n)
+\mathcal C_{{\rm in},\,{\rm odd}}^\mu(\mathbf n)
\right]
+\mathcal O\left(
\f{1}{r^\f{d-2}{2}u^\f{3d-10}{2}}
\right),
\\
A_{\rm acc}^\mu(x)\big|_{\substack{d=\mathrm{odd}\\u\to-\infty}}
&\simeq
\f{(-1)^\f{d-3}{2}\Gamma\left(\f{3d-10}{2}\right)}
{4\pi(2\pi)^\f{3(d-2)}{2}}
\f{\mathcal C_{{\rm in},\,{\rm odd}}^\mu(\mathbf n)}
{r^\f{d-2}{2}|u|^\f{3d-10}{2}} .
\end{aligned}}
\ee
The expressions for the three \(\mathcal C^\mu\)'s are given in
\eqref{eq:odd_hard_tensors}, and both limits in
\eqref{eq:A_acc_odd_late_early} are understood in the hierarchy
\eqref{eq:asymptotic_waveform_window}.  The correction displayed in the first line is
a non-logarithmic late-time term whose coefficient is not fixed here.  It does not
affect the universal coefficient of the dominant logarithmic tail.  The second line
is fixed by the negative-frequency discontinuity and cannot be modified by analytic
terms of the same soft order.  Since the mixed incoming-to-outgoing contribution
contains only \(\ln(\omega+i\epsilon)\), it contributes to the late-time logarithmic
tail but not to the early-time result.  The dimensionful notation \(\ln u\) means
\(\ln(u/u_0)\) for a fixed \(u_0\) of order \(\mathcal R\).  Changing \(u_0\)
only shifts the undetermined non-logarithmic late-time coefficient at the same
power of \(u\).

\subsection{Independent position-space derivation of the even-dimensional tail}
\label{S:position_space_even_tail}

The even-dimensional result above was obtained by first isolating the non-analytic
part of the current and then Fourier transforming the radiative field.  We now give
an independent derivation entirely in position space.  It starts from the
large-proper-time force, determines the corresponding trajectory, and localizes the
retarded field directly on the past light cone of the detector. 

\paragraph{Asymptotic trajectory:}
For even \(d\geq6\), the position-space retarded Green's function, with the
convention \(\Box G_R(x)=\delta^{(d)}(x)\) used in \eqref{eq:box_G}, is
\be\label{eq:position_even_green}
G_R(x)
=-\f{\Theta(x^0)}{2\pi^{(d-2)/2}}\,
\delta^{\left((d-4)/2\right)}
\left((x^0)^2-\vec x^{\,2}\right).
\ee
Here the superscript on the one-variable delta function denotes differentiation
with respect to its argument.  At sufficiently late time, the retarded root on
the source trajectory lies in the interior of \(\tau_a>0\); the lower endpoint
of that trajectory then contributes only a finite-time wavefront.  Convolving
\eqref{eq:position_even_green} with the straight outgoing source in
\eqref{eq:current_st} and differentiating gives the boosted Coulomb field
\be\label{eq:position_even_uniform_field}
F_{(a)}^{\rho\sigma}(x)
=
\f{q_a\Gamma\!\left(\f{d-1}{2}\right)}
{2\pi^{(d-1)/2}}\,
\f{v_a^\rho(x-r_a)^\sigma-v_a^\sigma(x-r_a)^\rho}
{\left[
\left(v_a.(x-r_a)\right)^2-v_a^2(x-r_a)^2
\right]^{(d-1)/2}} .
\ee
On the straight trajectory of particle \(b\), the numerator and denominator in
\eqref{eq:position_even_uniform_field} have the respective large-\(\tau_b\)
expansions
\be\label{eq:position_even_field_expansion}
\begin{aligned}
v_{b\sigma}
\left[v_a^\rho(r_{ba}+v_b\tau_b)^\sigma
-v_a^\sigma(r_{ba}+v_b\tau_b)^\rho\right]
&=
\tau_b\left[v_a^\rho v_b^2-(v_a.v_b)v_b^\rho\right]
+\mathcal O(\mathcal R),
\\
\left[v_a.(r_{ba}+v_b\tau_b)\right]^2
-v_a^2(r_{ba}+v_b\tau_b)^2
&=
\left[(v_a.v_b)^2-v_a^2v_b^2\right]\tau_b^2
+\mathcal O(\mathcal R\tau_b)+\mathcal O(\mathcal R^2).
\end{aligned}
\ee
Thus, for fixed non-degenerate relative velocities, the expansion applies for
\(\tau_b\) larger than a fixed multiple of \(\mathcal R\).  Using the gamma-function
identity
\be\label{eq:position_even_coulomb_normalization}
\f{\Gamma\!\left(\f{d-1}{2}\right)}{2\pi^{(d-1)/2}}
=
\f{(d-3)(d-4)!}
{(4\pi)^{(d-2)/2}\Gamma\!\left(\f{d-2}{2}\right)},
\ee
the leading contracted field is
\be\label{eq:position_even_contracted_field_tail}
\begin{aligned}
v_{b\sigma}F_{(a)}^{\rho\sigma}(r_b+v_b\tau_b)
&=
\f{q_a(d-3)(d-4)!}
{(4\pi)^{(d-2)/2}\Gamma\left(\f{d-2}{2}\right)}
\f{v_a^\rho v_b^2-(v_a.v_b)v_b^\rho}
{\left[(v_a.v_b)^2-v_a^2v_b^2\right]^{(d-1)/2}}
\f{1}{\tau_b^{d-2}}
\\
&\quad+\mathcal O\left(\tau_b^{-(d-1)}\right).
\end{aligned}
\ee
The endpoint positions enter only the displayed remainder.  Substitution into
\eqref{eq:Y_acceleration} gives the acceleration directly:
\be\label{eq:position_even_acceleration_tail}
\begin{aligned}
\ddot Y_b^\rho(\tau_b)
&=\f{q_b}{m_b}\,
\f{(d-3)(d-4)!}
{(4\pi)^\f{d-2}{2}\Gamma\!\left(\f{d-2}{2}\right)}
\sum_{\substack{a=1\\a\ne b}}^N q_a\,
\f{v_a^\rho v_b^2-(v_a.v_b)v_b^\rho}
{\left[(v_a.v_b)^2-v_a^2v_b^2\right]^\f{d-1}{2}}
\f{1}{\tau_b^{d-2}}
+\mathcal O\!\left(\tau_b^{-(d-1)}\right).
\end{aligned}
\ee
Integrating \eqref{eq:position_even_acceleration_tail} with the asymptotic boundary
condition in \eqref{eq:Y_acceleration} then yields
\be\label{eq:position_even_velocity_tail}
\begin{aligned}
\dot Y_b^\rho(\tau_b)
&=-\f{q_b}{m_b}\,
\f{(d-4)!}
{(4\pi)^\f{d-2}{2}\Gamma\!\left(\f{d-2}{2}\right)}
\sum_{\substack{a=1\\a\ne b}}^N q_a\,
\f{v_a^\rho v_b^2-(v_a.v_b)v_b^\rho}
{\left[(v_a.v_b)^2-v_a^2v_b^2\right]^\f{d-1}{2}}
\f{1}{\tau_b^{d-3}}
+\mathcal O\!\left(\tau_b^{-(d-2)}\right).
\end{aligned}
\ee
This tail is integrable for even \(d\geq6\).  Hence \(Y_b^\rho\) approaches a
constant with corrections of order \(\tau_b^{-(d-4)}\).  Absorbing the constant
into \(r_b^\mu\) gives the trajectory and tangent required below:
\be\label{eq:position_even_trajectory}
X_b^\rho(\tau_b)
=r_b^\rho+v_b^\rho\tau_b
+\f{q_b}{m_b}\,
\f{(d-5)!}
{(4\pi)^\f{d-2}{2}\Gamma\!\left(\f{d-2}{2}\right)}
\sum_{\substack{a=1\\a\ne b}}^N q_a\,
\f{v_a^\rho v_b^2-(v_a.v_b)v_b^\rho}
{\left[(v_a.v_b)^2-v_a^2v_b^2\right]^\f{d-1}{2}}
\f{1}{\tau_b^{d-4}}
+\mathcal O\!\left(\tau_b^{-(d-3)}\right).
\ee
Differentiating \eqref{eq:position_even_trajectory} gives
\be\label{eq:position_even_tangent}
\dot X_b^\rho(\tau_b)
=v_b^\rho
-\f{q_b}{m_b}\,
\f{(d-4)!}
{(4\pi)^\f{d-2}{2}\Gamma\!\left(\f{d-2}{2}\right)}
\sum_{\substack{a=1\\a\ne b}}^N q_a\,
\f{v_a^\rho v_b^2-(v_a.v_b)v_b^\rho}
{\left[(v_a.v_b)^2-v_a^2v_b^2\right]^\f{d-1}{2}}
\f{1}{\tau_b^{d-3}}
+\mathcal O\!\left(\tau_b^{-(d-2)}\right).
\ee

\paragraph{Light-cone localization:}
We follow the position-space method of section~3 of
\cite{Karan:2025allorder}, retaining the derivatives of the light-cone delta
function required in even \(d>4\).  At late positive retarded time only outgoing
asymptotic trajectories can meet the detector's past light cone.  Substitution of
their current into \eqref{eq:A_mu_gen} gives
\be\label{eq:position_even_green_convolution}
\begin{aligned}
A^\mu(x)\big|_{\rm out}
&=\f{1}{2\pi^\f{d-2}{2}}
\sum_{b=1}^{N}q_b\int_0^\infty d\tau_b\,
\dot X_b^\mu(\tau_b)\,
\Theta\!\left(x^0-X_b^0(\tau_b)\right)
\\[-1mm]
&\quad\times
\delta^{\left(\f{d-4}{2}\right)}
\!\left((x^0-X_b^0(\tau_b))^2
-|\vec x-\vec X_b(\tau_b)|^2\right).
\end{aligned}
\ee
For \(x^\mu=(u+r,r\hat x)\), the argument of the delta function and the leading
radiative field in the hierarchy \eqref{eq:asymptotic_waveform_window} become
\be\label{eq:position_even_radiation_zone}
\begin{aligned}
(x^0-X_b^0)^2-|\vec x-\vec X_b|^2
&=2r\left(u+\mathbf n.X_b\right)+\mathcal O(u^2),\\
A^\mu(x)\big|_{\rm out}
&\simeq
\f{1}{2(2\pi)^\f{d-2}{2}r^\f{d-2}{2}}
\sum_{b=1}^{N}q_b\int_0^\infty d\tau_b\,
\dot X_b^\mu(\tau_b)
\delta^{\left(\f{d-4}{2}\right)}
\!\left(u+\mathbf n.X_b(\tau_b)\right).
\end{aligned}
\ee
The omitted terms in the first line are suppressed by \(u/r\).  The normalization
in the second line follows from the scaling of the differentiated delta function
and agrees with the direct straight-line result \eqref{eq:A_st_even}.

Let \(\tau_b^{\rm sol}\) denote the unique late-time root on the outgoing
trajectory.  The null constraint gives
\be\label{eq:position_even_retarded_root}
\begin{aligned}
u+\mathbf n.X_b(\tau_b^{\rm sol})&=0,\\
\tau_b^{\rm sol}
&=-\f{u+\mathbf n.r_b}{\mathbf n.v_b}
+\mathcal O\!\left(u^{-(d-4)}\right)
\\
&=-\f{u}{\mathbf n.v_b}
\left[1+\mathcal O\!\left(\f{\mathcal R}{u}\right)\right]
+\mathcal O\!\left(u^{-(d-4)}\right).
\end{aligned}
\ee
Since \(\mathbf n.\dot X_b<0\), changing variables to the light-cone argument
gives, up to endpoint distributions supported at finite \(u\),
\be\label{eq:position_even_localization}
\int_0^\infty d\tau_b\,\dot X_b^\mu(\tau_b)
\delta^{\left(\f{d-4}{2}\right)}
\!\left(u+\mathbf n.X_b(\tau_b)\right)
=-\left(\f{d}{du}\right)^\f{d-4}{2}
\left.
\f{\dot X_b^\mu}{\mathbf n.\dot X_b}
\right|_{\tau_b=\tau_b^{\rm sol}} .
\ee
This is the even-dimensional extension of the worldline Jacobian in
\cite{Karan:2025allorder}.  Equations \eqref{eq:position_even_tangent} and
\eqref{eq:position_even_retarded_root} give
\be\label{eq:position_even_jacobian_tail}
\begin{aligned}
\left.\f{\dot X_b^\mu}{\mathbf n.\dot X_b}
\right|_{\tau_b=\tau_b^{\rm sol}}
&=\f{v_b^\mu}{\mathbf n.v_b}
+\f{q_b}{m_b}\,
\f{(d-4)!}
{(4\pi)^\f{d-2}{2}\Gamma\!\left(\f{d-2}{2}\right)}
\f{(\mathbf n.v_b)^{d-4}}{u^{d-3}}
\\
&\quad\times
\left(\delta^\mu_\rho
-\f{v_b^\mu\mathbf n_\rho}{\mathbf n.v_b}\right)
\sum_{\substack{a=1\\a\ne b}}^N q_a\,
\f{v_a^\rho v_b^2-(v_a.v_b)v_b^\rho}
{\left[(v_a.v_b)^2-v_a^2v_b^2\right]^\f{d-1}{2}}
+\mathcal O\!\left(u^{-(d-2)}\right).
\end{aligned}
\ee
The first term is annihilated by the derivatives in
\eqref{eq:position_even_localization}; its endpoint contribution is a distribution
supported at retarded times \(u=\mathcal O(\mathcal R)\), set by the hard-scattering
region.  The second term gives the long-time radiative tail,
using
\be\label{eq:position_even_power_derivative}
\left(\f{d}{du}\right)^\f{d-4}{2}
\f{1}{u^{d-3}}
=(-1)^\f{d-4}{2}
\f{\Gamma\!\left(\f{3d-10}{2}\right)}{\Gamma(d-3)}
\f{1}{u^\f{3d-10}{2}}.
\ee

\paragraph{Late- and early-time tails:}
Combining \eqref{eq:position_even_radiation_zone},
\eqref{eq:position_even_localization}, \eqref{eq:position_even_jacobian_tail} and
\eqref{eq:position_even_power_derivative} gives
\be\label{eq:position_even_tail_before_substitution}
\begin{aligned}
A_{\rm acc,\,log}^\mu(x)\big|_{u\to+\infty}^{d=\mathrm{even}}
&\simeq
(-1)^\f{d-2}{2}
\f{\Gamma\!\left(\f{3d-10}{2}\right)}
{2(2\pi)^\f{d-2}{2}(4\pi)^\f{d-2}{2}
\Gamma\!\left(\f{d-2}{2}\right)}
\f{1}{r^\f{d-2}{2}u^\f{3d-10}{2}}
\\
&\quad\times
\sum_{b=1}^{N}\f{q_b^2}{m_b}(\mathbf n.v_b)^{d-4}
\left(\delta^\mu_\rho
-\f{v_b^\mu\mathbf n_\rho}{\mathbf n.v_b}\right)
\sum_{\substack{a=1\\a\ne b}}^N q_a\,
\f{v_a^\rho v_b^2-(v_a.v_b)v_b^\rho}
{\left[(v_a.v_b)^2-v_a^2v_b^2\right]^\f{d-1}{2}} .
\end{aligned}
\ee
Using the contraction \eqref{eq:projected_tensor} and the momentum relation
\eqref{eq:velocity_momentum_invariant} gives
\be\label{eq:position_even_late_tail}
A_{\rm acc,\,log}^\mu(x)\big|_{u\to+\infty}^{d=\mathrm{even}}
\simeq
(-1)^\f{d-2}{2}
\f{\Gamma\!\left(\f{3d-10}{2}\right)}
{(2\pi)^\f{d-4}{2}(4\pi)^\f d2
\Gamma\!\left(\f{d-2}{2}\right)}
\f{\mathcal C_{{\rm out},\,{\rm even}}^\mu(\mathbf n)}
{r^\f{d-2}{2}u^\f{3d-10}{2}}.
\ee
This reproduces \eqref{eq:A_acc_log_late} without passing through frequency space.
For an incoming trajectory the retarded root lies at large negative proper time.
The primed calculation replaces \(u\) by \(|u|\), and
\be\label{eq:position_even_incoming_derivative}
\left(\f{d}{du}\right)^\f{d-4}{2}
\f{1}{|u|^{d-3}}
=
\f{\Gamma\!\left(\f{3d-10}{2}\right)}{\Gamma(d-3)}
\f{1}{|u|^\f{3d-10}{2}},
\qquad u<0.
\ee
The overall minus sign then comes from the localization identity
\eqref{eq:position_even_localization}.  Thus the direct calculation gives
\be\label{eq:position_even_time_tails}
A_{\rm acc,\,log}^\mu(x)\big|_{d=\mathrm{even}}
\simeq
\f{\Gamma\!\left(\f{3d-10}{2}\right)}
{(2\pi)^\f{d-4}{2}(4\pi)^\f d2
\Gamma\!\left(\f{d-2}{2}\right)}
\f{1}{r^\f{d-2}{2}}
\begin{cases}
\displaystyle
(-1)^\f{d-2}{2}
\f{\mathcal C_{{\rm out},\,{\rm even}}^\mu(\mathbf n)}
{u^\f{3d-10}{2}},
&u\to+\infty,\\[3mm]
\displaystyle
-\f{\mathcal C_{{\rm in},\,{\rm even}}^\mu(\mathbf n)}
{|u|^\f{3d-10}{2}},
&u\to-\infty .
\end{cases}
\ee
The vectors are defined in \eqref{eq:hard_tensors}.  Since
\eqref{eq:position_even_green} has support only on the light cone, these are source
tails generated by the indefinitely decaying long-range acceleration.  Moreover, the endpoint positions and the next term in
\eqref{eq:position_even_velocity_tail} are suppressed by one additional inverse
power of proper time.  This gives a position-space explanation for why the source
phase and the non-marginal terms discarded in
\eqref{eq:source_phase_expansion}--\eqref{eq:radial_endpoint_evaluation} cannot change the
leading logarithmic coefficient.
Appendix~\ref{app:position_space_integral_check} makes
the same statement precise in frequency space: the finite-time part gives analytic
terms, while the subleading trajectory remainder has enough moments to exclude an
\(\omega^{d-4}\ln\omega\) contribution.

\section{Quantum logarithmic soft-photon theorem in higher dimensions}
\label{S:quantum_soft}
The retarded classical analysis in section~\ref{S:log_correction} identifies a
logarithmic contribution to the frequency-space waveform arising from the
acceleration-induced current, while
subsection~\ref{S:position_space_even_tail} derives its even-dimensional tail
independently in position space.  Since the same long-range electromagnetic
interaction is represented by the low-energy region of the quantum amplitude, the
classical result should have a counterpart in the one-loop soft factor.  Deriving
that factor provides an independent check of the waveform and clarifies how its
retarded contribution is embedded in the complete quantum amplitude.

Logarithmic soft terms have previously been associated primarily with
four-dimensional scattering, where long-range interactions and infrared
divergences obstruct an expansion in integer powers of the soft energy
\cite{Laddha:2018myi,Sahoo:2018lxl,Krishna:2023log}.  In dimensional
regularization, the four-dimensional \(\ln\omega\) term can be anticipated by
commuting the universal part of the tree-level subleading soft-photon operator
through the infrared-divergent conventional eikonal factor and treating the soft
energy \(\omega\) as the infrared cutoff, thereby replacing the resulting
\(1/\epsilon_{\rm IR}\) pole by \(\ln\omega\)
\cite{Sahoo:2018lxl,Krishna:2023log}.  The same infrared correspondence appears in
eikonal analyses and in Ward-identity interpretations of the logarithmic soft
photon and graviton theorems
\cite{Alessio:2024eikonalsoft,Agrawal:2023logward,Campiglia:2019logward,
Bhatkar:2019logward,Choi:2024logsymmetry,Choi:2024triangle}.  This reasoning has
no analogue for \(d>4\), where that conventional eikonal factor is infrared finite
and supplies no pole that can be converted into \(\ln\omega\).

We show instead that the one-loop soft factor in general \(d>4\) contains a
universal term of order \(\omega^{d-4}\ln\omega\), arising from the same
loop-momentum region as in the four-dimensional case, which must be isolated
directly from the one-loop scattering amplitude involving a soft photon.  This extends the four-dimensional
analysis of \cite{Sahoo:2018lxl}, establishes logarithmic soft behavior beyond four
dimensions, and provides the amplitude counterpart of the classical electromagnetic
waveform derived in section~\ref{S:log_correction}.  Since the \(S\)-matrix is
infrared finite for \(d>4\), the virtual photon propagator may be kept in its
standard Feynman form and the Grammer-Yennie \(K/G\) decomposition
\cite{Grammer:1973gy} is unnecessary.
The comparison between the quantum soft photon factor and the logarithmic
electromagnetic waveform is especially informative in odd dimensions, where the
outgoing, incoming, and mixed classical terms have different causal origins.  In
subsection~\ref{S:quantum_classical_comparison} we trace these terms to the
corresponding pole contributions, assemble the even- and odd-dimensional classical
soft factors explicitly in the all-outgoing notation, and distinguish the retarded
result from the additional Feynman photon-pole contribution.

\subsection{Scattering setup}

The scattering process considered here is the quantum counterpart of the classical
scattering configuration introduced in section~\ref{S:setup}.  In both descriptions,
\(M\) incoming and \(N\) outgoing massive charged particles interact within a compact
region and emit a soft photon.  The asymptotic worldlines
\eqref{eq:asymptotic_trajectory} are now replaced by external one-particle scalar states,
while the dynamics inside the compact interaction region is encoded in the hard
scattering amplitude generated by \(\mathcal L_{\rm hard}\) as described below.  Correspondingly, the
classical radiative field measured near future null infinity is replaced by an outgoing
soft-photon state of momentum \(k^\mu\) and polarization \(\varepsilon^\mu\).

More explicitly, the \(N\) outgoing quantum states carry the same physical data
\((m_a,p_a^\mu,q_a)\) as the outgoing particles in
\eqref{eq:hard_momenta}, whereas the \(M\) incoming quantum states carry
\((m'_a,p_a^{\prime\mu},q'_a)\).  Thus the quantum process contains
\(n=N+M\) hard external scalar states in addition to the soft photon.  We order the
outgoing states first and combine the primed and unprimed asymptotic data into the
all-outgoing scattering parameters
\bea\label{eq:quantum_eta_convention}
\eta_a&=&
\begin{cases}
+1,&1\leq a\leq N,\\
-1,&N+1\leq a\leq N+M,
\end{cases}\qquad \mathfrak m_a\equiv
\begin{cases}
m_a,&1\leq a\leq N,\\
m'_{a-N},&N+1\leq a\leq N+M,
\end{cases}
\non\\
P_a^\mu&\equiv&
\begin{cases}
p_a^\mu,&1\leq a\leq N,\\
-p_{a-N}^{\prime\mu},&N+1\leq a\leq N+M,
\end{cases}
\qquad
Q_a\equiv
\begin{cases}
q_a,&1\leq a\leq N,\\
-q'_{a-N},&N+1\leq a\leq N+M.
\end{cases}
\eea
The signed momentum \(P_a^\mu\) therefore has positive energy for an outgoing particle
and negative energy for an incoming particle, while
\(\eta_aP_a^\mu\) is always the corresponding future-directed physical momentum.
Similarly, \(Q_a\) is the charge in the all-outgoing convention, and
\(\eta_aQ_a\) is the physical charge of the corresponding asymptotic particle.  This
identification keeps the classical derivation of
section~\ref{S:setup}--\ref{S:log_correction} and the quantum
Feynman-diagram derivation of the present section in direct parallel.  All hard lines can
now be treated on the same footing:
\be\label{eq:quantum_all_outgoing_constraints}
P_a^2=-\mathfrak m_a^2,\qquad
\sum_{a=1}^{N+M}P_a^\mu+k^\mu=0,\qquad
\sum_{a=1}^{N+M}Q_a=0,\qquad k^2=0 .
\ee
The soft photon is outgoing, with \(k^\mu=\omega\mathbf n^\mu\), \(\omega>0\), and
polarization \(\varepsilon^\mu\) satisfying
\be\label{eq:quantum_soft_polarization}
k.\varepsilon=0 .
\ee

For definiteness, we take the hard scattering to be generated by a non-derivative contact
interaction involving all \(N+M\) scalar fields.  The gauge-fixed Lagrangian is
\bea\label{eq:scalar_qed_lagrangian}
\mathcal L
&=&-\f14F_{\mu\nu}F^{\mu\nu}
-\f12(\p_\mu A^\mu)^2
-\sum_{a=1}^{N+M}\left[
(D_\mu\phi_a)^*D^\mu\phi_a+\mathfrak m_a^2\phi_a^*\phi_a
\right]
+\mathcal L_{\rm hard},
\non\\
D_\mu\phi_a&\equiv&(\p_\mu-iQ_aA_\mu)\phi_a,
\quad
\mathcal L_{\rm hard}
=\lambda\,\phi_1\phi_2\cdots\phi_{N+M}
+\lambda\,\phi_1^*\phi_2^*\cdots\phi_{N+M}^*.
\eea
Here \(\phi_a\) is the scalar field with mass
\(\mathfrak m_a\) and all-outgoing charge \(Q_a\) as defined in
\eqref{eq:quantum_eta_convention}, and \(\lambda\) is taken to be real.  Charge conservation makes
\(\mathcal L_{\rm hard}\) gauge invariant.  Since the interaction contains no
derivatives, it introduces no additional photon vertex, and the hard tree amplitude
is \(\mathcal M_n^{(0)}=i\lambda\).  The choice of a contact interaction only
simplifies the bookkeeping; the non-analytic contribution derived below depends
exclusively on the external scalar lines.
The gauge-fixing term in \eqref{eq:scalar_qed_lagrangian} puts the photon in Feynman
gauge.  The momentum assignments and the corresponding Feynman rules are displayed
together in \eqref{eq:scalar_qed_feynman_rules}:
\be\label{eq:scalar_qed_feynman_rules}
\begin{array}{c@{\;\;}c@{\qquad\qquad}c@{\;\;}c}
\vcenter{\hbox{\begin{tikzpicture}[
  x=1cm,y=1cm,
  photon/.style={line width=0.32mm,decorate,
    decoration={snake,amplitude=0.55mm,segment length=2.2mm},line cap=round},
  momentum/.style={-{Stealth[length=1.5mm,width=1.15mm]},line width=0.23mm}
]
\draw[photon] (-1.05,0) -- (1.05,0);
\draw[momentum] (-0.30,0.27) -- (0.30,0.27);
\node[above] at (0,0.30) {\(\scriptstyle \ell\)};
\node[left] at (-1.05,0) {\(\scriptstyle \mu\)};
\node[right] at (1.05,0) {\(\scriptstyle \nu\)};
\end{tikzpicture}}}
&=\displaystyle\f{-i\eta_{\mu\nu}}{\ell^2-i\epsilon}
&
\vcenter{\hbox{\begin{tikzpicture}[
  x=1cm,y=1cm,
  scalar/.style={line width=0.58mm,line cap=round},
  charge flow/.style={postaction={decorate},
    decoration={markings,mark=at position 0.82 with
      {\arrow{Stealth[length=1.6mm,width=1.25mm]}}}},
  momentum/.style={-{Stealth[length=1.5mm,width=1.15mm]},line width=0.23mm}
]
\draw[scalar,charge flow] (-1.05,0) -- (1.05,0);
\draw[momentum] (-0.30,0.27) -- (0.30,0.27);
\node[above] at (0,0.30) {\(\scriptstyle r\)};
\node[below] at (0,-0.10) {\(\scriptstyle \phi_a\)};
\end{tikzpicture}}}
&=\displaystyle\f{-i}{r^2+\mathfrak m_a^2-i\epsilon}
\\[1.2cm]
\vcenter{\hbox{\begin{tikzpicture}[
  x=1cm,y=1cm,
  scalar/.style={line width=0.58mm,line cap=round},
  charge flow/.style={postaction={decorate},
    decoration={markings,mark=at position 0.82 with
      {\arrow{Stealth[length=1.6mm,width=1.25mm]}}}},
  photon/.style={line width=0.32mm,decorate,
    decoration={snake,amplitude=0.55mm,segment length=2.2mm},line cap=round},
  momentum/.style={-{Stealth[length=1.5mm,width=1.15mm]},line width=0.23mm}
]
\draw[scalar,charge flow] (-1.15,0) -- (1.15,0);
\draw[photon] (0,0) -- (0,1.05);
\fill (0,0) circle (0.055);
\draw[momentum] (-0.92,0.25) -- (-0.42,0.25);
\draw[momentum] (0.42,0.25) -- (0.92,0.25);
\draw[momentum] (0.23,0.36) -- (0.23,0.82);
\node[above] at (-0.67,0.28) {\(\scriptstyle r\)};
\node[above] at (0.67,0.28) {\(\scriptstyle r'\)};
\node[right] at (0.24,0.70) {\(\scriptstyle q\)};
\node[above] at (0,1.07){\(\mu\)};
\node[below] at (0,-0.17) {\(\scriptstyle r=r'+q\)};
\end{tikzpicture}}}
&=iQ_a(r+r')^\mu
&
\vcenter{\hbox{\begin{tikzpicture}[
  x=1cm,y=1cm,
  scalar/.style={line width=0.58mm,line cap=round},
  charge flow/.style={postaction={decorate},
    decoration={markings,mark=at position 0.82 with
      {\arrow{Stealth[length=1.6mm,width=1.25mm]}}}},
  photon/.style={line width=0.32mm,decorate,
    decoration={snake,amplitude=0.55mm,segment length=2.2mm},line cap=round},
  momentum/.style={-{Stealth[length=1.5mm,width=1.15mm]},line width=0.23mm}
]
\draw[scalar,charge flow] (-1.05,0) -- (1.05,0);
\draw[photon] (0,0) -- (-0.62,0.92);
\draw[photon] (0,0) -- (0.62,0.92);
\fill (0,0) circle (0.055);
\node[left] at (-0.62,0.92) {\(\scriptstyle \mu\)};
\node[right] at (0.62,0.92) {\(\scriptstyle \nu\)};
\end{tikzpicture}}}
&=-2iQ_a^2\eta^{\mu\nu}
\end{array}
\ee
The arrows on the solid scalar lines in
\eqref{eq:scalar_qed_feynman_rules} indicate charge flow, while the displaced arrows
define the displayed momentum flow.  At the three-point vertex, \(r\) enters along
the scalar line, while \(r'\) and \(q\) leave the vertex, so that \(r=r'+q\).

Let \(\mathcal M_n\) denote the amplitude without the soft photon and
\(\mathcal M_{n+1}(k,\varepsilon)\) the amplitude with it.  At tree level their leading
soft relation is
\bea\label{eq:quantum_tree_soft_factor}
\mathcal M_{n+1}^{(0)}(k,\varepsilon)
&=&
\left[
\sum_{a=1}^{N+M}Q_a\,\f{\varepsilon.P_a}{k.P_a}
\right]\mathcal M_n^{(0)}
+\mathcal O(\omega^0).
\eea
In the next subsection, we compute the leading non-analytic contribution to the
one-loop soft photon factor as \(\omega\to0\).  Here ``leading non-analytic'' refers
only to the non-analytic sector of the soft expansion.  Non-analyticity is understood in the sense of
footnote~\ref{fn:nonanalytic_definition}: the contribution does not admit a Taylor
expansion in a neighbourhood of \(\omega=0\). Between the leading
\(1/\omega\) pole and the term of order \(\omega^{d-4}\ln\omega\), analytic terms
may occur at every integer power from \(\omega^0\) through \(\omega^{d-5}\), together
with a non-logarithmic term at order \(\omega^{d-4}\); none of these coefficients is
determined here. 

\subsection{One-loop diagrams and the logarithmic region}

\begin{figure}[t]
\centering
\begin{tikzpicture}[
  x=1cm,y=1cm,
  scalar/.style={line width=0.65mm,line cap=round},
  outward/.style={postaction={decorate},
    decoration={markings,mark=at position 0.84 with
      {\arrow{Stealth[length=1.8mm,width=1.4mm]}}}},
  photon/.style={line width=0.35mm,decorate,
    decoration={snake,amplitude=0.65mm,segment length=2.5mm},line cap=round},
  hard region/.style={pattern=north east lines,line width=0.4mm},
  momentum arrow/.style={-{Stealth[length=1.6mm,width=1.25mm]},
    line width=0.25mm}
]
\begin{scope}[shift={(-5.1,0)}]
\draw[hard region] (0,0) circle (0.27);
\draw[scalar,outward] (0,0) -- (-1.25,2.7);
\draw[scalar,outward] (0,0) -- (1.25,2.7);
\draw[photon] (-0.63,1.36) -- (0.63,1.36);
\draw[momentum arrow] (-0.16,1.16) -- (0.18,1.16);
\draw[photon] (0.91,1.98) -- (2.10,2.35);
\draw[momentum arrow] (1.27,2.27) -- (1.57,2.36);
\node[above left] at (-1.25,2.7) {$P_a$};
\node[above right] at (1.25,2.7) {$P_b$};
\node[above] at (0,1.48) {$\ell$};
\node[right] at (2.10,2.35) {$k$};
\node[left] at (-0.29,0.72) {\scriptsize \(P_a+\ell\)};
\node[right] at (0.22,0.65) {\scriptsize \(P_b+k-\ell\)};
\node[right] at (0.70,1.63) {\scriptsize \(P_b+k\)};
\node[below] at (0,-0.40) {(a)};
\end{scope}
\begin{scope}[shift={(0,0)}]
\draw[hard region] (0,0) circle (0.27);
\draw[scalar,outward] (0,0) -- (-1.25,2.7);
\draw[scalar,outward] (0,0) -- (1.25,2.7);
\draw[photon] (-0.82,1.77) -- (0.82,1.77);
\draw[momentum arrow] (-0.16,1.56) -- (0.18,1.56);
\draw[photon] (0.42,0.90) -- (1.62,0.53);
\draw[momentum arrow] (0.90,0.93) -- (1.20,0.84);
\node[above left] at (-1.25,2.7) {$P_a$};
\node[above right] at (1.25,2.7) {$P_b$};
\node[above] at (0,1.89) {$\ell$};
\node[right] at (1.62,0.53) {$k$};
\node[left] at (-0.34,0.78) {\scriptsize \(P_a+\ell\)};
\node[right] at (0.2,0.25) {\scriptsize \(P_b+k-\ell\)};
\node[right] at (0.60,1.33) {\scriptsize \(P_b-\ell\)};
\node[below] at (0,-0.40) {(b)};
\end{scope}
\begin{scope}[shift={(5.1,0)}]
\draw[hard region] (0,0) circle (0.27);
\draw[scalar,outward] (0,0) -- (-1.25,2.7);
\draw[scalar,outward] (0,0) -- (1.25,2.7);
\coordinate (V) at (0.68,1.47);
\fill (V) circle (0.075);
\draw[photon] (-0.68,1.47) -- (V);
\draw[momentum arrow] (-0.14,1.26) -- (0.20,1.26);
\draw[photon] (V) -- (1.90,1.08);
\draw[momentum arrow] (1.15,1.50) -- (1.45,1.40);
\node[above left] at (-1.25,2.7) {$P_a$};
\node[above right] at (1.25,2.7) {$P_b$};
\node[above] at (-0.05,1.59) {$\ell$};
\node[right] at (1.90,1.08) {$k$};
\node[left] at (-0.30,0.73) {\scriptsize \(P_a+\ell\)};
\node[right] at (0.20,0.66) {\scriptsize \(P_b+k-\ell\)};
\node[below] at (0,-0.40) {(c)};
\end{scope}
\end{tikzpicture}
\caption{One-loop diagrams associated with an ordered pair of hard scalar lines
\((a,b)\).  Every hard momentum \(P_a\) is directed out of the hatched hard-interaction disk;
for a crossed incoming particle this outgoing momentum has negative energy, as specified
in \eqref{eq:quantum_eta_convention}.
The parallel arrow beside the virtual wavy line defines \(\ell\) to flow from line \(a\)
to line \(b\); the parallel arrow beside the second wavy line defines the observed soft
momentum \(k\) to be outgoing. Diagrams (a) and (b) differ by the ordering of the two
photon vertices on line \(b\), while diagram (c) is the scalar-QED seagull graph.}
\label{f:quantum_pair_diagrams}
\end{figure}
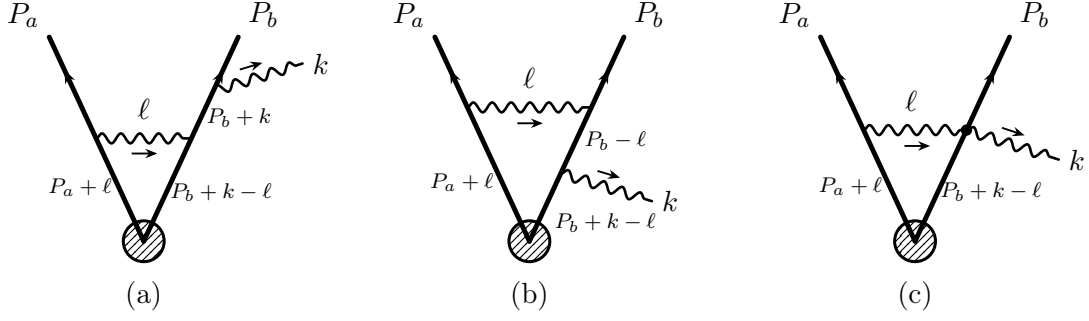

The universal one-loop contribution comes from a virtual photon exchanged between two
different external scalar lines.  For a fixed ordered pair \((a,b)\), let the observed soft
photon be emitted from line \(b\).  The three gauge-invariantly related diagrams relevant for our analysis are shown
in figure~\ref{f:quantum_pair_diagrams}.

Self-energy insertions with both ends of the virtual photon on the same external line,
together with the corresponding mass counterterm, do not contribute to the soft factor
in the on-shell scheme.  Their sum is proportional to a double external pole and its
coefficient vanishes by the on-shell mass-renormalization condition, precisely as in the
four-dimensional analysis of \cite{Sahoo:2018lxl}.  Wave-function factors multiply both
\(\mathcal M_n\) and \(\mathcal M_{n+1}\) and therefore cancel in their ratio.  We
therefore restrict the ordered sum below to \(a\ne b\).

We first write the three diagrams before making a soft or eikonal approximation, with
the outgoing photon leg contracted with its polarization \(\varepsilon^\mu\).  The
momentum routing is exactly the one displayed in
figure~\ref{f:quantum_pair_diagrams}. Diagram (a), in which the virtual photon is absorbed by line \(b\) before the soft
photon is emitted, gives
\bea\label{eq:quantum_diagram_a}
\mathcal M^{(1)}_{(a),ab}(k,\varepsilon)
&=&
(i\lambda)(-iQ_aQ_b^2)
\int\f{d^d\ell}{(2\pi)^d}\,
\f{\varepsilon.(2P_b+k)
(2P_a+\ell).(2P_b+2k-\ell)}
{\ell^2-i\epsilon}
\non\\
&&\times
\f{1}
{\left[(P_a+\ell)^2+\mathfrak m_a^2-i\epsilon\right]
\left[(P_b+k-\ell)^2+\mathfrak m_b^2-i\epsilon\right]
\left[(P_b+k)^2+\mathfrak m_b^2-i\epsilon\right]} .
\eea
In diagram (b), the soft photon is emitted first and the virtual photon is subsequently
absorbed.  Its contribution is
\bea\label{eq:quantum_diagram_b}
\mathcal M^{(1)}_{(b),ab}(k,\varepsilon)
&=&
(i\lambda)(-iQ_aQ_b^2)
\int\f{d^d\ell}{(2\pi)^d}\,
\f{\varepsilon.(2P_b+k-2\ell)
(2P_a+\ell).(2P_b-\ell)}
{\ell^2-i\epsilon}
\non\\
&&\times
\f{1}
{\left[(P_a+\ell)^2+\mathfrak m_a^2-i\epsilon\right]
\left[(P_b+k-\ell)^2+\mathfrak m_b^2-i\epsilon\right]
\left[(P_b-\ell)^2+\mathfrak m_b^2-i\epsilon\right]} .
\eea
Finally, the seagull vertex in diagram (c) gives
\bea\label{eq:quantum_diagram_c}
\mathcal M^{(1)}_{(c),ab}(k,\varepsilon)
&=&
(i\lambda)(-iQ_aQ_b^2)
\int\f{d^d\ell}{(2\pi)^d}\,
\f{-2\varepsilon.(2P_a+\ell)}
{\ell^2-i\epsilon}
\non\\
&&\times
\f{1}
{\left[(P_a+\ell)^2+\mathfrak m_a^2-i\epsilon\right]
\left[(P_b+k-\ell)^2+\mathfrak m_b^2-i\epsilon\right]} .
\eea
Equations
\eqref{eq:quantum_diagram_a}--\eqref{eq:quantum_diagram_c} also make clear why all three
graphs are required: diagram (a) contains the external soft pole, while diagrams (b) and
(c) complete its gauge-invariant subleading part when one sums over all pairs of legs and uses charge conservation.

For comparison, the virtual exchange between the same ordered pair in the amplitude
without the soft photon is
\bea\label{eq:quantum_nonradiative_pair}
\left.\mathcal M_n^{(1)}\right|_{ab}
&=&
(i\lambda)(-iQ_aQ_b)
\int\f{d^d\ell}{(2\pi)^d}
\f{(2P_a+\ell).(2P_b-\ell)}
{\ell^2-i\epsilon}
\non\\
&&\times
\f{1}
{\left[(P_a+\ell)^2+\mathfrak m_a^2-i\epsilon\right]
\left[(P_b-\ell)^2+\mathfrak m_b^2-i\epsilon\right]} .
\eea
Indeed, the \(k\to0\) singular part of \eqref{eq:quantum_diagram_a} is
\(Q_b\varepsilon.P_b/(k.P_b)\) times
\eqref{eq:quantum_nonradiative_pair}.  After summing over all virtual exchanges and
emitting legs, this shows that the leading soft relation
\eqref{eq:quantum_tree_soft_factor} remains valid at one loop, with
\(\mathcal M_{n+1}^{(0)}\) and \(\mathcal M_n^{(0)}\) replaced by
\(\mathcal M_{n+1}^{(1)}\) and \(\mathcal M_n^{(1)}\), respectively.

\subsubsection{Expansion in the logarithmic region}

For each pair \(a\ne b\), let \(\Lambda\) denote a characteristic hard-energy scale of
order \(\sqrt{-\eta_a\eta_b P_a.P_b}\).  The separation of momentum regions follows
the same logic as in the classical analysis surrounding
\eqref{eq:logarithmic_overlap_region}--\eqref{eq:higher_phase_log_order}.  For
loop momentum of order \(\Lambda\) or larger, the propagators admit a Taylor
expansion in \(k\), and hence contribute only analytic powers of the soft momentum.
The region \(0<|\ell|\lesssim\omega\) is most transparently counted after the
leading-soft term
\(Q_b\varepsilon.P_b\left.\mathcal M_n^{(1)}\right|_{ab}/(k.P_b)\), identified
by comparing \eqref{eq:quantum_diagram_a} with
\eqref{eq:quantum_nonradiative_pair}, has been separated.  This term contains a
single factor \(1/(k.P_b)\), while the non-radiative amplitude multiplying it is
independent of \(k\); it therefore cannot generate a stronger soft pole.  In the
remaining integrand displayed below in
\eqref{eq:quantum_three_diagram_decomposition}, the numerator is linear in \(\ell\).
In the subregion \(|\ell|\ll\omega\), the hard-particle propagator in the soft
approximation, \(1/((\ell-k).P_b+i\epsilon)\), may be expanded in \(\ell/k\).
The term of order \(n\), integrated up to any fixed fraction
\(c\omega\) with \(0<c<1\), has the radial behavior
\be\label{eq:quantum_small_loop_radial_behavior}
\f{1}{\omega^{n+1}}\int_0^{c\omega}
d|\ell|\,|\ell|^{d-4+n}\ \propto\ \omega^{d-4}.
\ee
For \(d\geq4\), the radial exponent is never \(-1\), and every apparent inverse
power of \(\omega\) is compensated by the endpoint of the loop integral.  At
\(|\ell|\sim\omega\), where this expansion is not valid, the rescaling
\(\ell=\omega L\) applied to the same degree-\(-4\) integrand again gives an overall
factor \(\omega^{d-4}\) multiplying an integral over \(L\) of order unity.  Thus,
apart from the leading-soft term written above, the full region
\(0<|\ell|\lesssim\omega\) produces neither \(\ln\omega\) nor an additional inverse
power of \(\omega\), although it may contribute to the non-logarithmic coefficient
at order \(\omega^{d-4}\).  The universal logarithmic coefficient can therefore
arise only from the intermediate region
\be\label{eq:quantum_logarithmic_region}
\omega\ll|\ell|\ll\Lambda .
\ee
In what follows, \(\int_\omega^\Lambda d^d\ell\) denotes integration restricted to
this shell.  We expand the massive scalar propagators in soft momentum over the hard
scale, retaining every term of the required homogeneous degree, but do not yet expand
the dependence on \(P_b.(\ell-k)\) in \(k/\ell\).  After separating the leading-soft
term from diagram (a), we combine the three diagrams and organize the remaining
propagators and numerators under
\((\ell,k)\to\tau(\ell,k)\).  The leading-soft term has degree
\(-5\), while the first contribution beyond the leading-soft factor capable of producing the
\(\omega^{d-4}\ln\omega\) term has degree \(-4\).  Expanding the factor
\(1/((\ell-k).P_b+i\epsilon)\) in this sector produces, at order
\((k.P_b)^{d-4}\), the scale-invariant
radial measure \(d|\ell|/|\ell|\).  Terms of higher degree give analytic
contributions or logarithms suppressed by additional powers of \(\omega/\Lambda\).
We therefore retain the degree \(-4\) terms, for which the three diagrams have the
common denominator
\be\label{eq:quantum_common_denominator}
(\ell^2-i\epsilon)(P_a.\ell-i\epsilon)(P_b.\ell+i\epsilon)
\big((\ell-k).P_b+i\epsilon\big).
\ee
Since this denominator scales as \(\tau^5\), the numerators in the degree
\(-4\) sector scale as \(\tau\).  For diagram (a) the leading-soft pole is first
separated, whereas for the seagull graph the missing factor
\(P_b.\ell+i\epsilon\) is inserted in both numerator and denominator.  The
separation of the leading-soft pole leaves a degree \(-5\) term in diagram (a)
which cancels its counterpart in diagram (b).  The expansion also
generates equal-and-opposite degree \(-4\) terms in these two diagrams.  The
relations below are therefore shell expansions rather than exact identities:
\(\simeq\) denotes restriction to
\eqref{eq:quantum_logarithmic_region}, with the complete homogeneous
contributions of degrees \(-5\) and \(-4\) retained as applicable.  Denoting the
complete cancelling contribution assigned to diagram (a) by
\(\mathcal M^{(1)}_{{\rm extra},ab}\), and hence the contribution assigned to
diagram (b) by \(-\mathcal M^{(1)}_{{\rm extra},ab}\), we obtain
\bea\label{eq:quantum_diagram_a_region}
\mathcal M^{(1)}_{(a),ab}(k,\varepsilon)
&\simeq&
\f{Q_b\varepsilon.P_b}{k.P_b}
\left.\mathcal M_n^{(1)}\right|_{ab}
\non\\
&&+(i\lambda)(-iQ_aQ_b^2)
\int_\omega^\Lambda\f{d^d\ell}{(2\pi)^d}\,
\f{1}{\ell^2-i\epsilon}\f{1}{P_a.\ell-i\epsilon}
\non\\
&&\quad\times
\f{1}{P_b.\ell+i\epsilon}\f{1}{(\ell-k).P_b+i\epsilon}
\left\{-\f{\varepsilon.P_b}{k.P_b}
\left[(P_a.k)(P_b.\ell)-(P_a.P_b)k.\ell\right]\right\}
\non\\
&&\quad+\mathcal M^{(1)}_{{\rm extra},ab}(k,\varepsilon),
\eea
\bea\label{eq:quantum_diagram_b_region}
\mathcal M^{(1)}_{(b),ab}(k,\varepsilon)
&\simeq&
(i\lambda)(-iQ_aQ_b^2)
\int_\omega^\Lambda\f{d^d\ell}{(2\pi)^d}\,
\f{1}{\ell^2-i\epsilon}\f{1}{P_a.\ell-i\epsilon}
\non\\
&&\quad\times
\f{1}{P_b.\ell+i\epsilon}\f{1}{(\ell-k).P_b+i\epsilon}
\left[-(P_a.P_b)\varepsilon.\ell\right]
\non\\
&&\quad-\mathcal M^{(1)}_{{\rm extra},ab}(k,\varepsilon),
\eea
\bea\label{eq:quantum_diagram_c_region}
\mathcal M^{(1)}_{(c),ab}(k,\varepsilon)
&\simeq&
(i\lambda)(-iQ_aQ_b^2)
\int_\omega^\Lambda\f{d^d\ell}{(2\pi)^d}\,
\f{1}{\ell^2-i\epsilon}\f{1}{P_a.\ell-i\epsilon}
\non\\
&&\quad\times
\f{1}{P_b.\ell+i\epsilon}\f{1}{(\ell-k).P_b+i\epsilon}
\left[(\varepsilon.P_a)P_b.\ell\right].
\eea

For completeness, the cancelling term retained in this homogeneous expansion
and appearing in
\eqref{eq:quantum_diagram_a_region} and
\eqref{eq:quantum_diagram_b_region} is defined as
\bea\label{eq:quantum_extra_terms}
\mathcal M^{(1)}_{{\rm extra},ab}(k,\varepsilon)
&\equiv&
-(i\lambda)(-iQ_aQ_b^2)
\int_\omega^\Lambda\f{d^d\ell}{(2\pi)^d}\,
\f{1}{\ell^2-i\epsilon}\f{1}{P_a.\ell-i\epsilon}
\non\\
&&\quad\times
\f{1}{P_b.\ell+i\epsilon}
\f{1}{(\ell-k).P_b+i\epsilon}
\,
(\varepsilon.P_b)(P_a.P_b)
\non\\
&&\quad\times
\left\{
1+\f{P_b.\ell-P_a.\ell}{2P_a.P_b}
-\f{\ell^2}{2(P_a.\ell-i\epsilon)}
\right.
\non\\
&&\qquad\left.
\mathord{}+\f{\ell^2-2k.\ell}{2((\ell-k).P_b+i\epsilon)}
\mathord{}+\f{\ell^2}{2(P_b.\ell+i\epsilon)}
\right\}.
\eea
The unit term in braces is the complete degree \(-5\) part, while the remaining
terms form the complete degree \(-4\) part.  Thus
\eqref{eq:quantum_extra_terms} is the cancelling contribution at the retained
homogeneous orders, not an exact rewriting of either full integral
\eqref{eq:quantum_diagram_a} or \eqref{eq:quantum_diagram_b}.  Since it enters
\eqref{eq:quantum_diagram_a_region} and
\eqref{eq:quantum_diagram_b_region} with opposite signs, it cancels before the
three diagrams are summed.

The sum of the three diagrams is therefore
\bea\label{eq:quantum_three_diagram_decomposition}
\sum_{X=a,b,c}\mathcal M^{(1)}_{(X),ab}(k,\varepsilon)
&=&
\f{Q_b\,\varepsilon.P_b}{k.P_b}
\left.\mathcal M_n^{(1)}\right|_{ab}\non\\
&&+(i\lambda)(-iQ_aQ_b^2)
\left(\varepsilon_\rho-\f{\varepsilon.P_b}{k.P_b}k_\rho\right)
\int_\omega^\Lambda\f{d^d\ell}{(2\pi)^d}
\f{1}{\ell^2-i\epsilon}
\f{1}{P_a.\ell-i\epsilon}
\f{1}{P_b.\ell+i\epsilon}
\non\\
&&\quad\times \f{1}{(\ell-k).P_b+i\epsilon}
\left(P_a^\rho P_b.\ell-P_a.P_b\,\ell^\rho\right)+\text{remainder}
\non\\
&\equiv&
\f{Q_b\,\varepsilon.P_b}{k.P_b}
\left.\mathcal M_n^{(1)}\right|_{ab}
+i\lambda\,\mathcal S_{ab}(\varepsilon,k)
+\text{remainder},
\eea
where the second equality defines the pair contribution
\(\mathcal S_{ab}(\varepsilon,k)\).  Equation
\eqref{eq:quantum_three_diagram_decomposition} displays both the leading soft factor
multiplying the non-radiative one-loop amplitude and the integral defining
\(\mathcal S_{ab}(\varepsilon,k)\).  The remainder contains terms analytic in \(k\),
including a possible non-logarithmic contribution at order \(\omega^{d-4}\), as
well as possible higher-order non-analytic terms
\(\omega^p\ln\omega\) with \(p>d-4\), none of which affects the coefficient extracted
below.

To isolate the logarithm, we expand the factor
\(1/((\ell-k).P_b+i\epsilon)\) in
\eqref{eq:quantum_three_diagram_decomposition}:
\bea\label{eq:quantum_soft_denominator_expansion}
\f{1}{(\ell-k).P_b+i\epsilon}
&=&
\sum_{n=0}^{\infty}
\f{(k.P_b)^n}{(P_b.\ell+i\epsilon)^{n+1}} .
\eea
The factor being expanded originates from a massive-particle propagator after the
soft approximation has reduced its denominator to \((\ell-k).P_b+i\epsilon\).
The higher powers of \(P_b.\ell+i\epsilon\) are Taylor coefficients of this shifted
factor, not terms in the conventional exponentiated QED eikonal factor.  Selecting
\(n=d-4\) gives the power \(d-3\) for the expanded factor; together with the
pre-existing factor \(1/(P_b.\ell+i\epsilon)\), it gives the power \(d-2\) in the
vector integral below.  Equivalently, the scalar master integral has power \(d-3\),
and its momentum derivative produces the power \(d-2\).
At order \(n\), the radial dependence is
\(d|\ell|\,|\ell|^{d-n-5}\).  Hence the logarithm arises uniquely from \(n=d-4\),
giving
\bea\label{eq:quantum_pair_kernel_log}
\mathcal S_{ab}(\varepsilon,k)\big|_{\omega^{d-4}\ln\omega}
&=&
-i Q_aQ_b^2(k.P_b)^{d-4}
\left(\varepsilon_\rho-\f{\varepsilon.P_b}{k.P_b}k_\rho\right)
\non\\
&&\times
\int_\omega^\Lambda\f{d^d\ell}{(2\pi)^d}
\f{1}{\ell^2-i\epsilon}
\f{P_a^\rho\,P_b.\ell-P_a.P_b\,\ell^\rho}
{(P_a.\ell-i\epsilon)(P_b.\ell+i\epsilon)^{d-2}} .
\eea
All terms with \(n<d-4\) are sensitive to the hard scale \(\Lambda\) and are analytic in
the soft momentum.  Terms with \(n>d-4\) are dominated by the lower endpoint, where
the expansion in \(k/\ell\) is not ordered.  After resummation they may contribute
to the non-logarithmic coefficient at order \(\omega^{d-4}\), but cannot modify the
logarithmic coefficient selected by \(n=d-4\).  We denote the complete one-loop soft contribution with scaling
\(\omega^{d-4}\ln\omega\) by
\(\mathcal S_{\rm em}^{\ln}(\varepsilon,k)\).  The superscript \(\ln\) labels this
non-analytic contribution rather than the conventional subleading soft order.  For
example, in \(d=6\) it is of order \(\omega^2\ln\omega\), whereas the usual
subleading soft factor is of order \(\omega^0\).  Its definition is the ordered-pair
sum
\bea\label{eq:quantum_soft_factor_pair_sum}
\mathcal S_{\rm em}^{\ln}(\varepsilon,k)
&=&
\sum_{b=1}^{N+M}
\sum_{\substack{a=1\\a\ne b}}^{N+M}
\mathcal S_{ab}(\varepsilon,k)
\big|_{\omega^{d-4}\ln\omega}.
\eea

\subsection{Reduction to a scalar Feynman integral}
The vector integral present in \eqref{eq:quantum_pair_kernel_log} can be translated to a momentum derivative of a scalar integral in the following way:\footnote{The differential structure in \eqref{eq:quantum_soft_from_master} may suggest extending the four-dimensional observation of \cite[pp.~33--34]{Sahoo:2018lxl}, where the ordinary subleading angular-momentum soft operator acts on the eikonal factor with the momentum integral restricted to the logarithmic region, to the higher-order projected soft identities of \cite{Hamada:2018soft,Li:2018soft}.  At order \(k^{d-4}\), however, those projected identities involve \(d-3\) hard-momentum derivatives in total.  In the angular-momentum form, these comprise \(d-4\) factors of \(k.\partial_{P_b}\) in addition to the single derivative contained in \(J_b^{\mu\nu}\).  The repeated-derivative part already gives \((k.\partial_{P_b})^{d-4}(P_b.\ell+i\epsilon)^{-1}=(-1)^{d-4}(d-4)!(k.\ell)^{d-4}(P_b.\ell+i\epsilon)^{-(d-3)}\), before the derivative in \(J_b^{\mu\nu}\) is applied, and therefore generates additional powers of \(\ell\) in the numerator.  This is not the structure of the present logarithmic term.  Here \((k.P_b)^{d-4}\) and \((P_b.\ell+i\epsilon)^{-(d-3)}\) arise together from the \(n=d-4\) term of the propagator expansion \eqref{eq:quantum_soft_denominator_expansion}; the single subsequent \(P_b\)-derivative in \eqref{eq:quantum_soft_from_master} produces the vector numerator.  Thus the higher-dimensional result cannot be obtained by applying the higher-order operators of \cite{Hamada:2018soft,Li:2018soft} to the conventional exponentiated eikonal factor in $d>4$.}
\bea\label{eq:quantum_soft_from_master}
\mathcal S_{ab}(\varepsilon,k)\big|_{\omega^{d-4}\ln\omega}
&=&
\f{i Q_aQ_b^2}{d-3}(k.P_b)^{d-4}
\left(\varepsilon_\rho-\f{\varepsilon.P_b}{k.P_b}k_\rho\right)
\non\\
&&\times
\left(P_a^\rho P_{b\sigma}-P_a.P_b\,\delta^\rho_\sigma\right)
\f{\p\mathcal K^{\rm F}_{ab}}{\p P_{b\sigma}} ,
\eea
where
\be\label{eq:quantum_feynman_master}
\mathcal K^{\rm F}_{ab}\equiv
\int_\omega^\Lambda\f{d^d\ell}{(2\pi)^d}
\f{1}{\ell^2-i\epsilon}
\f{1}{(P_a.\ell-i\epsilon)(P_b.\ell+i\epsilon)^{d-3}} .
\ee
Here \(\partial/\partial P_{b\sigma}\) is taken with \(P_a\) and \(P_b\) independent
and unconstrained, with \(\eta_a\) and \(\eta_b\) fixed.  The physical masses enter
only when \(P_i^2=-\mathfrak m_i^2\) is imposed after differentiation.
The integral \eqref{eq:quantum_feynman_master} is evaluated in
appendix~\ref{app:feynman_integral}, with the final result given in
\eqref{eq:feynman_master_result_app}.  To quote it here, define
\bea\label{eq:quantum_pair_invariants}
\Delta_{ab}
&\equiv&
(P_a.P_b)^2-P_a^2P_b^2,
\non\\
s_{ab}
&\equiv&
-\eta_a\eta_b\,\f{P_a.P_b}{\mathfrak m_a\mathfrak m_b},
\qquad
\Delta_{ab}=\mathfrak m_a^2\mathfrak m_b^2(s_{ab}^2-1).
\eea
For distinct non-collinear massive lines, \(s_{ab}>1\).  The function entering the
photon-pole coefficient is
\bea\label{eq:quantum_H_function}
\mathcal H_d(s)
&\equiv&
\f{1}{s}\,
\mathrm B\left(\f12,\f{d-2}{2}\right)
{}_2F_1\left(
1,\f12;\f{d-1}{2};1-\f{1}{s^2}
\right).
\eea
The matter-pole coefficient is
\be\label{eq:quantum_matter_coefficient}
\mathcal C_d\equiv
\begin{cases}
\displaystyle
\f{(-1)^{(d-4)/2}}
{(4\pi)^{(d-2)/2}\Gamma\left(\f{d-2}{2}\right)},
&d\ \text{even},\\[4mm]
\displaystyle
\f{i\,\Gamma\left(\f{4-d}{2}\right)}
{2^{d-2}\pi^{d/2}},
&d\ \text{odd}.
\end{cases}
\ee
With these definitions, and discarding the hard-scale constant \(\ln\Lambda\)
under the \(\simeq\) prescription, \eqref{eq:feynman_master_result_app} gives
\be\label{eq:quantum_feynman_master_result}
\boxed{
\begin{aligned}
\mathcal K^{\rm F}_{ab}\big|_{\ln}
&\simeq
\ln\omega\Bigg[
-\f{1+\eta_a\eta_b}{2}\,
\mathcal C_d\,
\f{\mathfrak m_a^{d-4}}{\Delta_{ab}^{(d-3)/2}}
\\[-1mm]
&\hspace{2.0cm}
-\f{i(-1)^d\eta_a\eta_b}
{2^{d-1}\pi^{d/2}\Gamma\left(\f{d-2}{2}\right)}
\f{\mathcal H_d(s_{ab})}{\mathfrak m_a\mathfrak m_b^{d-3}}
\Bigg].
\end{aligned}
}
\ee
The two terms in square brackets in
\eqref{eq:quantum_feynman_master_result} are generated by the scalar and Feynman
photon poles, respectively; their contour evaluation is given in
appendix~\ref{app:feynman_integral}.

Equation \eqref{eq:quantum_feynman_master_result} is the on-shell form of the
logarithmic coefficient and must not be differentiated directly.  For timelike
\(P_a\) and \(P_b\), with their time orientations held fixed, the required
Lorentz-covariant homogeneous continuation is
\bea\label{eq:quantum_feynman_master_off_shell}
\left.\mathcal K^{\rm F}_{ab}\right|_{\ln,\,\mathrm{off}}
&\simeq&
\ln\omega\Bigg[
-\f{1+\eta_a\eta_b}{2}\,
\mathcal C_d\,
\f{(-P_a^2)^{(d-4)/2}}{\Delta_{ab}^{(d-3)/2}}
\non\\
&&\hspace{1.2cm}
-\f{i(-1)^d\eta_a\eta_b}
{2^{d-1}\pi^{d/2}\Gamma\left(\f{d-2}{2}\right)}
\f{
\mathcal H_d\left(
-\eta_a\eta_b
\f{P_a.P_b}{\sqrt{(-P_a^2)(-P_b^2)}}
\right)}
{\sqrt{-P_a^2}\,(-P_b^2)^{(d-3)/2}}
\Bigg].
\eea
It has degree \(-1\) in \(P_a\) and degree \(-(d-3)\) in \(P_b\) analogous to the starting integral in \eqref{eq:quantum_feynman_master}, and reduces to
\eqref{eq:quantum_feynman_master_result} after imposing
\(P_i^2=-\mathfrak m_i^2\).

\subsection{Final one-loop soft factor}

Applying the differential operator in \eqref{eq:quantum_soft_from_master} to the
off-shell expression \eqref{eq:quantum_feynman_master_off_shell} and only then
imposing \(P_i^2=-\mathfrak m_i^2\) gives a vector
proportional to \(P_a^\rho P_b^2-P_a.P_b\,P_b^\rho\).  Its contraction with the
transverse factor in \eqref{eq:quantum_soft_from_master} is
\be\label{eq:quantum_transverse_contraction}
\left(\varepsilon_\rho-\f{\varepsilon.P_b}{k.P_b}k_\rho\right)
\left(P_a^\rho P_b^2-P_a.P_b\,P_b^\rho\right)
=P_b^2\left(
\varepsilon.P_a-\f{k.P_a}{k.P_b}\varepsilon.P_b
\right).
\ee
Differentiating the photon-pole term produces
\((d-3)\mathcal H_d(s_{ab})+s_{ab}\mathcal H'_d(s_{ab})\), while the
matter-pole term gives
\(\mathfrak m_a^{d-2}\mathfrak m_b^2/\Delta_{ab}^{(d-1)/2}\).
For even \(d\), the recurrence
\be\label{eq:quantum_H_even_recurrence}
\mathcal H_d(s)=
\f{s}{s^2-1}\,
\mathrm B\left(\f12,\f{d-4}{2}\right)
-\f{\mathcal H_{d-2}(s)}{s^2-1},
\qquad
\mathcal H_4(s)=\f{2\operatorname{arccosh}s}{\sqrt{s^2-1}},
\ee
terminates after finitely many steps.  For odd \(d\), the finite polynomial in
\eqref{eq:odd_mix_H_function} terminates directly.

Substituting these two representations, performing the ordered-pair sum
\eqref{eq:quantum_soft_factor_pair_sum}, and writing the answer only in terms of the
hard data and \(\Delta_{ab}\) gives
\begin{subequations}\label{eq:quantum_soft_factor_general_d}
\begin{align}
\boxed{\begin{aligned}
\left.\mathcal S_{\rm em}^{\ln}(\varepsilon,k)\right|_{d=\mathrm{even}}
&=\ln\omega
\sum_{b=1}^{N+M}\sum_{\substack{a=1\\a\ne b}}^{N+M}
Q_aQ_b^2(k.P_b)^{d-4}
\left(\varepsilon.P_a-\f{k.P_a}{k.P_b}\varepsilon.P_b\right)
\\[-1mm]
&\quad\times
\Bigg[
-\f{i(-1)^{(d-4)/2}(1+\eta_a\eta_b)}
{2^{d-1}\pi^{(d-2)/2}\Gamma\left(\f{d-2}{2}\right)}
\f{\mathfrak m_a^{d-2}\mathfrak m_b^2}
{\Delta_{ab}^{(d-1)/2}}
\\[-1mm]
&\qquad
+\f{1}{(d-3)2^{d-1}\pi^{d/2}
\Gamma\left(\f{d-2}{2}\right)}
\Bigg\{
\f{2(-1)^{(d-4)/2}(P_a.P_b)\mathfrak m_a^{d-4}}
{\Delta_{ab}^{(d-2)/2}}
\\[-1mm]
&\qquad
+\sum_{j=1}^{(d-4)/2}
\f{(-1)^{j-1}\sqrt{\pi}\,
\Gamma\left(\f{d-2}{2}-j\right)}
{\Gamma\left(\f{d-1}{2}-j\right)}
\f{(P_a.P_b)\mathfrak m_a^{2j-2}}
{\mathfrak m_b^{d-2j-2}\Delta_{ab}^{j+1}}
\\[-1mm]
&\qquad\quad\times
\left[(d-2-2j)(P_a.P_b)^2
-(d-2)\mathfrak m_a^2\mathfrak m_b^2\right]
\\[-1mm]
&\qquad
+\f{2(-1)^{(d-4)/2}(d-3)\eta_a\eta_b
\mathfrak m_a^{d-2}\mathfrak m_b^2}
{\Delta_{ab}^{(d-1)/2}}
\operatorname{arccosh}\left(
-\eta_a\eta_b\f{P_a.P_b}{\mathfrak m_a\mathfrak m_b}
\right)
\Bigg\}
\Bigg].
\end{aligned}}
\label{eq:quantum_soft_factor_even_d}\displaybreak[1]
\\[-2mm]
\boxed{\begin{aligned}
\left.\mathcal S_{\rm em}^{\ln}(\varepsilon,k)\right|_{d=\mathrm{odd}}
&=\ln\omega
\sum_{b=1}^{N+M}\sum_{\substack{a=1\\a\ne b}}^{N+M}
Q_aQ_b^2(k.P_b)^{d-4}
\left(\varepsilon.P_a-\f{k.P_a}{k.P_b}\varepsilon.P_b\right)
\\[-1mm]
&\quad\times
\Bigg[
\f{(1+\eta_a\eta_b)\Gamma\left(\f{4-d}{2}\right)}
{2^{d-1}\pi^{d/2}}
\f{\mathfrak m_a^{d-2}\mathfrak m_b^2}
{\Delta_{ab}^{(d-1)/2}}
\\[-1mm]
&\qquad
+\f{\eta_a\eta_b}
{(d-3)2^{2d-4}\pi^{(d-2)/2}
\Gamma\left(\f{d-2}{2}\right)
\mathfrak m_a\mathfrak m_b^{d-3}}
\Bigg\{
(d-3)\binom{d-3}{\frac{d-3}{2}}
\\[-1mm]
&\qquad
+2\sum_{j=1}^{(d-3)/2}
\binom{d-3}{\frac{d-3}{2}-j}
\left(
\f{\mathfrak m_a\mathfrak m_b+\eta_a\eta_bP_a.P_b}
{\mathfrak m_a\mathfrak m_b-\eta_a\eta_bP_a.P_b}
\right)^j
\\[-1mm]
&\qquad\quad\times
\left[
d-3-\f{2j\eta_a\eta_b(P_a.P_b)\mathfrak m_a\mathfrak m_b}
{\Delta_{ab}}
\right]
\Bigg\}
\Bigg].
\end{aligned}}
\label{eq:quantum_soft_factor_odd_d}
\end{align}
\end{subequations}
Using the same amplitude-factorization convention as the tree-level relation
\eqref{eq:quantum_tree_soft_factor}, the soft factor
\eqref{eq:quantum_soft_factor_general_d} gives the logarithmic one-loop theorem
\bea\label{eq:quantum_soft_theorem_general_d}
\mathcal M^{(1)}_{n+1}(k,\varepsilon)
\big|_{\omega^{d-4}\ln\omega}
&=&
\mathcal S_{\rm em}^{\ln}(\varepsilon,k)
\mathcal M_n^{(0)} .
\eea
Equation \eqref{eq:quantum_soft_factor_general_d} is gauge invariant, since it vanishes
under \(\varepsilon^\mu\to k^\mu\).  The first term in square brackets has support only for
\(\eta_a\eta_b=1\) and is generated by the on-shell scalar pole, while the second is generated
by the on-shell pole of the virtual Feynman photon.

For even \(d\geq6\) the expression is a finite rational series plus one inverse
hyperbolic cosine, whereas for odd \(d\geq5\) it is a finite rational binomial
series.  Thus neither final result requires \(s_{ab}\), \(\mathcal H_d\), or
\(\mathcal C_d\).  The retarded classical waveform is organized differently
because it resolves the causal branch assignments, which in odd dimensions lead to
the distinct outgoing, incoming and mixed functions in \eqref{eq:odd_F_main}.  In
the next subsection we reconcile these organizations by extracting the explicit
retarded classical soft factors and tracing their matter- and photon-pole
representations.

\subsection{Classical limit and comparison with the frequency-space waveform}
\label{S:quantum_classical_comparison}

We define the classical soft factor by
\(S_{\rm cl}(\varepsilon,k)=i\,\varepsilon_\mu\widehat J^\mu(k)\).
For a positive-energy outgoing photon, combining this definition with the radiative
relation \eqref{eq:AJ_relation} gives
\bea\label{eq:classical_waveform_soft_relation}
\varepsilon_\mu\wt A^\mu(\omega,\vec x)
&\simeq&
\f{e^{i\omega r}}{2(\omega+i\epsilon)}
\left(\f{\omega+i\epsilon}{2\pi i r}\right)^{\f{d-2}{2}}
S_{\rm cl}(\varepsilon,k),
\qquad
S_{\rm cl}(\varepsilon,k)
=i\,\varepsilon_\mu\widehat J^\mu(k).
\eea
This normalization agrees with the general relation between classical radiation and
soft factors in \cite{Laddha:2018rle,Laddha:2019yaj}.

Two boundary-condition statements must be kept distinct in making the comparison.
The quantum amplitude is evaluated for an outgoing photon with \(\omega>0\), for
which \(\ln(\omega+i\epsilon)=\ln(\omega-i\epsilon)=\ln\omega\).  It therefore fixes
the logarithmic coefficients but not their continuation to negative frequency.  The
classical field instead uses the retarded photon propagator, and the outgoing and
incoming proper-time ranges in \eqref{eq:Ib_def} fix the respective boundary values
\(\ln(\omega+i\epsilon)\) and \(\ln(\omega-i\epsilon)\).  We retain these boundary
values in the classical soft factors below so that their substitution into
\eqref{eq:classical_waveform_soft_relation} gives the retarded waveform on the full
frequency axis.

To isolate the classical part of the one-loop result, the Feynman propagator in
\eqref{eq:quantum_feynman_master} must be compared with the retarded propagator used
in section~\ref{S:log_correction}.  After converting the incoming lines according to
\eqref{eq:quantum_eta_convention}, both integrals contain the same factor
\((P_a.\ell-i\epsilon)^{-1}(P_b.\ell+i\epsilon)^{-(d-3)}\) and differ only in
the boundary condition of the photon propagator.  They may be evaluated
in a common half-plane containing no retarded photon pole.  The retarded integral is
then given by the enclosed matter-particle residues, whereas the Feynman integral
contains the same matter poles and one additional on-shell photon pole.  The latter
is the intrinsically quantum contribution.  Thus the separation is made by comparing
the complete Feynman and retarded integrals with the same contour: the residues
present in the retarded integral give the classical term, while the additional
Feynman photon residue gives the quantum correction.  The label attached to a
residue in another contour representation is not sufficient for this purpose,
because deforming the contour can rewrite the same classical retarded integral in
terms of a different set of enclosed poles.
The corresponding four-dimensional comparison of Feynman and retarded
prescriptions was given in \cite{Sahoo:2018lxl}.  The classical limit of
amplitude-based radiation was analyzed in
\cite{Kosower:2018adc,Manu:2020soft,Akhtar:2025large,Paul:2026logsoft}.

For even \(d\), the two retarded photon-pole residues cancel and mixed retarded
contributions vanish.  The classical logarithmic soft factor is consequently the
same-branch matter-pole term of
\eqref{eq:quantum_soft_factor_even_d}, with the exact relative signs and mass
factors fixed by \eqref{eq:feynman_matter_relation_even}.  Using
\eqref{eq:quantum_eta_convention} gives
\bea\label{eq:quantum_classical_part}
\left.S_{\rm cl}(\varepsilon,k)
\right|_{\substack{d=\mathrm{even}\\\omega^{d-4}\ln}}
&\simeq&
-i
\sum_{b=1}^{N+M}\sum_{\substack{a=1\\a\ne b}}^{N+M}
Q_aQ_b^2(k.P_b)^{d-4}
\non\\
&&\times
\mathcal C_d\,
\f{\mathfrak m_a^{d-2}\mathfrak m_b^2}
{\Delta_{ab}^{(d-1)/2}}
\left(\varepsilon.P_a-\f{k.P_a}{k.P_b}\varepsilon.P_b\right)
\non\\
&&\times
\left[
\f{(1+\eta_a)(1+\eta_b)}{4}\ln(\omega+i\epsilon)
+\f{(1-\eta_a)(1-\eta_b)}{4}\ln(\omega-i\epsilon)
\right].
\eea
The two projectors in \eqref{eq:quantum_classical_part} select the outgoing and
incoming same-branch assignments separately.  At positive frequency their sum
reduces to \((1+\eta_a\eta_b)/2\), and
\eqref{eq:quantum_classical_part} becomes precisely the matter-pole part of
\eqref{eq:quantum_soft_factor_even_d}.  With the two retarded boundary values kept
distinct, it agrees with the classical current \eqref{eq:Jacc_even_log}.

For odd \(d\), the support of the retarded Green's function inside the light cone
makes the classical contribution depend on the time orientations of the two lines.
Applying the all-outgoing replacements \eqref{eq:quantum_eta_convention} to
\eqref{eq:Jacc_odd_log} and using the definition in
\eqref{eq:classical_waveform_soft_relation} gives the explicit retarded classical
soft factor
\bea\label{eq:quantum_classical_part_odd}
\left.S_{\rm cl}(\varepsilon,k)
\right|_{\substack{d=\mathrm{odd}\\\omega^{d-4}\ln}}
&\simeq&
\f{1}{(2\pi)^{d-1}}
\sum_{b=1}^{N+M}\sum_{\substack{a=1\\a\ne b}}^{N+M}
\f{Q_aQ_b^2(k.P_b)^{d-4}}
{\mathfrak m_a\mathfrak m_b^{d-3}}
\left(\varepsilon.P_a-\f{k.P_a}{k.P_b}\varepsilon.P_b\right)
\non\\[-1mm]
&&\times\Bigg\{
-\ln(\omega+i\epsilon)
\Bigg[
\f{(1+\eta_a)(1+\eta_b)}{4}\,
\mathcal F_{\rm out}^{(d)}
\left(\f{P_a.P_b}{\mathfrak m_a\mathfrak m_b}\right)
\non\\[-1mm]
&&\hspace{4.4cm}
+\f{(1-\eta_a)(1+\eta_b)}{4}\,
\mathcal F_{\rm mix}^{(d)}
\left(\f{P_a.P_b}{\mathfrak m_a\mathfrak m_b}\right)
\Bigg]
\non\\[-1mm]
&&\hspace{1.2cm}
+\ln(\omega-i\epsilon)
\f{(1-\eta_a)(1-\eta_b)}{4}\,
\mathcal F_{\rm in}^{(d)}
\left(\f{P_a.P_b}{\mathfrak m_a\mathfrak m_b}\right)
\Bigg\}.
\eea
The three functions in \eqref{eq:quantum_classical_part_odd} are given explicitly
in \eqref{eq:odd_F_main}.  The first projector selects two outgoing lines, the
second selects an incoming source acting on an outgoing line, and the third selects
two incoming lines.  There is no projector for an outgoing source acting on an
incoming line because that ordering vanishes by retarded causality.  The common
kinematic factor in \eqref{eq:quantum_classical_part} and
\eqref{eq:quantum_classical_part_odd} follows from
\(k^\mu=\omega\mathbf n^\mu\).  It is transverse, consistently with
\eqref{eq:quantum_soft_polarization}, because it vanishes when
\(\varepsilon^\mu\) is replaced by \(k^\mu\).  The relative minus signs of the
outgoing and mixed odd-dimensional terms arise when
\((-p_b.\mathbf n)^{d-4}\) in \eqref{eq:odd_hard_tensors} is written as
\((k.P_b)^{d-4}\); the power \(d-4\) is odd.

The pole decomposition explains how
\eqref{eq:quantum_classical_part_odd} is contained in the complete quantum result.
In the contour convention of \eqref{eq:quantum_feynman_master_result},
\(\mathcal F_{\rm out}^{(d)}\) contains the matter-pole term plus twice the
retarded photon-pole term, \(\mathcal F_{\rm in}^{(d)}\) contains only the
matter-pole term, and \(\mathcal F_{\rm mix}^{(d)}\) contains twice the retarded
photon-pole term.  These relations are established in
\eqref{eq:feynman_retarded_odd_relation} and
\eqref{eq:feynman_retarded_odd_mixed_relation}.  Equivalently, the first term in
\eqref{eq:quantum_soft_factor_odd_d} supplies the same-branch matter contribution,
while twice its finite-series term supplies the remaining outgoing contribution and
the complete incoming-to-outgoing contribution.  It supplies neither the incoming
nor the reverse mixed term.  The occurrence of retarded photon-pole residues in this
representation does not make them intrinsically quantum.  As explained above, when
the Feynman and retarded integrals are evaluated with the same contour containing no
retarded photon pole, the complete retarded result is represented by matter-particle
residues and only the additional Feynman photon residue is the quantum correction.

Equations \eqref{eq:quantum_classical_part} and
\eqref{eq:quantum_classical_part_odd} are therefore the all-outgoing forms of
\(i\varepsilon_\mu\widehat J_{\rm acc}^{\mu}(k)\) in
\eqref{eq:Jacc_even_log} and \eqref{eq:Jacc_odd_log}.  At \(\omega>0\), replacing
both logarithmic boundary values by \(\ln\omega\) gives the classical parts selected
from the corresponding one-loop soft factors.  Keeping the retarded prescriptions
before using \eqref{eq:classical_waveform_soft_relation} instead reproduces the
even-dimensional waveform \eqref{eq:A_acc_log_frequency} and the full
odd-dimensional waveform \eqref{eq:A_acc_odd_frequency}, including its causal mixed
term and the two boundary values required for the early- and late-time tails.

\section{Outlook}\label{S:outlook}

The present analysis determines the first logarithmic term generated by the
long-range electromagnetic interaction in \(d>4\).  In four dimensions, the
position-space iteration of the asymptotic equations of motion and the wave equation
generates the complete sequence of logarithmic terms in the classical
frequency-space waveform \cite{Karan:2025allorder}.
Subsection~\ref{S:position_space_even_tail} implements the corresponding
position-space analysis for the first logarithmic term in even \(d\geq6\),
while appendix~\ref{app:position_space_integral_check} proves directly that the
same trajectory tail yields the logarithmic term in the proper-time integral.
Generalizing that
analysis by iterating the higher-dimensional trajectory and field equations should
classify the complete sequence of powers and logarithms fixed by the asymptotic
scattering data, together with their retarded-time tails.  The quantum
problem is to identify the corresponding logarithms at higher loops, separate
their universal factorizing parts from process-dependent terms, and compare their
classical limits order by order.  Extending the one-loop scalar-QED calculation to
particles with spin and gauge-invariant non-minimal couplings is also necessary to
test whether these coefficients are universal and one-loop exact in the sense of
\cite{Krishna:2023log}.

The same strategy should extend to gravity.  Four-dimensional logarithmic
soft-graviton terms and their waveform tails were analyzed in
\cite{Laddha:2018myi,Laddha:2018logtail,Sahoo:2018lxl,Saha:2019tub}.  The
leading and subleading relation between soft factors and gravitational radiation
in higher dimensions was established in
\cite{Laddha:2018rle,Laddha:2019yaj,Manu:2020soft}.  The momentum-region analysis
and the comparison of Feynman and retarded prescriptions developed here should
allow one to derive the expected \(\omega^{d-4}\ln\omega\) soft-graviton factor
and its classical waveform.  The gravitational calculation must additionally
include emission from the exchanged gravitational field and nonlinear corrections
to the classical source.

A third direction concerns a Ward-identity interpretation.  In four dimensions,
logarithmic soft-photon and soft-graviton theorems have been related to
loop-corrected superphaserotation and superrotation charges
\cite{Campiglia:2019logward,Bhatkar:2019logward,Choi:2024logsymmetry,
Agrawal:2023logward,Choi:2024triangle}.  For higher-dimensional massless QED, the
leading soft photon theorem has been recast as a large-gauge Ward identity, while the
subleading theorem gives electric and magnetic Ward identities
\cite{Kapec:2014evenQED,He:2019higherDleading,He:2019higherDsubleading,
He:2019higherDmagnetic}. The logarithmic theorem derived here poses a different problem: its soft factor
contains the ordered double sum
\(\sum_b\sum_{a\ne b}\).
The same multiparticle feature motivated the introduction of additional celestial
fields in the four-dimensional high-energy analysis of
\cite{Banerjee:2026celestial} to interpret the \(\mathcal O(\ln\omega)\) soft photon theorem as a local Ward identity.  It remains to determine whether a suitable
high-energy limit, possibly together with additional celestial fields, can
reorganize the pairwise soft factor derived here into a local Ward identity on
\(S^{d-2}\), and, if so, whether the associated operator admits an interpretation
as a higher-dimensional analogue of the dipole current.

\acknowledgments
We are grateful to Ana-Maria Raclariu and Binsong Xiong for valuable
discussions during a closely related project that inspired the present work, and
for their comments on an earlier version of this manuscript. This work was
partially supported by the Science Technology \& Facilities Council (STFC) under
grant ST/X000753/1.

We acknowledge the use of OpenAI Codex as an assistive tool in preparing
this manuscript, primarily for refining the presentation and for cross-checking
algebraic manipulations, mass dimensions, normalization factors and internal
consistency.  We developed and verified the scientific ideas, analytical strategy,
derivations and conclusions.  We critically reviewed all AI-assisted suggestions
before inclusion and take full responsibility for
the accuracy and integrity of the work.

\appendix

\section{Odd-dimensional Fourier transforms}\label{app:odd_fourier_transform}

Here we evaluate the three branch-sensitive Fourier integrals used in the
odd-dimensional waveform analysis.

\paragraph{Integral for the straight-line waveform:}
The first integral is the one introduced in \eqref{eq:odd_tail_integral_def} in the
odd-dimensional straight-line waveform discussion.  The boundary value on the real axis
is
\be\label{eq:odd_branch_real_axis}
\lim_{\epsilon\to0^+}(\omega+i\epsilon)^\alpha=
\begin{cases}
\omega^\alpha, & \omega>0,\\[2mm]
|\omega|^\alpha e^{i\pi\alpha}, & \omega<0 .
\end{cases}
\ee
For finite positive $\epsilon$, the branch cut is displaced below the negative real
$\omega$ axis.  With the factor $e^{-i\omega s}$, the contour closes in the upper
half-plane for $s<0$.  No singularity is enclosed, so
\be
{\cal I}_\alpha(s)=0,\qquad s<0 .
\ee
For $s>0$ the contour closes in the lower half-plane and the non-zero answer is the
branch-cut contribution.  Keeping the support factor explicit, and using
\eqref{eq:odd_branch_real_axis} to split the real line into negative and positive
frequencies,
\bea
{\cal I}_\alpha(s)
&=&\Theta(s)\left[
\f{1}{2\pi}\int_0^\infty d\omega\, \omega^\alpha e^{-i\omega s}
+\f{e^{i\pi\alpha}}{2\pi}\int_{-\infty}^0 d\omega\, |\omega|^\alpha e^{-i\omega s}
\right]
\non\\
&=&\Theta(s)\left[
\f{1}{2\pi}\int_0^\infty d\omega\, \omega^\alpha e^{-i\omega s}
+\f{e^{i\pi\alpha}}{2\pi}\int_0^\infty d\omega\, \omega^\alpha e^{i\omega s}
\right].
\eea
The two oscillatory integrals are defined with the usual convergence regulator.  For
$s>0$ and ${\rm Re}\,\alpha>-1$,
\be\label{eq:odd_oscillatory_integrals}
\int_0^\infty d\omega\, \omega^\alpha e^{i\omega s}
=e^{i\pi(\alpha+1)/2}\f{\Gamma(\alpha+1)}{s^{\alpha+1}},\qquad
\int_0^\infty d\omega\, \omega^\alpha e^{-i\omega s}
=e^{-i\pi(\alpha+1)/2}\f{\Gamma(\alpha+1)}{s^{\alpha+1}}.
\ee
Thus, away from contact terms at \(s=0\),
\bea\label{eq:odd_power_transform}
{\cal I}_\alpha(s)
&=&\f{\Gamma(\alpha+1)}{2\pi}
\f{\Theta(s)}{s^{\alpha+1}}
\left(e^{-i\pi(\alpha+1)/2}
+e^{i\pi\alpha}e^{i\pi(\alpha+1)/2}\right)
\non\\
&=&
\f{e^{i\pi\alpha/2}}{\Gamma(-\alpha)}
\f{\Theta(s)}{s^{\alpha+1}} .
\eea
In the second equality we used
\(\Gamma(-\alpha)\Gamma(\alpha+1)=-\pi/\sin(\pi\alpha)\).

\paragraph{Integral for the logarithmic waveform:}
The second integral appears in the retarded-time transform of
\eqref{eq:A_acc_odd_frequency}.  Since
\(\partial_\alpha(\omega+i\epsilon)^\alpha
=(\omega+i\epsilon)^\alpha\ln(\omega+i\epsilon)\),
differentiating \eqref{eq:odd_power_transform} with respect to \(\alpha\) gives
\bea\label{eq:odd_log_power_transform}
\int_{-\infty}^{\infty}\f{d\omega}{2\pi}e^{-i\omega s}
(\omega+i\epsilon)^\alpha\ln(\omega+i\epsilon)
&=&
\f{e^{i\pi\alpha/2}}{\Gamma(-\alpha)}
\f{\Theta(s)}{s^{\alpha+1}}
\left[
\f{\Gamma'(-\alpha)}{\Gamma(-\alpha)}
+\f{i\pi}{2}
-\ln s
\right].
\eea

\paragraph{Integral with the negative-frequency projector:}
The third integral is generated by the second term in
\eqref{eq:odd_log_prescription_relation} when transforming
\eqref{eq:A_acc_odd_frequency} to retarded time.  Using
\eqref{eq:odd_branch_real_axis} and first using \(\Theta(-\omega)\) to restrict the
integration range gives
\be\label{eq:odd_negative_frequency_intermediate}
\int_{-\infty}^{\infty}\f{d\omega}{2\pi}\,
e^{-i\omega s}(\omega+i\epsilon)^\alpha\Theta(-\omega)
=
\f{e^{i\pi\alpha}}{2\pi}\int_0^\infty d\omega\,\omega^\alpha e^{i\omega s},
\ee
where we changed variables from \(\omega\) to \(-\omega\).  For \(s>0\), the integral on
the right-hand side is given directly by the first relation in
\eqref{eq:odd_oscillatory_integrals};
for \(s<0\), writing \(s=-|s|\) converts it to the second oscillatory integral in
\eqref{eq:odd_oscillatory_integrals}.  Therefore
\be\label{eq:odd_negative_frequency_piecewise}
\int_{-\infty}^{\infty}\f{d\omega}{2\pi}\,
e^{-i\omega s}(\omega+i\epsilon)^\alpha\Theta(-\omega)
=
\f{\Gamma(\alpha+1)}{2\pi}\,e^{i\pi\alpha}
\begin{cases}
e^{i\pi(\alpha+1)/2}\,s^{-(\alpha+1)}, & s>0,\\[2mm]
e^{-i\pi(\alpha+1)/2}\,|s|^{-(\alpha+1)}, & s<0.
\end{cases}
\ee
Combining the two cases gives, for
non-integer \(\alpha\):
\bea\label{eq:odd_negative_frequency_transform}
\int_{-\infty}^{\infty}\f{d\omega}{2\pi}\,
e^{-i\omega s}(\omega+i\epsilon)^\alpha\Theta(-\omega)
&=&
\f{\Gamma(\alpha+1)}{2\pi}\,e^{i\pi\alpha}
\left[
e^{i\pi(\alpha+1)/2}\f{\Theta(s)}{s^{\alpha+1}}
+
e^{-i\pi(\alpha+1)/2}\f{\Theta(-s)}{|s|^{\alpha+1}}
\right].
\eea

\section{Evaluation of the acceleration integral and position-space check}
\label{app:acceleration_integral}

\begin{figure}[t]
\centering
\begin{tikzpicture}[
  x=0.88cm,y=0.88cm,
  contour/.style={line width=0.8pt,postaction={decorate},
    decoration={markings,mark=at position 0.58 with {\arrow{Stealth}}}},
  greenpole/.style={red!75!black,line width=0.9pt},
  simple/.style={blue!70!black,fill=blue!70!black},
  higher/.style={blue!70!black,line width=0.8pt}
]
\begin{scope}[xshift=-3.4cm]
  \draw[->] (-2.8,0)--(2.9,0) node[right] {\(\operatorname{Re}\ell^0\)};
  \draw[->] (0,-2.70)--(0,2.70) node[above] {\(\operatorname{Im}\ell^0\)};
  \draw[contour] (-2.55,0)--(2.55,0);
  \draw[contour] (2.55,0) arc[start angle=0,end angle=-180,radius=2.55];
  \draw[greenpole] (-2.10,-0.31)--(-1.92,-0.13);
  \draw[greenpole] (-2.10,-0.13)--(-1.92,-0.31);
  \draw[greenpole] (1.92,-0.31)--(2.10,-0.13);
  \draw[greenpole] (1.92,-0.13)--(2.10,-0.31);
  \fill[simple] (-0.65,-0.22) circle (2.1pt);
  \draw[higher] (0.72,0.22) rectangle +(0.14,0.14);
  \node[below left] at (-2.01,-0.31) {\(\ell^0_{\gamma,-}\)};
  \node[below right] at (2.01,-0.31) {\(\ell^0_{\gamma,+}\)};
  \node[below] at (-0.65,-0.30) {\(\beta_aq-i\epsilon\)};
  \node[above] at (0.79,0.40) {\(\beta_bq+i\epsilon\)};
  \node[font=\small] at (0,-2.98) {(a) outgoing pair};
\end{scope}
\begin{scope}[xshift=3.4cm]
  \draw[->] (-2.8,0)--(2.9,0) node[right] {\(\operatorname{Re}\ell^0\)};
  \draw[->] (0,-2.70)--(0,2.70) node[above] {\(\operatorname{Im}\ell^0\)};
  \draw[contour] (-2.55,0)--(2.55,0);
  \draw[contour] (2.55,0) arc[start angle=0,end angle=180,radius=2.55];
  \draw[greenpole] (-2.10,-0.31)--(-1.92,-0.13);
  \draw[greenpole] (-2.10,-0.13)--(-1.92,-0.31);
  \draw[greenpole] (1.92,-0.31)--(2.10,-0.13);
  \draw[greenpole] (1.92,-0.13)--(2.10,-0.31);
  \fill[simple] (-0.65,0.22) circle (2.1pt);
  \draw[higher] (0.72,-0.36) rectangle +(0.14,0.14);
  \node[below left] at (-2.01,-0.31) {\(\ell^0_{\gamma,-}\)};
  \node[below right] at (2.01,-0.31) {\(\ell^0_{\gamma,+}\)};
  \node[above] at (-0.65,0.30) {\(\beta'_aq+i\epsilon\)};
  \node[below] at (0.79,-0.44) {\(\beta'_bq-i\epsilon\)};
  \node[font=\small] at (0,-2.98) {(b) incoming pair};
\end{scope}
\end{tikzpicture}
\caption{Retarded \(\ell^0\)-plane contours used in
appendix~\ref{app:acceleration_integral}.  Red crosses denote the two retarded Green's function poles
\(\ell^0_{\gamma,\pm}=\pm L-i\epsilon\), the filled circle denotes the simple
\(a\)-particle pole, and the square denotes the order-\((d-3)\)
\(b\)-particle pole.  The outgoing contour in panel (a) closes clockwise below and
encloses the two retarded Green's function poles and the \(v_a\) pole.  The incoming contour in panel
(b) closes counterclockwise above and encloses only the \(v'_a\) pole.  These pole
configurations apply in both even and odd dimensions; whether \(d\) is even or odd determines whether the
residues of the two outgoing Green's function poles cancel or add after \(q\to-q\).}
\label{fig:retarded_l0_contours}
\end{figure}
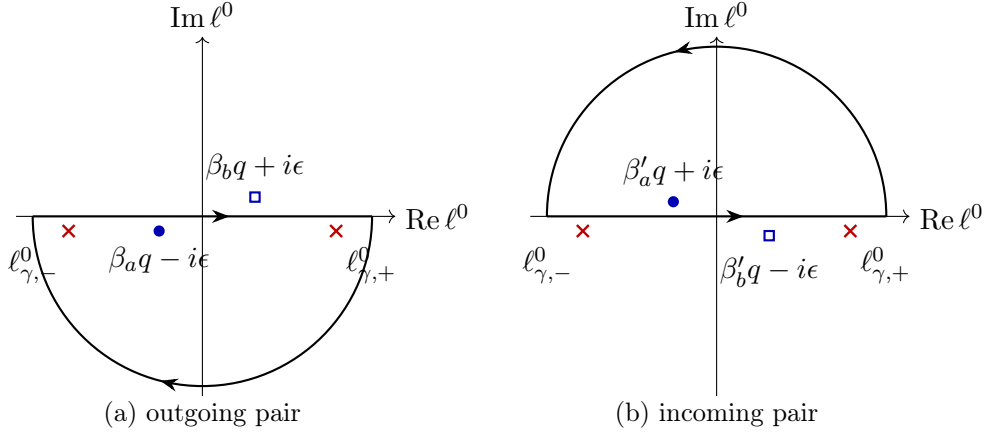

We now evaluate the momentum integral in \eqref{eq:Ib_log_integral}.  Differentiating the
scalar denominator with respect to $v_b$ supplies the factor of $\ell^\rho$ in the tensor
numerator and reduces the calculation to
\bea\label{eq:IJ}
I_b^\rho(k)\big|_{\ln}
&\simeq&
-\f{q_b}{m_b}\f{(v_b.k)^{d-4}}{d-3}
\sum_{\substack{a=1\\a\ne b}}^N
q_a
\left(v_a^\rho v_{b\sigma}-v_a.v_b\,\delta^\rho_\sigma\right)
\f{\p\mathcal J^{\rm class}_{ab}(v_a,v_b)}{\p v_{b\sigma}},
\eea
where
\be\label{eq:Jab_def}
\mathcal J^{\rm class}_{ab}(v_a,v_b)\equiv
\int_\omega\f{d^d\ell}{(2\pi)^d}
\f{1}{(\ell^0+i\epsilon)^2-\vec\ell^2}
\f{1}{(\ell.v_a-i\epsilon)(\ell.v_b+i\epsilon)^{d-3}} .
\ee
As explained below \eqref{eq:Ib_log_integral}, \(\int_\omega\) restricts the
radial integration to scales above a lower cutoff of order \(|\omega|\).  The
appropriate complex boundary value is restored from the proper-time
\(i\epsilon\) prescription after the logarithmic coefficient is evaluated.
The derivative in \eqref{eq:IJ} is an ordinary component derivative in the ambient
vector space: \(v_a\) and \(v_b\) are treated as independent, unconstrained vectors
while differentiating.  One may use unit-normalized velocities to evaluate the scalar
integral, but its separately homogeneous covariant extension, with all dependence on
\(v_a^2\) and \(v_b^2\) restored, must be used in \eqref{eq:IJ}.  The physical
conditions \(v_a^2=v_b^2=-1\) are imposed only after the derivative and its contraction
have been evaluated.  Imposing them earlier would replace the required ambient
derivative by an unspecified derivative of a function known only on the unit hyperboloid
and could omit derivatives of the explicit norm factors.  The same convention applies to
the primed and mixed velocity pairs below.
We display the velocity arguments in the final covariant formulas and suppress them
in intermediate contour formulas when no ambiguity can arise.
The superscript emphasizes that
\(\mathcal J^{\rm class}_{ab}\) is defined using the retarded Green's function that
enters the classical current.  It is Lorentz invariant under proper orthochronous
transformations.  We evaluate its logarithmic
coefficient in a frame where the two velocities are collinear:
\be\label{eq:collinear_velocities}
v_a=\gamma_a(1,\beta_a,\vec 0_{d-2}),\qquad
v_b=\gamma_b(1,\beta_b,\vec 0_{d-2}),\qquad
\gamma_i=(1-\beta_i^2)^{-1/2}.
\ee
We keep all frame-dependent quantities below directly in terms of
\(\beta_a,\beta_b,\gamma_a,\gamma_b\); the even- and odd-dimensional cases will ultimately be written in
terms of Lorentz invariants.
Let \(q\) be the spatial component of \(\vec\ell\) in the common direction,
\(\vec\ell_\perp\) the remaining \(d-2\) components, and
\(L=(q^2+\vec\ell_\perp^{\,2})^{1/2}\).  With mostly-plus signature,
\(\ell.v_i=-\gamma_i(\ell^0-\beta_iq)\).  The retarded Green's function poles and the
\((\ell.v_a-i\epsilon)^{-1}\) pole lie in the lower half-plane, whereas
\((\ell.v_b+i\epsilon)^{-(d-3)}\) has an order-\((d-3)\) pole in the upper half-plane at
\(\ell^0=\beta_bq+i\epsilon\).  The large-semicircle contribution vanishes, so either
closure gives the same real-axis integral after the contour orientation is included.
Closing in the upper half-plane picks only the higher-order pole from
\((\ell.v_b+i\epsilon)^{-(d-3)}\) and gives a
$(d-4)$-th derivative representation.  To distinguish the even- and odd-dimensional
cases below, it is more useful to
close in the lower half-plane.  The contour is then clockwise and encloses the simple
pole from \((\ell.v_a-i\epsilon)^{-1}\) and the two retarded Green's function poles, as shown in
Figure~\ref{fig:retarded_l0_contours}(a).  The reversed prescriptions for two incoming
lines give Figure~\ref{fig:retarded_l0_contours}(b), where the upper contour encloses
only the simple \(v'_a\) pole.

Including the clockwise orientation in
Figure~\ref{fig:retarded_l0_contours}(a), the sum of the enclosed residues gives
\bea\label{eq:Jab_collinear_log}
\mathcal J^{\rm class}_{ab}\big|_{\mathrm{out},\,\ln}
&=&
\f{i}{\gamma_a\gamma_b^{d-3}}
\int_\omega\f{d^{d-2}\vec\ell_\perp}{(2\pi)^{d-2}}
\int\f{dq}{2\pi}\Bigg[
-\f{1}{q^2(1-\beta_a^2)+\vec\ell_\perp^{\,2}}\,
\f{1}{(q(\beta_b-\beta_a)+i\epsilon)^{d-3}}
\non\\
&&\hspace{2.0cm}
+\f{1}{2L(L-\beta_aq)}
\f{1}{(\beta_bq-L+i\epsilon)^{d-3}}
\non\\
&&\hspace{2.0cm}
+\f{1}{2L(L+\beta_aq)}
\f{1}{(\beta_bq+L+i\epsilon)^{d-3}}
\Bigg].
\eea
Each term inside the square bracket of \eqref{eq:Jab_collinear_log} scales as $\lambda^{-(d-1)}$ under
$(q,\vec\ell_\perp)\to\lambda(q,\vec\ell_\perp)$.  Since the remaining spatial measure
has dimension $d-1$, this is precisely the logarithmic radial scaling.

The singular velocity denominator is treated using the general distribution identity
\be\label{eq:velocity_distribution_identity}
\f{1}{\left((\beta_b-\beta_a)q+i\epsilon\right)^{d-3}}
=\mathop{\rm FP}\f{1}{\left((\beta_b-\beta_a)q\right)^{d-3}}
-\f{i\pi(-1)^{d-4}}{(d-4)!}
\delta^{(d-4)}\left((\beta_b-\beta_a)q\right).
\ee
Here \(\mathop{\rm FP}\) denotes the Hadamard finite-part distribution used throughout
this appendix; for \(d=4\), it reduces to the ordinary Cauchy principal value.  For a
smooth function \(g(x)\) and \(j\in\mathbb Z_{>0}\), the finite part needed below is
defined by subtracting the divergent even Taylor coefficients of \(g\) about \(x=0\):
\bea\label{eq:hadamard_finite_part_evaluation}
\mathop{\rm FP}\int_{-1}^{1}dx\,\f{g(x)}{x^{2j}}
&\equiv&
\lim_{\eta\to0^+}
\left[
\int_{-1}^{-\eta}dx\,\f{g(x)}{x^{2j}}
+\int_{\eta}^{1}dx\,\f{g(x)}{x^{2j}}
\right.
\non\\
&&\left.\hspace{1.0cm}
-2\sum_{r=0}^{j-1}
\f{g^{(2r)}(0)}
{(2r)!(2j-2r-1)\eta^{2j-2r-1}}
\right].
\eea

\subsection{Even spacetime dimensions}
For even $d$, $d-3$ is odd.  Under $q\to-q$, the last two terms of
\eqref{eq:Jab_collinear_log}, generated by the two retarded Green's function poles enclosed in
Figure~\ref{fig:retarded_l0_contours}(a), cancel.  Indeed \(L\) is even in \(q\), and
\(\beta_bq\pm L\) never vanishes for the massive velocity \(|\beta_b|<1\), so the
\(i\epsilon\) prescription does not obstruct this cancellation under \(q\to-q\).  The finite-part
contribution of the first term is also odd.  Since
\((-1)^{d-4}=1\), only the delta-function term in
\eqref{eq:velocity_distribution_identity} survives.
Since $d-4$ is even,
\be\label{eq:even_delta_scaling}
\delta^{(d-4)}\left((\beta_b-\beta_a)q\right)
=\f{1}{|\beta_b-\beta_a|^{d-3}}\delta^{(d-4)}(q).
\ee
Substitution into \eqref{eq:Jab_collinear_log} gives
\bea
\mathcal J^{\rm class}_{ab}
\big|_{\substack{d=\mathrm{even}\\ \mathrm{out},\,\ln}}
&=&
-\f{1}{2\gamma_a\gamma_b^{d-3}|\beta_b-\beta_a|^{d-3}(d-4)!}
\int_\omega\f{d^{d-2}\vec\ell_\perp}{(2\pi)^{d-2}}
\int dq\,
\f{\delta^{(d-4)}(q)}
{q^2(1-\beta_a^2)+\vec\ell_\perp^{\,2}} .
\eea
The one-dimensional distributional integral is fixed without any further contour
choice:
\bea\label{eq:even_delta_derivative}
\int dq\,
\f{\delta^{(d-4)}(q)}
{q^2(1-\beta_a^2)+\vec\ell_\perp^{\,2}}
&=&
(-1)^{d-4}\left(\f{\p}{\p q}\right)^{d-4}
\f{1}{q^2(1-\beta_a^2)+\vec\ell_\perp^{\,2}}
\Bigg|_{q=0}
\non\\
&=&
(-1)^\f{d-4}{2}(d-4)!\,
\f{(1-\beta_a^2)^\f{d-4}{2}}{|\ell_\perp|^{d-2}} .
\eea
Combining \eqref{eq:even_delta_derivative} with the preceding expression gives
\bea
\mathcal J^{\rm class}_{ab}
\big|_{\substack{d=\mathrm{even}\\ \mathrm{out},\,\ln}}
&=&
-\f{(-1)^\f{d-4}{2}}
{2\left(\gamma_a\gamma_b|\beta_a-\beta_b|\right)^{d-3}}
\int_\omega\f{d^{d-2}\vec\ell_\perp}{(2\pi)^{d-2}}
\f{1}{|\ell_\perp|^{d-2}} .
\eea
The remaining transverse integral is
\be\label{eq:even_radial_log}
\int_\omega\f{d^{d-2}\vec\ell_\perp}{(2\pi)^{d-2}}
\f{1}{|\ell_\perp|^{d-2}}
\simeq
-\f{2\pi^\f{d-2}{2}}{(2\pi)^{d-2}\Gamma\left(\f{d-2}{2}\right)}
\ln\omega .
\ee
Only the logarithm of the lower limit is retained in
\eqref{eq:even_radial_log}.  Its branch is
fixed by the outgoing proper-time integral in \eqref{eq:Ib_def}: the convergence factor is
\(e^{-i(k.v_b-i\epsilon)\tau_b}\) with \(\tau_b>0\), or equivalently the denominator after
the \(\tau_b\) integration is \(((\ell-k).v_b+i\epsilon)^{-1}\).  Hence the logarithmic
cutoff is \(-k.v_b+i\epsilon=\omega(-\mathbf n.v_b)+i\epsilon\).  Since
\(-\mathbf n.v_b>0\), the factor \(\ln(-\mathbf n.v_b)\) is independent of \(\omega\) and
is discarded under the \(\simeq\) prescription.  Using the proper-time normalization
\(v_a^2=v_b^2=-1\), one also has
\be\label{eq:relative_velocity_invariant}
\gamma_a\gamma_b(1-\beta_a\beta_b)=-v_a.v_b,\qquad
\left[\gamma_a\gamma_b(\beta_a-\beta_b)\right]^2
=(v_a.v_b)^2-v_a^2v_b^2.
\ee
Restoring the retarded prescription, the result is
\bea\label{eq:Jab_even_result}
\mathcal J^{\rm class}_{ab}(v_a,v_b)
\big|_{\substack{d=\mathrm{even}\\ \mathrm{out},\,\ln}}
&\simeq&
\ln(\omega+i\epsilon)\,
\f{(-1)^\f{d-4}{2}}
{(4\pi)^\f{d-2}{2}\Gamma\left(\f{d-2}{2}\right)}
\non\\
&&\times
\f{(-v_a^2)^\f{d-4}{2}}
{\left[(v_a.v_b)^2-v_a^2v_b^2\right]^\f{d-3}{2}} .
\eea
The factor \((-v_a^2)^{(d-4)/2}\) in \eqref{eq:Jab_even_result} restores the degree
\(-1\) in \(v_a\) required by \eqref{eq:Jab_def}; it equals one for a unit-normalized
velocity and is independent of \(v_b\).  To recover the tensor integral, differentiate
the off-shell covariant expression \eqref{eq:Jab_even_result}.  Before imposing either
normalization, one has
\bea\label{eq:Jab_derivative_even}
\left(v_a^\rho v_{b\sigma}-v_a.v_b\,\delta^\rho_\sigma\right)
\f{\p}{\p v_{b\sigma}}
\f{1}{\left[(v_a.v_b)^2-v_a^2v_b^2\right]^\f{d-3}{2}}
&=&
(d-3)v_a^2
\f{v_a^\rho v_b^2-v_a.v_b\,v_b^\rho}
{\left[(v_a.v_b)^2-v_a^2v_b^2\right]^\f{d-1}{2}} .
\eea
Only now setting \(v_a^2=v_b^2=-1\), the factor \(v_a^2=-1\) in
\eqref{eq:Jab_derivative_even} supplies the minus sign that cancels the explicit minus
sign in \eqref{eq:IJ}.  Substituting \eqref{eq:Jab_even_result} into \eqref{eq:IJ}
therefore gives
\bea\label{eq:Ib_log_result_app}
I_b^\rho(k)\big|_{\substack{d=\mathrm{even}\\ \mathrm{out},\,\ln}}
&\simeq&
\omega^{d-4}\ln(\omega+i\epsilon)\,
\f{q_b}{m_b}(v_b.\mathbf{n})^{d-4}\,
\f{(-1)^\f{d-4}{2}}
{(4\pi)^\f{d-2}{2}\Gamma\left(\f{d-2}{2}\right)}
\non\\
&&\times
\sum_{\substack{a=1\\a\ne b}}^N q_a
\f{v_a^\rho v_b^2-v_a.v_b v_b^\rho}
{\left[(v_a.v_b)^2-v_a^2 v_b^2\right]^\f{d-1}{2}} .
\eea
Equation \eqref{eq:Ib_log_result_app} is the result quoted in
\eqref{eq:Ib_log_result}.  At the scalar-integral level, reversing both proper-time
ranges gives
\bea\label{eq:Jab_even_in_result}
\mathcal J^{\prime\,\rm class}_{ab}(v'_a,v'_b)
\big|_{\substack{d=\mathrm{even}\\ \mathrm{in},\,\ln}}
&\simeq&
\ln(\omega-i\epsilon)\,
\f{(-1)^\f{d-4}{2}}
{(4\pi)^\f{d-2}{2}\Gamma\left(\f{d-2}{2}\right)}
\f{(-v_a^{\prime2})^\f{d-4}{2}}
{\left[(v'_a.v'_b)^2-v_a^{\prime2}v_b^{\prime2}\right]^\f{d-3}{2}} .
\eea
The incoming tensor result follows from \eqref{eq:Jab_even_in_result} by the primed
counterpart of \eqref{eq:IJ}; it is obtained from
\eqref{eq:Ib_log_result_app} by the primed replacement and
\(\ln(\omega+i\epsilon)\to\ln(\omega-i\epsilon)\).  Its upper-half-plane contour is
shown in Figure~\ref{fig:retarded_l0_contours}(b).

\subsection{Proper-time check from the position-space trajectory}
\label{app:position_space_integral_check}

The contour evaluation above can be checked without expanding the momentum
integrand.  We instead insert the position-space velocity tail
\eqref{eq:position_even_velocity_tail} directly into the proper-time integral
\eqref{eq:Ib_def}.  First split the exact integral at a matching time \(T_b\):
\be\label{eq:position_check_split_integral}
I_b^\rho(k)
=
\int_0^{T_b}d\tau_b\,
e^{-i(k.v_b-i\epsilon)\tau_b}\dot Y_b^\rho(\tau_b)
+
\int_{T_b}^{\infty}d\tau_b\,
e^{-i(k.v_b-i\epsilon)\tau_b}\dot Y_b^\rho(\tau_b).
\ee
The first term is entire in \(k\).  The choice of \(T_b\) is otherwise
irrelevant: it is any time of order \(\mathcal R\) for which
\eqref{eq:position_even_velocity_tail} is valid and
\(\omega T_b\ll1\).  The leading term in the second integral is therefore
\be\label{eq:position_check_leading_tail}
\begin{aligned}
I_b^\rho(k)\big|_{\rm leading\ tail}
&=-\f{q_b}{m_b}\,
\f{(d-4)!}
{(4\pi)^\f{d-2}{2}\Gamma\!\left(\f{d-2}{2}\right)}
\sum_{\substack{a=1\\a\ne b}}^N q_a\,
\f{v_a^\rho v_b^2-(v_a.v_b)v_b^\rho}
{\left[(v_a.v_b)^2-v_a^2v_b^2\right]^\f{d-1}{2}}
\int_{T_b}^{\infty}d\tau_b\,
\f{e^{-i(k.v_b-i\epsilon)\tau_b}}{\tau_b^{d-3}} .
\end{aligned}
\ee
The remaining integral is exact:
\be\label{eq:position_check_incomplete_gamma}
\int_{T_b}^{\infty}d\tau_b\,
\f{e^{-i(k.v_b-i\epsilon)\tau_b}}{\tau_b^{d-3}}
=
\left[i(k.v_b-i\epsilon)\right]^{d-4}
\Gamma\!\left(4-d,i(k.v_b-i\epsilon)T_b\right).
\ee
The recurrence relation for the incomplete gamma function, terminating at
\(\Gamma(0,z)\), now determines the small-\(k\) structure without assuming its
form:
\be\label{eq:position_check_tail_transform}
\begin{aligned}
&\int_{T_b}^{\infty}d\tau_b\,
\f{e^{-i(k.v_b-i\epsilon)\tau_b}}{\tau_b^{d-3}}
=-\f{\left[i(k.v_b-i\epsilon)\right]^{d-4}}{(d-4)!}
\ln\!\left[i(k.v_b-i\epsilon)T_b\right]
\\
&\quad+
\f{1}{(d-4)!}\Bigg\{
\left[i(k.v_b-i\epsilon)\right]^{d-4}
\left[-\gamma_{\rm E}
-\sum_{r=1}^{\infty}
\f{\left[-i(k.v_b-i\epsilon)T_b\right]^r}{r\,r!}\right]
\\
&\hspace{2.8cm}
-e^{-i(k.v_b-i\epsilon)T_b}
\sum_{j=1}^{d-4}(-1)^{j-1}(j-1)!\,
\left[i(k.v_b-i\epsilon)\right]^{d-4-j}T_b^{-j}
\Bigg\}.
\end{aligned}
\ee
Here \(\gamma_{\rm E}\) is the Euler--Mascheroni constant.
The second and third lines of \eqref{eq:position_check_tail_transform} are an
entire function of \(k.v_b\), since the displayed series has infinite radius of
convergence.  The logarithm in the first line has therefore not been assumed: it
is forced by the exact transform, equivalently by the failure of the
\((d-4)\)-th moment of the leading velocity tail to exist.
Substitution in \eqref{eq:position_check_split_integral} gives
\be\label{eq:position_check_Ib_log}
\begin{aligned}
I_b^\rho(k)\big|_{\ln}
&=
\f{q_b}{m_b}\,
\f{1}
{(4\pi)^\f{d-2}{2}\Gamma\!\left(\f{d-2}{2}\right)}
\sum_{\substack{a=1\\a\ne b}}^N q_a\,
\f{v_a^\rho v_b^2-(v_a.v_b)v_b^\rho}
{\left[(v_a.v_b)^2-v_a^2v_b^2\right]^\f{d-1}{2}}
\\
&\quad\times
\left[i(k.v_b-i\epsilon)\right]^{d-4}
\ln\!\left[i(k.v_b-i\epsilon)T_b\right].
\end{aligned}
\ee

It remains to verify that the omitted trajectory terms cannot change this
coefficient.  Equation \eqref{eq:position_even_velocity_tail} states precisely
that
\be\label{eq:position_check_remainder_bound}
\begin{aligned}
&\dot Y_b^\rho(\tau_b)
+\f{q_b}{m_b}\,
\f{(d-4)!}
{(4\pi)^\f{d-2}{2}\Gamma\!\left(\f{d-2}{2}\right)}
\sum_{\substack{a=1\\a\ne b}}^N q_a\,
\f{v_a^\rho v_b^2-(v_a.v_b)v_b^\rho}
{\left[(v_a.v_b)^2-v_a^2v_b^2\right]^\f{d-1}{2}}
\f{1}{\tau_b^{d-3}}
=\mathcal O\!\left(\tau_b^{-(d-2)}\right).
\end{aligned}
\ee
Consequently, all the moments
\be\label{eq:position_check_remainder_moments}
\begin{aligned}
&\int_{T_b}^{\infty}d\tau_b\,\tau_b^j
\Bigg[
\dot Y_b^\rho(\tau_b)
+\f{q_b}{m_b}\,
\f{(d-4)!}
{(4\pi)^\f{d-2}{2}\Gamma\!\left(\f{d-2}{2}\right)}
\sum_{\substack{a=1\\a\ne b}}^N q_a\,
\f{v_a^\rho v_b^2-(v_a.v_b)v_b^\rho}
{\left[(v_a.v_b)^2-v_a^2v_b^2\right]^\f{d-1}{2}}
\f{1}{\tau_b^{d-3}}
\Bigg]
\quad\hbox{exist},
\\
&\hspace{7cm}0\leq j\leq d-4 .
\end{aligned}
\ee
Taylor's theorem for the Laplace transform then gives
\be\label{eq:position_check_remainder_transform}
\begin{aligned}
&\int_{T_b}^{\infty}d\tau_b\,
e^{-i(k.v_b-i\epsilon)\tau_b}
\Bigg[
\dot Y_b^\rho(\tau_b)
\\[-1mm]
&\hspace{1.4cm}
+\f{q_b}{m_b}\,
\f{(d-4)!}
{(4\pi)^\f{d-2}{2}\Gamma\!\left(\f{d-2}{2}\right)}
\sum_{\substack{a=1\\a\ne b}}^N q_a\,
\f{v_a^\rho v_b^2-(v_a.v_b)v_b^\rho}
{\left[(v_a.v_b)^2-v_a^2v_b^2\right]^\f{d-1}{2}}
\f{1}{\tau_b^{d-3}}
\Bigg]
\\
&=
\sum_{j=0}^{d-4}
\f{\left[-i(k.v_b-i\epsilon)\right]^j}{j!}
\int_{T_b}^{\infty}d\tau_b\,\tau_b^j
\Bigg[
\dot Y_b^\rho(\tau_b)
\\[-1mm]
&\hspace{1.4cm}
+\f{q_b}{m_b}\,
\f{(d-4)!}
{(4\pi)^\f{d-2}{2}\Gamma\!\left(\f{d-2}{2}\right)}
\sum_{\substack{a=1\\a\ne b}}^N q_a\,
\f{v_a^\rho v_b^2-(v_a.v_b)v_b^\rho}
{\left[(v_a.v_b)^2-v_a^2v_b^2\right]^\f{d-1}{2}}
\f{1}{\tau_b^{d-3}}
\Bigg]
\\
&\quad+o\!\left(\left|k.v_b-i\epsilon\right|^{d-4}\right),
\\
&\hspace{5.3cm}k.v_b-i\epsilon\to0 .
\end{aligned}
\ee
The remainder need not be analytic without a complete asymptotic expansion, but
\eqref{eq:position_check_remainder_transform} is sufficient to exclude an
\(\left[k.v_b-i\epsilon\right]^{d-4}\ln(k.v_b-i\epsilon)\) term.
Using \(k=\omega\mathbf n\) in
\eqref{eq:position_check_Ib_log} and retaining the non-analytic boundary value gives
\be\label{eq:position_check_match_Ib}
\begin{aligned}
I_b^\rho(k)\big|_{d=\mathrm{even},\,\ln}
&\simeq
\omega^{d-4}\ln(\omega+i\epsilon)\,
\f{q_b}{m_b}(\mathbf n.v_b)^{d-4}
\f{(-1)^\f{d-4}{2}}
{(4\pi)^\f{d-2}{2}\Gamma\!\left(\f{d-2}{2}\right)}
\\[-1mm]
&\quad\times
\sum_{\substack{a=1\\a\ne b}}^N q_a\,
\f{v_a^\rho v_b^2-(v_a.v_b)v_b^\rho}
{\left[(v_a.v_b)^2-v_a^2v_b^2\right]^\f{d-1}{2}} .
\end{aligned}
\ee
Equation \eqref{eq:position_check_match_Ib} reproduces
\eqref{eq:Ib_log_result_app}.  Changing \(T_b\) changes only the analytic monomial
\(\left[i(k.v_b-i\epsilon)\right]^{d-4}\ln(T_b'/T_b)\), which is cancelled by
the other pieces of the exact
split.  This proves directly that the finite proper-time region, the endpoint
positions appearing through the source phase, and the subleading trajectory tail
cannot modify the logarithmic coefficient selected by the marginal momentum-space
term.  The incoming calculation is obtained by reversing the proper-time range and
gives \(\ln(\omega-i\epsilon)\).

\subsection{Odd spacetime dimensions}
\paragraph{Outgoing contribution:}
For odd \(d\geq5\), \(d-3\) is even.  We continue directly from
\eqref{eq:Jab_collinear_log}, with the velocities kept in the form
\eqref{eq:collinear_velocities}.  Figure~\ref{fig:retarded_l0_contours}(a) shows
the simple \(v_a\) pole and the two retarded Green's function poles below the real axis,
and the order-\((d-3)\) \(v_b\) pole above it.  The clockwise lower contour encloses
the three lower-half-plane poles.  In
\eqref{eq:velocity_distribution_identity}, odd \(d\) gives
\((-1)^{d-4}=-1\), so the delta-function term has the opposite sign from the
even-dimensional case.  Since \(d-4\) is odd, this term is odd in \(q\) and integrates to zero
against the remaining even factor.  In the last two terms of
\eqref{eq:Jab_collinear_log}, the denominators \(\beta_bq\pm L\) do not vanish for
\(|\beta_b|<1\), so their \(i\epsilon\) prescriptions may be omitted.  These two
terms are exchanged by \(q\to-q\); hence their integrals over the symmetric
\(q\)-range are equal and can be combined.  Equation \eqref{eq:Jab_collinear_log}
therefore becomes
\bea\label{eq:Jab_odd_finite_part}
\mathcal J^{\rm class}_{ab}(v_a,v_b)
\big|_{\substack{d=\mathrm{odd}\\ \mathrm{out},\,\ln}}
&=&
\f{i}{\gamma_a\gamma_b^{d-3}}
\int_\omega\f{d^{d-2}\vec\ell_\perp}{(2\pi)^{d-2}}
\int\f{dq}{2\pi}\,
\Bigg[
-\f{1}{L^2-\beta_a^2q^2}\,
\mathop{\rm FP}
\f{1}{\left((\beta_b-\beta_a)q\right)^{d-3}}
\non\\
&&\hspace{1.0cm}
+\f{1}{L(L-\beta_aq)(L-\beta_bq)^{d-3}}
\Bigg].
\eea
Setting \(x=q/L\), the spatial measure becomes
\be\label{eq:odd_spatial_measure}
d^{d-2}\vec\ell_\perp\,dq
=
\f{2\pi^\f{d-2}{2}}{\Gamma\left(\f{d-2}{2}\right)}
L^{d-2}dL\,(1-x^2)^\f{d-4}{2}dx .
\ee
Both terms in \eqref{eq:Jab_odd_finite_part} scale as \(L^{-(d-1)}\), so
\eqref{eq:odd_spatial_measure} leaves the common radial factor \(dL/L\) and gives
\bea\label{eq:Jab_odd_angular_integral}
\mathcal J^{\rm class}_{ab}
\big|_{\substack{d=\mathrm{odd}\\ \mathrm{out},\,\ln}}
&=&
\f{i}{2^{d-2}\pi^\f d2
\Gamma\left(\f{d-2}{2}\right)\gamma_a\gamma_b^{d-3}}
\int_\omega\f{dL}{L}
\non\\
&&\times
\int_{-1}^{1}dx\,
(1-x^2)^\f{d-4}{2}
\left[
-\f{1}{1-\beta_a^2x^2}\,
\mathop{\rm FP}\f{1}
{\left((\beta_b-\beta_a)x\right)^{d-3}}
\right.
\non\\
&&\left.\hspace{4.0cm}
+\f{1}{(1-\beta_ax)(1-\beta_bx)^{d-3}}
\right].
\eea
We now evaluate the two angular terms in \eqref{eq:Jab_odd_angular_integral}
separately.  Since \(d-3\) is even, the finite-part integrand is even.  Setting
\(t=x^2\) and using the analytically continued Euler integral prescribed by
\eqref{eq:hadamard_finite_part_evaluation} gives
\bea\label{eq:odd_fp_angular_piece}
&&-\f{1}{(\beta_b-\beta_a)^{d-3}}
\mathop{\rm FP}\int_{-1}^{1}dx\,
\f{(1-x^2)^\f{d-4}{2}}
{x^{d-3}(1-\beta_a^2x^2)}
=-\f{\Gamma\left(\f{4-d}{2}\right)
\Gamma\left(\f{d-2}{2}\right)}
{(\beta_b-\beta_a)^{d-3}\gamma_a^{d-4}}.
\eea
For the regular term, set \(y=(x-\beta_b)/(1-\beta_bx)\), which maps
\([-1,1]\) to itself.  After the substitution, averaging the integrand under
\(y\to-y\) removes its odd part and gives
\bea\label{eq:odd_regular_even_reduction}
&&\int_{-1}^{1}dx\,
\f{(1-x^2)^\f{d-4}{2}}
{(1-\beta_ax)(1-\beta_bx)^{d-3}}
=\f{\gamma_b^{d-4}}{1-\beta_a\beta_b}
\int_{-1}^{1}dy\,
\f{(1-y^2)^\f{d-4}{2}}
{1-\left(\f{\beta_b-\beta_a}{1-\beta_a\beta_b}\right)^2y^2}.
\eea
The last integral is reduced by writing one factor of \(1-y^2\) as a multiple of its
denominator.  This gives the recursion
\bea\label{eq:odd_regular_recursion}
\int_{-1}^{1}dy\,
\f{(1-y^2)^\f{d-4}{2}}
{1-\left(\f{\beta_b-\beta_a}{1-\beta_a\beta_b}\right)^2y^2}
&=&
\left(\f{1-\beta_a\beta_b}{\beta_b-\beta_a}\right)^2
\int_{-1}^{1}dy\,(1-y^2)^\f{d-6}{2}
\non\\
&&+
\left[1-\left(\f{1-\beta_a\beta_b}{\beta_b-\beta_a}\right)^2\right]
\int_{-1}^{1}dy\,
\f{(1-y^2)^\f{d-6}{2}}
{1-\left(\f{\beta_b-\beta_a}{1-\beta_a\beta_b}\right)^2y^2}.
\eea
Iterating \eqref{eq:odd_regular_recursion} until the exponent is \(-1/2\) and
performing the resulting Euler beta-function integrals gives
\bea\label{eq:odd_regular_iteration}
\int_{-1}^{1}dy\,
\f{(1-y^2)^\f{d-4}{2}}
{1-\left(\f{\beta_b-\beta_a}{1-\beta_a\beta_b}\right)^2y^2}
&=&
\Gamma\left(\f{4-d}{2}\right)
\Gamma\left(\f{d-2}{2}\right)
\f{\left[1-\left(\f{\beta_b-\beta_a}
{1-\beta_a\beta_b}\right)^2\right]^\f{d-4}{2}}
{\left(\f{\beta_b-\beta_a}{1-\beta_a\beta_b}\right)^{d-3}}
\non\\
&&-\Gamma\left(\f{d-2}{2}\right)
\sum_{j=1}^{\f{d-3}{2}}
\f{\Gamma\left(\f12-j\right)}
{\Gamma\left(\f{d-1}{2}-j\right)}
\left(\f{1-\beta_a\beta_b}{\beta_a-\beta_b}\right)^{2j}.
\eea
Since
\(1-[(\beta_b-\beta_a)/(1-\beta_a\beta_b)]^2
=1/[\gamma_a^2\gamma_b^2(1-\beta_a\beta_b)^2]\), the first term in
\eqref{eq:odd_regular_iteration}, after including the prefactor in
\eqref{eq:odd_regular_even_reduction}, cancels \eqref{eq:odd_fp_angular_piece}.
The remaining finite sum yields
\bea\label{eq:odd_angular_sum}
&&\int_{-1}^{1}dx\,
(1-x^2)^\f{d-4}{2}
\left[
-\f{1}{1-\beta_a^2x^2}\,
\mathop{\rm FP}\f{1}
{\left((\beta_b-\beta_a)x\right)^{d-3}}
+\f{1}{(1-\beta_ax)(1-\beta_bx)^{d-3}}
\right]
\non\\
&&\qquad=
-\f{\Gamma\left(\f{d-2}{2}\right)\gamma_b^{d-4}}
{1-\beta_a\beta_b}
\sum_{j=1}^{\f{d-3}{2}}
\f{\Gamma\left(\f12-j\right)}
{\Gamma\left(\f{d-1}{2}-j\right)}
\left(\f{1-\beta_a\beta_b}{\beta_a-\beta_b}\right)^{2j}.
\eea
The radial integral in \eqref{eq:Jab_odd_angular_integral} is
\be\label{eq:odd_radial_log}
\int_\omega\f{dL}{L}\simeq-\ln\omega ,
\ee
where the outgoing proper-time range fixes the prescription as
\(\ln(\omega+i\epsilon)\).  Substituting \eqref{eq:odd_angular_sum} and
\eqref{eq:odd_radial_log} into \eqref{eq:Jab_odd_angular_integral} gives
\bea\label{eq:Jab_odd_collinear_result}
\mathcal J^{\rm class}_{ab}
\big|_{\substack{d=\mathrm{odd}\\ \mathrm{out},\,\ln}}
&\simeq&
\f{i\ln(\omega+i\epsilon)}
{2^{d-2}\pi^\f d2\gamma_a\gamma_b(1-\beta_a\beta_b)}
\sum_{j=1}^{\f{d-3}{2}}
\f{\Gamma\left(\f12-j\right)}
{\Gamma\left(\f{d-1}{2}-j\right)}
\left(\f{1-\beta_a\beta_b}{\beta_a-\beta_b}\right)^{2j}.
\eea
For the normalized velocities in \eqref{eq:collinear_velocities},
\bea\label{eq:odd_collinear_invariant_identities}
\f{1}{\gamma_a\gamma_b(1-\beta_a\beta_b)}
\left(\f{1-\beta_a\beta_b}{\beta_a-\beta_b}\right)^{2j}
&=&
-\f{(v_a.v_b)^{2j-1}}
{\left[(v_a.v_b)^2-v_a^2v_b^2\right]^j},
\non\\
\f{\Gamma\left(\f12-j\right)}
{\Gamma\left(\f{d-1}{2}-j\right)}
&=&
\sqrt{\pi}\,
\f{(-4)^j j!}
{(2j)!\left(\f{d-3}{2}-j\right)!}.
\eea
Using \eqref{eq:odd_collinear_invariant_identities} in
\eqref{eq:Jab_odd_collinear_result} fixes the dependence on \(v_a.v_b\) and
\((v_a.v_b)^2-v_a^2v_b^2\) for normalized velocities.  Since
\eqref{eq:Jab_def} is homogeneous of degree \(-1\) in \(v_a\) and degree
\(-(d-3)\) in \(v_b\), restoring these degrees gives
\bea\label{eq:Jab_odd_result}
\mathcal J^{\rm class}_{ab}(v_a,v_b)
\big|_{\substack{d=\mathrm{odd}\\ \mathrm{out},\,\ln}}
&\simeq&
-\f{i\ln(\omega+i\epsilon)}
{2^{d-2}\pi^\f{d-1}{2}}\,
\f{1}{(-v_b^2)^\f{d-4}{2}}
\non\\
&&\times
\sum_{j=1}^{\f{d-3}{2}}
\f{(-4)^j j!}
{(2j)!\left(\f{d-3}{2}-j\right)!}\,
\f{(v_a.v_b)^{2j-1}}
{\left[(v_a.v_b)^2-v_a^2v_b^2\right]^j}.
\eea
In accordance with the rule stated after \eqref{eq:Jab_def}, the normalized expression
\eqref{eq:Jab_odd_collinear_result} is not differentiated directly.  Its homogeneous
off-shell extension \eqref{eq:Jab_odd_result} must be differentiated before imposing
\(v_a^2=v_b^2=-1\).  The contraction in \eqref{eq:IJ} annihilates derivatives of
\(v_a.v_b\); differentiating the remaining two invariants and only then imposing the
normalizations gives
\bea\label{eq:Jab_derivative_odd}
\left.
\left(v_a^\rho v_{b\sigma}-v_a.v_b\,\delta^\rho_\sigma\right)
\f{\p\mathcal J^{\rm class}_{ab}
|_{\substack{d=\mathrm{odd}\\ \mathrm{out},\,\ln}}}{\p v_{b\sigma}}
\right|_{v_a^2=v_b^2=-1}
&\simeq&
\f{i\ln(\omega+i\epsilon)}
{2^{d-2}\pi^\f{d-1}{2}}
\sum_{j=1}^{\f{d-3}{2}}
\f{(-4)^j j!}
{(2j)!\left(\f{d-3}{2}-j\right)!}
\non\\
&&\times
\f{(v_a.v_b)^{2j-1}}
{\left[(v_a.v_b)^2-1\right]^j}
\left[-(d-4)+\f{2j}{(v_a.v_b)^2-1}\right]
\non\\
&&\times
\left(v_a^\rho v_b^2-v_a.v_b\,v_b^\rho\right).
\eea
For \(k=\omega\mathbf n\),
\(v_b.k=-\omega(-\mathbf n.v_b)\).  Since \(d-4\) is odd, the resulting minus sign
cancels the overall minus sign in \eqref{eq:IJ}.  Substitution of
\eqref{eq:Jab_derivative_odd} therefore gives
\bea\label{eq:odd_I_out_covariant}
I_b^\rho(k)\big|_{\substack{d=\mathrm{odd}\\ \mathrm{out},\,\ln}}
&\simeq&
\f{i\,q_b}{m_b}
\f{\omega^{d-4}\ln(\omega+i\epsilon)}{(2\pi)^{d-1}}
\sum_{\substack{a=1\\a\ne b}}^N q_a
(-\mathbf n.v_b)^{d-4}
\non\\
&&\times
\mathcal F_{\rm out}^{(d)}(v_a.v_b)
\left(v_a^\rho v_b^2-v_a.v_b\,v_b^\rho\right),
\eea
where
\bea\label{eq:odd_F_out}
\mathcal F_{\rm out}^{(d)}(z)
&\equiv&
\f{2\pi^\f{d-1}{2}}{d-3}
\sum_{j=1}^{\f{d-3}{2}}
\f{(-4)^j j!}
{(2j)!\left(\f{d-3}{2}-j\right)!}\,
\f{z^{2j-1}}{(z^2-1)^j}
\left[-(d-4)+\f{2j}{z^2-1}\right]
\non\\
&=&
\f{2\pi^\f{d-2}{2}z}{z^2-1}
\left[
\f{\Gamma\left(\f12\right)}
{\Gamma\left(\f{d-1}{2}\right)}
+\sum_{j=1}^{\f{d-3}{2}}
\f{\Gamma\left(\f12-j\right)}
{\Gamma\left(\f{d-1}{2}-j\right)}
\f{z^{2j-2}}{(z^2-1)^j}
\right].
\eea
The second form follows from the gamma-function identity displayed before
\eqref{eq:Jab_odd_result} and an elementary rearrangement of the finite sum.

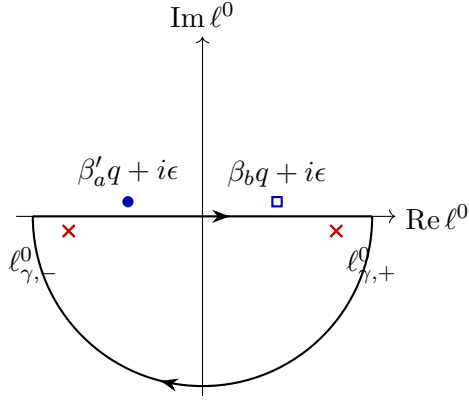
\begin{figure}[t]
\centering
\begin{tikzpicture}[
  x=0.88cm,y=0.88cm,
  contour/.style={line width=0.8pt,postaction={decorate},
    decoration={markings,mark=at position 0.58 with {\arrow{Stealth}}}},
  greenpole/.style={red!75!black,line width=0.9pt},
  simple/.style={blue!70!black,fill=blue!70!black},
  higher/.style={blue!70!black,line width=0.8pt}
]
  \draw[->] (-2.8,0)--(2.9,0) node[right] {\(\operatorname{Re}\ell^0\)};
  \draw[->] (0,-2.70)--(0,2.70) node[above] {\(\operatorname{Im}\ell^0\)};
  \draw[contour] (-2.55,0)--(2.55,0);
  \draw[contour] (2.55,0) arc[start angle=0,end angle=-180,radius=2.55];
  \draw[greenpole] (-2.10,-0.31)--(-1.92,-0.13);
  \draw[greenpole] (-2.10,-0.13)--(-1.92,-0.31);
  \draw[greenpole] (1.92,-0.31)--(2.10,-0.13);
  \draw[greenpole] (1.92,-0.13)--(2.10,-0.31);
  \fill[simple] (-1.12,0.22) circle (2.1pt);
  \draw[higher] (1.05,0.15) rectangle +(0.14,0.14);
  \node[below left] at (-2.01,-0.31) {\(\ell^0_{\gamma,-}\)};
  \node[below right] at (2.01,-0.31) {\(\ell^0_{\gamma,+}\)};
  \node[above] at (-1.12,0.30) {\(\beta'_aq+i\epsilon\)};
  \node[above] at (1.12,0.30) {\(\beta_bq+i\epsilon\)};
\end{tikzpicture}
\caption{Retarded \(\ell^0\)-plane contour for an incoming source acting on an
outgoing trajectory.  The symbols have the same meaning as in
Figure~\ref{fig:retarded_l0_contours}.  Both velocity poles lie above the real axis,
while the clockwise lower contour encloses only the two retarded Green's function poles.}
\label{fig:retarded_mixed_l0_contour}
\end{figure}

\paragraph{Incoming source acting on an outgoing trajectory:}
As shown in \eqref{eq:incoming_outgoing_timelike} and the discussion following it, an
incoming asymptotic point is timelike separated from a sufficiently late point on an
outgoing trajectory.  Since the odd-dimensional retarded Green's function has support at
timelike separation, every incoming asymptotic trajectory contributes to the late-time
outgoing acceleration.  The relevant scalar integral is
\be\label{eq:Jab_mix_def}
\mathcal J^{{\rm class},\,{\rm mix}}_{ab}(v'_a,v_b)\equiv
\int_\omega\f{d^d\ell}{(2\pi)^d}
\f{1}{(\ell^0+i\epsilon)^2-\vec\ell^2}
\f{1}{(\ell.v'_a+i\epsilon)(\ell.v_b+i\epsilon)^{d-3}} .
\ee
We evaluate it in the frame
\be\label{eq:mix_collinear_velocities}
v'_a=\gamma'_a(1,\beta'_a,\vec0_{d-2}),\qquad
v_b=\gamma_b(1,\beta_b,\vec0_{d-2}),\qquad
\gamma_i=(1-\beta_i^2)^{-1/2}.
\ee
Figure~\ref{fig:retarded_mixed_l0_contour} shows the pole positions.  The simple pole
from \((\ell.v'_a+i\epsilon)^{-1}\) and the order-\((d-3)\) pole from
\((\ell.v_b+i\epsilon)^{-(d-3)}\) both lie above the real axis, whereas the two
retarded Green's function poles lie below it.  The clockwise lower contour therefore
encloses only the two retarded Green's function poles.

Including the clockwise orientation in Figure~\ref{fig:retarded_mixed_l0_contour}
gives
\bea\label{eq:Jab_mix_contour}
\mathcal J^{{\rm class},\,{\rm mix}}_{ab}(v'_a,v_b)
\big|_{\substack{d=\mathrm{odd}\\ \mathrm{mix},\,\ln}}
&=&
\f{i}{\gamma'_a\gamma_b^{d-3}}
\int_\omega\f{d^{d-2}\vec\ell_\perp}{(2\pi)^{d-2}}
\int\f{dq}{2\pi}\,
\Bigg[
\f{1}{2L(L-\beta'_aq)(L-\beta_bq)^{d-3}}
\non\\
&&\hspace{2.2cm}
+\f{1}{2L(L+\beta'_aq)(L+\beta_bq)^{d-3}}
\Bigg].
\eea
The two terms in \eqref{eq:Jab_mix_contour} are exchanged by \(q\to-q\).  Setting
\(x=q/L\) and using \eqref{eq:odd_spatial_measure} therefore reduces their sum to the
regular, second angular term in \eqref{eq:Jab_odd_angular_integral}, with
\(\beta_a\) replaced by \(\beta'_a\).  Its evaluation is already given in
\eqref{eq:odd_regular_even_reduction}--\eqref{eq:odd_regular_iteration}.  Here the
full result in \eqref{eq:odd_regular_iteration} must be retained: in the outgoing
calculation its first term cancels \eqref{eq:odd_fp_angular_piece}, whereas
\eqref{eq:Jab_mix_contour} contains no finite-part term.

Including the velocity prefactor in \eqref{eq:Jab_mix_contour}, the result depends on
\be\label{eq:odd_mix_invariant}
z\equiv \gamma'_a\gamma_b(1-\beta'_a\beta_b)=-v'_a.v_b,\qquad z>1 .
\ee
Using \eqref{eq:odd_regular_even_reduction}--\eqref{eq:odd_regular_iteration} and
expressing the same Euler integral in hypergeometric form then gives
\bea\label{eq:odd_mix_H_function}
&&\f{1}{\gamma'_a\gamma_b^{d-3}}
\int_{-1}^{1}dx\,
\f{(1-x^2)^{(d-4)/2}}
{(1-\beta'_ax)(1-\beta_bx)^{d-3}}
=
\mathcal H_d(z),
\non\\
\mathcal H_d(z)
&\equiv &
\f{1}{z}\,
\mathrm B\left(\f12,\f{d-2}{2}\right)
{}_2F_1\left(
1,\f12;\f{d-1}{2};1-\f{1}{z^2}
\right)
\non\\
&=&
\f{\pi}{2^{d-3}}
\left[
\binom{d-3}{\frac{d-3}{2}}
+2\sum_{j=1}^{\frac{d-3}{2}}(-1)^j
\binom{d-3}{\frac{d-3}{2}-j}
\left(\f{z-1}{z+1}\right)^j
\right],
\qquad d\ \mathrm{odd}.
\eea
The last line is a finite polynomial in \((z-1)/(z+1)\).  
The radial integral is again \(\int_\omega dL/L\simeq-\ln\omega\), with the outgoing
proper-time prescription \(\ln(\omega+i\epsilon)\).  Therefore, for normalized
velocities,
\bea\label{eq:Jab_mix_collinear_result}
\mathcal J^{{\rm class},\,{\rm mix}}_{ab}(v'_a,v_b)
\big|_{\substack{d=\mathrm{odd}\\ \mathrm{mix},\,\ln\\
v_a^{\prime2}=v_b^2=-1}}
&\simeq&
-\f{i\ln(\omega+i\epsilon)}
{2^{d-2}\pi^\f d2\Gamma\left(\f{d-2}{2}\right)}
\mathcal H_d(-v'_a.v_b).
\eea
Since \eqref{eq:Jab_mix_def} is homogeneous of degree \(-1\) in \(v'_a\) and degree
\(-(d-3)\) in \(v_b\), its covariant form is
\bea\label{eq:Jab_mix_result}
\mathcal J^{{\rm class},\,{\rm mix}}_{ab}(v'_a,v_b)
\big|_{\substack{d=\mathrm{odd}\\ \mathrm{mix},\,\ln}}
&\simeq&
-\f{i\ln(\omega+i\epsilon)}
{2^{d-2}\pi^\f d2\Gamma\left(\f{d-2}{2}\right)}
\f{1}{\sqrt{-v_a^{\prime2}}\,(-v_b^2)^\f{d-3}{2}}
\non\\
&&\times
\mathcal H_d\left(
-\f{v'_a.v_b}{\sqrt{(-v_a^{\prime2})(-v_b^2)}}
\right).
\eea
To differentiate \eqref{eq:Jab_mix_result}, let
\be\label{eq:odd_mix_z_def}
z\equiv -\f{v'_a.v_b}{\sqrt{(-v_a^{\prime2})(-v_b^2)}} .
\ee
Then
\bea\label{eq:odd_mix_derivative_contractions}
\left(v_a^{\prime\rho}v_{b\sigma}-v'_a.v_b\,\delta^\rho_\sigma\right)
\f{\p(-v_b^2)}{\p v_{b\sigma}}
&=&
-2\left(v_a^{\prime\rho}v_b^2-v'_a.v_b\,v_b^\rho\right),
\non\\
\left(v_a^{\prime\rho}v_{b\sigma}-v'_a.v_b\,\delta^\rho_\sigma\right)
\f{\p z}{\p v_{b\sigma}}
&=&
-\f{z}{v_b^2}
\left(v_a^{\prime\rho}v_b^2-v'_a.v_b\,v_b^\rho\right).
\eea
Applying \eqref{eq:odd_mix_derivative_contractions} to \eqref{eq:Jab_mix_result} and
only then imposing \(v_a^{\prime2}=v_b^2=-1\), we find
\bea\label{eq:Jab_mix_derivative}
\left.
\begin{aligned}
&\left(v_a^{\prime\rho}v_{b\sigma}
-v'_a.v_b\,\delta^\rho_\sigma\right)\\[-1mm]
&\qquad\times
\f{\p \mathcal J^{{\rm class},\,{\rm mix}}_{ab}
|_{\substack{d=\mathrm{odd}\\ \mathrm{mix},\,\ln}}}
{\p v_{b\sigma}}
\end{aligned}
\right|_{v_a^{\prime2}=v_b^2=-1}
&\simeq&
-\f{i\ln(\omega+i\epsilon)}
{2^{d-2}\pi^\f d2\Gamma\left(\f{d-2}{2}\right)}
\non\\
&&\times
\left[
(d-3)\mathcal H_d(-v'_a.v_b)
+(-v'_a.v_b)\mathcal H_d'(-v'_a.v_b)
\right]
\non\\
&&\times
\left(v_a^{\prime\rho}v_b^2-v'_a.v_b\,v_b^\rho\right).
\eea
The incoming source carries the opposite proper-time orientation from an outgoing
source:
\be\label{eq:mixed_proper_time_orientations}
\int_{-\infty}^{0}d\tau'_a\,
e^{-i(\ell.v'_a+i\epsilon)\tau'_a}
=\f{i}{\ell.v'_a+i\epsilon},
\qquad
\int_{0}^{\infty}d\tau_a\,
e^{-i(\ell.v_a-i\epsilon)\tau_a}
=-\f{i}{\ell.v_a-i\epsilon}.
\ee
Consequently, the mixed counterpart of \eqref{eq:IJ} has the opposite overall sign
from the outgoing-source expression.  Replacing
\(\mathcal J^{\rm class}_{ab}(v_a,v_b)\) by
\(\mathcal J^{{\rm class},\,{\rm mix}}_{ab}(v'_a,v_b)\), the mixed contribution is
therefore proportional to
\be\label{eq:mixed_classical_contribution_structure}
+\f{q_b}{m_b}\f{(v_b.k)^{d-4}}{d-3}
\sum_{a=1}^{M}q'_a
\left(v_a^{\prime\rho}v_{b\sigma}
-v'_a.v_b\,\delta^\rho_\sigma\right)
\f{\partial\mathcal J^{{\rm class},\,{\rm mix}}_{ab}}
{\partial v_{b\sigma}} .
\ee
Using \eqref{eq:Jab_mix_derivative} together with
\((v_b.k)^{d-4}=-\omega^{d-4}(-\mathbf n.v_b)^{d-4}\) therefore gives
\bea\label{eq:odd_I_mix_covariant}
I_{b,\,{\rm mix}}^\rho(k)\big|_{\substack{d=\mathrm{odd}\\ \mathrm{mix},\,\ln}}
&\simeq&
\f{i\,q_b}{m_b}
\f{\omega^{d-4}\ln(\omega+i\epsilon)}{(2\pi)^{d-1}}
\sum_{a=1}^{M} q'_a
(-\mathbf n.v_b)^{d-4}
\non\\
&&\times
\mathcal F_{\rm mix}^{(d)}(-v'_a.v_b)
\left(v_a^{\prime\rho}v_b^2-v'_a.v_b\,v_b^\rho\right),
\eea
with
\bea\label{eq:odd_F_mix}
\mathcal F_{\rm mix}^{(d)}(z)
&\equiv&
\f{2\pi^{\frac d2-1}}
{(d-3)\Gamma\left(\f{d-2}{2}\right)}
\left[
(d-3)\mathcal H_d(z)+z\,\mathcal H_d'(z)
\right].
\eea
There is no corresponding contribution to an incoming trajectory from an outgoing
source.  The velocity factors in the would-be contour integral are
\((\ell.v_a-i\epsilon)^{-1}\) for the outgoing source and
\((\ell.v'_b-i\epsilon)^{-(d-3)}\) for the incoming trajectory.  Since
\(\ell.v_i=-\gamma_i(\ell^0-\beta_iq)\), both particle poles lie below the real
\(\ell^0\) axis; the two retarded Green's function poles
\(\ell^0=\pm L-i\epsilon\) lie there as well.  The upper half-plane therefore contains
no poles.  Moreover, the integrand falls as \((\ell^0)^{-d}\), so the contribution from
the upper semicircle vanishes.  Closing the contour in the upper half-plane thus gives
zero, which is the momentum-space statement that the retarded field of a future
outgoing source cannot influence the earlier incoming branch.

\paragraph{Incoming contribution:}
For an incoming trajectory the proper-time ranges reverse the \(i\epsilon\)
prescriptions of the velocity denominators.
Introduce the scalar integral
\be\label{eq:Jab_in_def}
\mathcal J^{\prime\,\rm class}_{ab}(v'_a,v'_b)\equiv
\int_\omega\f{d^d\ell}{(2\pi)^d}
\f{1}{(\ell^0+i\epsilon)^2-\vec\ell^2}
\f{1}{(\ell.v'_a+i\epsilon)(\ell.v'_b-i\epsilon)^{d-3}} .
\ee
Using the primed counterpart of \eqref{eq:collinear_velocities} in
\eqref{eq:Jab_in_def}, Figure~\ref{fig:retarded_l0_contours}(b) shows the two
retarded Green's function poles and the order-\((d-3)\) \(v'_b\) pole below the real
axis, and the simple \(v'_a\) pole above it.  The counterclockwise upper contour
therefore encloses only the simple pole at
\(\ell^0=\beta'_aq+i\epsilon\).  Its residue supplies a factor
\(-1/\gamma'_a\), while the photon propagator evaluated at the pole supplies
\(-1/(L^2-\beta_a^{\prime2}q^2)\); these two minus signs cancel.  The contour integral
therefore gives
\bea\label{eq:Jab_odd_in_contour}
\mathcal J^{\prime\,\rm class}_{ab}(v'_a,v'_b)
\big|_{\substack{d=\mathrm{odd}\\ \mathrm{in},\,\ln}}
&=&
\f{i}{\gamma'_a\gamma_b^{\prime\,d-3}}
\int_\omega\f{d^{d-2}\vec\ell_\perp}{(2\pi)^{d-2}}
\int\f{dq}{2\pi}\,
\f{1}{L^2-\beta_a^{\prime2}q^2}
\f{1}{\left((\beta'_a-\beta'_b)q+i\epsilon\right)^{d-3}}.
\eea
The primed counterpart of \eqref{eq:velocity_distribution_identity} applies to the singular
factor in \eqref{eq:Jab_odd_in_contour}.  Its delta-function term again integrates to
zero because it is an odd derivative acting on an even function of \(q\).  Using
\(x=q/L\) and \eqref{eq:odd_spatial_measure} therefore gives
\bea\label{eq:Jab_odd_in_angular_integral}
\mathcal J^{\prime\,\rm class}_{ab}
\big|_{\substack{d=\mathrm{odd}\\ \mathrm{in},\,\ln}}
&=&
\f{i}{2^{d-2}\pi^\f d2
\Gamma\left(\f{d-2}{2}\right)
\gamma'_a\gamma_b^{\prime\,d-3}(\beta'_a-\beta'_b)^{d-3}}
\int_\omega\f{dL}{L}
\non\\
&&\times
\mathop{\rm FP}\int_{-1}^{1}dx\,
\f{(1-x^2)^\f{d-4}{2}}
{x^{d-3}(1-\beta_a^{\prime2}x^2)}.
\eea
The finite-part angular integral in \eqref{eq:Jab_odd_in_angular_integral} is
\bea\label{eq:odd_incoming_angular}
\mathop{\rm FP}\int_{-1}^{1}dx\,
\f{(1-x^2)^\f{d-4}{2}}
{x^{d-3}(1-\beta_a^{\prime2}x^2)}
&=&
\mathrm B\left(\f{4-d}{2},\f{d-2}{2}\right)
{}_2F_1\left(
1,\f{4-d}{2};1;
\beta_a^{\prime2}
\right)
\non\\
&=&
(-1)^\f{d-3}{2}\pi
\left(1-\beta_a^{\prime2}\right)^\f{d-4}{2}.
\eea
The first equality in \eqref{eq:odd_incoming_angular} is its analytically continued
Euler representation.  This is the same finite-part integral already evaluated in
\eqref{eq:odd_fp_angular_piece}: removing the common factor
\(-1/(\beta_b-\beta_a)^{d-3}\) there and replacing \(\beta_a\) by \(\beta'_a\)
reproduces the same result.  The equivalence of the two lines follows from the
hypergeometric and gamma-function identities displayed in
\eqref{eq:odd_incoming_angular}.  Substituting
\eqref{eq:odd_incoming_angular} and \eqref{eq:odd_radial_log} into
\eqref{eq:Jab_odd_in_angular_integral}, with the incoming prescription
\(\ln(\omega-i\epsilon)\), gives
\bea\label{eq:Jab_odd_in_collinear_result}
\mathcal J^{\prime\,\rm class}_{ab}
\big|_{\substack{d=\mathrm{odd}\\ \mathrm{in},\,\ln}}
&\simeq&
-\f{i(-1)^\f{d-3}{2}\ln(\omega-i\epsilon)}
{2^{d-2}\pi^\f{d-2}{2}\Gamma\left(\f{d-2}{2}\right)}
\f{1}{
\left[\gamma'_a\gamma'_b(\beta'_a-\beta'_b)\right]^{d-3}}.
\eea
For the normalized velocities in the primed counterpart of
\eqref{eq:collinear_velocities},
\be\label{eq:odd_incoming_collinear_invariant}
\left[\gamma'_a\gamma'_b(\beta'_a-\beta'_b)\right]^2
=(v'_a.v'_b)^2-v_a^{\prime2}v_b^{\prime2}.
\ee
Using \eqref{eq:odd_incoming_collinear_invariant} in
\eqref{eq:Jab_odd_in_collinear_result} and restoring the degree \(-1\) in \(v'_a\)
and degree \(-(d-3)\) in \(v'_b\) required by \eqref{eq:Jab_in_def} gives
\bea\label{eq:Jab_odd_in_result}
\mathcal J^{\prime\,\rm class}_{ab}(v'_a,v'_b)
\big|_{\substack{d=\mathrm{odd}\\ \mathrm{in},\,\ln}}
&\simeq&
-\f{i(-1)^\f{d-3}{2}\ln(\omega-i\epsilon)}
{2^{d-2}\pi^\f{d-2}{2}\Gamma\left(\f{d-2}{2}\right)}
\f{(-v_a^{\prime2})^\f{d-4}{2}}
{\left[(v'_a.v'_b)^2-v_a^{\prime2}v_b^{\prime2}\right]^\f{d-3}{2}}.
\eea
Only the denominator in \eqref{eq:Jab_odd_in_result} depends on \(v'_b\).
Differentiating it before imposing
\(v_a^{\prime2}=v_b^{\prime2}=-1\) gives
\bea\label{eq:odd_incoming_derivative}
\left.
\left(v_a^{\prime\rho}v'_{b\sigma}
-v'_a.v'_b\,\delta^\rho_\sigma\right)
\f{\p\mathcal J^{\prime\,\rm class}_{ab}
|_{\substack{d=\mathrm{odd}\\ \mathrm{in},\,\ln}}}
{\p v'_{b\sigma}}
\right|_{v_a^{\prime2}=v_b^{\prime2}=-1}
&\simeq&
\f{i(-1)^\f{d-3}{2}(d-3)\ln(\omega-i\epsilon)}
{2^{d-2}\pi^\f{d-2}{2}\Gamma\left(\f{d-2}{2}\right)}
\non\\
&&\times
\f{v_a^{\prime\rho}v_b^{\prime2}-v'_a.v'_b\,v_b^{\prime\rho}}
{\left[(v'_a.v'_b)^2-1\right]^\f{d-1}{2}}.
\eea
Substituting \eqref{eq:odd_incoming_derivative} into the primed version of
\eqref{eq:IJ} and using the odd-\(d\) identity
\((v'_b.k)^{d-4}=-\omega^{d-4}(-\mathbf n.v'_b)^{d-4}\) gives
\bea\label{eq:odd_I_in_covariant}
I_b^{\prime\rho}(k)\big|_{\substack{d=\mathrm{odd}\\ \mathrm{in},\,\ln}}
&\simeq&
\f{i\,q'_b}{m'_b}
\f{\omega^{d-4}\ln(\omega-i\epsilon)}{(2\pi)^{d-1}}
\sum_{\substack{a=1\\a\ne b}}^M q'_a
(-\mathbf n.v'_b)^{d-4}
\non\\
&&\times
\mathcal F_{\rm in}^{(d)}(v'_a.v'_b)
\left(v_a^{\prime\rho}v_b^{\prime2}
-v'_a.v'_b\,v_b^{\prime\rho}\right),
\eea
with
\bea\label{eq:odd_F_in}
\mathcal F_{\rm in}^{(d)}(z)
&\equiv&
\f{2(-1)^\f{d-3}{2}\pi^\f d2}
{\Gamma\left(\f{d-2}{2}\right)
(z^2-1)^\f{d-1}{2}}.
\eea

\paragraph{Comparison and causal interpretation:}
For normalized velocities, the three scalar integrals in five dimensions reduce to
\bea\label{eq:odd_d5_out_in_comparison}
\mathcal J^{\rm class}_{ab}(v_a,v_b)
\big|_{\substack{d=5\\ \mathrm{out},\,\ln}}
&\simeq&
\f{i\,v_a.v_b}
{4\pi^2\left[(v_a.v_b)^2-1\right]}
\ln(\omega+i\epsilon),
\non\\
\mathcal J^{\prime\,\rm class}_{ab}(v'_a,v'_b)
\big|_{\substack{d=5\\ \mathrm{in},\,\ln}}
&\simeq&
\f{i}
{4\pi^2\left[(v'_a.v'_b)^2-1\right]}
\ln(\omega-i\epsilon),
\non\\
\mathcal J^{{\rm class},\,{\rm mix}}_{ab}(v'_a,v_b)
\big|_{\substack{d=5\\ \mathrm{mix},\,\ln}}
&\simeq&
-\f{i}{4\pi^2\left(1-v'_a.v_b\right)}
\ln(\omega+i\epsilon).
\eea
The mixed line follows from \eqref{eq:Jab_mix_result} using
\(\mathcal H_5(z)=\pi/(z+1)\).  The three lines in
\eqref{eq:odd_d5_out_in_comparison} are the three causal branch orderings allowed by the
retarded field: outgoing-to-outgoing, incoming-to-incoming, and
incoming-to-outgoing.  The reverse mixed ordering vanishes by the contour argument
preceding \eqref{eq:Jab_in_def}.  These coefficients therefore describe the causal
ordering of one long-range interaction, not three independent interactions.

With the \(\ell^0\) contour chosen throughout this appendix to avoid the
order-\((d-3)\) particle pole, their pole content makes the distinction precise in
this contour representation:
\(\mathcal F_{\rm out}^{(d)}\) receives the source-worldline pole and the retarded
propagator poles in \eqref{eq:Jab_collinear_log};
\(\mathcal F_{\rm mix}^{(d)}\) receives only the retarded propagator poles in
\eqref{eq:Jab_mix_contour}; and \(\mathcal F_{\rm in}^{(d)}\) receives only the
source-worldline pole in \eqref{eq:Jab_odd_in_contour}.  The propagator-pole part is the
odd-dimensional tail: support of the retarded Green's function inside the past light
cone permits an incoming source to affect a later outgoing trajectory
\cite{Satishchandran:2017pek,Laddha:2019yaj}.  In even dimensions the retarded Green's function has
only null support, so this mixed term vanishes because the two asymptotic points are
timelike separated, as shown in \eqref{eq:incoming_outgoing_timelike}.  Exchanging the
incoming and outgoing branches while keeping the Green's function retarded is not a time
reversal; a full time reversal would also replace the retarded Green's function by the
advanced one.

Equations \eqref{eq:odd_I_out_covariant}, \eqref{eq:odd_I_mix_covariant} and
\eqref{eq:odd_I_in_covariant} are the odd-dimensional results used in the
odd-dimensional part of section~\ref{S:log_correction}.

\section{Feynman master integral for the quantum soft factor}
\label{app:feynman_integral}

In this appendix we evaluate the scalar integral
\eqref{eq:quantum_feynman_master} over the logarithmic region
\eqref{eq:quantum_logarithmic_region}.  We repeat it for convenience:
\be\label{eq:feynman_master_app}
\mathcal K^{\rm F}_{ab}=
\int_\omega^\Lambda\f{d^d\ell}{(2\pi)^d}
\f{1}{\ell^2-i\epsilon}
\f{1}{(P_a.\ell-i\epsilon)(P_b.\ell+i\epsilon)^{d-3}} .
\ee
As discussed below \eqref{eq:quantum_soft_from_master}, the momentum derivative is
an ambient component derivative.  We therefore use the homogeneous off-shell form
\eqref{eq:feynman_master_off_shell} and impose the mass-shell conditions only after
differentiation.  Since the integrand in \eqref{eq:feynman_master_app} has degree
\(-d\), the radial integral is logarithmic; the contour calculation below determines
its angular coefficient and its dependence on the incoming or outgoing assignments.
The \(d=4\) result was obtained in \cite{Sahoo:2018lxl}; here we evaluate it for
arbitrary \(d\).

\subsection{Pole locations in the two-velocity frame}

According to \eqref{eq:quantum_eta_convention}, the future-directed physical momenta
are \(\eta_aP_a\) and \(\eta_bP_b\).  Choose a frame in which they are collinear, as
in appendix~\ref{app:acceleration_integral}:
\bea\label{eq:feynman_collinear_momenta}
\eta_aP_a&=&\mathfrak m_a\gamma_a(1,\beta_a,\vec0_{d-2}),
\qquad
\eta_bP_b=\mathfrak m_b\gamma_b(1,\beta_b,\vec0_{d-2}),\qquad
\gamma_i=(1-\beta_i^2)^{-1/2}.
\eea
Let \(q\) be the component of \(\vec\ell\) along the common spatial direction,
\(\vec\ell_\perp\) the remaining \(d-2\) components, and
\be\label{eq:feynman_spatial_radius}
L\equiv\sqrt{q^2+\vec\ell_\perp^{\,2}} .
\ee
The three denominators containing \(\ell^0\) factor as
\bea\label{eq:feynman_pole_factorization}
\ell^2-i\epsilon
&=&-(\ell^0-L+i\epsilon)(\ell^0+L-i\epsilon),
\non\\
P_a.\ell-i\epsilon
&=&-\eta_a\mathfrak m_a\gamma_a
\left(\ell^0-\beta_aq+i\eta_a\epsilon\right),
\non\\
P_b.\ell+i\epsilon
&=&-\eta_b\mathfrak m_b\gamma_b
\left(\ell^0-\beta_bq-i\eta_b\epsilon\right).
\eea
Consequently, the positive-energy photon pole lies below the real axis and the
negative-energy photon pole lies above it.  The poles in the complex \(\ell^0\) plane
are
\bea\label{eq:feynman_pole_locations}
\ell^0_a&=&\beta_aq-i\eta_a\epsilon,
\qquad
\ell^0_b=\beta_bq+i\eta_b\epsilon,
\non\\
\ell^0_{\gamma,+}&=&L-i\epsilon,
\qquad
\ell^0_{\gamma,-}=-L+i\epsilon .
\eea
In \eqref{eq:feynman_pole_factorization} and
\eqref{eq:feynman_pole_locations}, positive factors multiplying the infinitesimal
\(\epsilon\) have been absorbed into its definition; only the side of the real axis
on which each pole lies is used below.

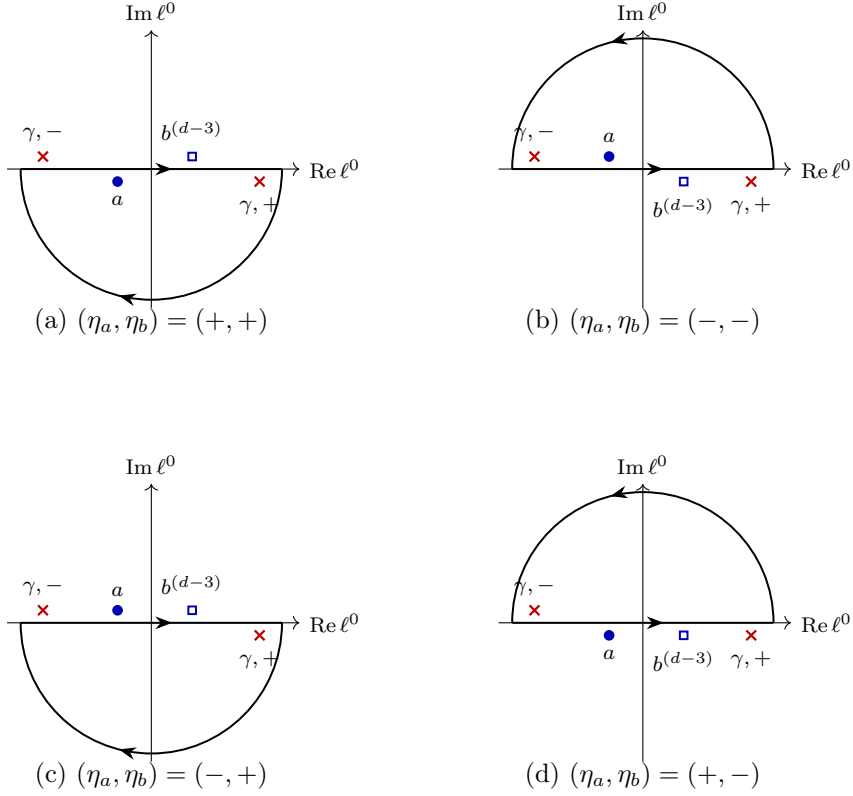
\begin{figure}[t]
\centering
\begin{tikzpicture}[
  x=0.72cm,y=0.72cm,
  contour/.style={line width=0.75pt,postaction={decorate},
    decoration={markings,mark=at position 0.58 with {\arrow{Stealth}}}},
  photon/.style={red!75!black,line width=0.85pt},
  simple/.style={blue!70!black,fill=blue!70!black},
  higher/.style={blue!70!black,line width=0.75pt},
  polelabel/.style={font=\scriptsize}
]
\begin{scope}[xshift=-3.25cm,yshift=3.00cm]
  \draw[->] (-2.65,0)--(2.72,0) node[right,polelabel] {\(\operatorname{Re}\ell^0\)};
  \draw[->] (0,-2.55)--(0,2.55) node[above,polelabel] {\(\operatorname{Im}\ell^0\)};
  \draw[contour] (-2.40,0)--(2.40,0);
  \draw[contour] (2.40,0) arc[start angle=0,end angle=-180,radius=2.40];
  \draw[photon] (-2.08,0.14)--(-1.90,0.32);
  \draw[photon] (-2.08,0.32)--(-1.90,0.14);
  \draw[photon] (1.90,-0.32)--(2.08,-0.14);
  \draw[photon] (1.90,-0.14)--(2.08,-0.32);
  \fill[simple] (-0.62,-0.23) circle (2pt);
  \draw[higher] (0.68,0.16) rectangle +(0.14,0.14);
  \node[polelabel,above] at (-1.99,0.33) {\(\gamma,-\)};
  \node[polelabel,below] at (1.99,-0.33) {\(\gamma,+\)};
  \node[polelabel,below] at (-0.62,-0.31) {\(a\)};
  \node[polelabel,above] at (0.75,0.34) {\(b^{(d-3)}\)};
  \node[font=\small] at (0,-2.80) {(a) \((\eta_a,\eta_b)=(+,+)\)};
\end{scope}
\begin{scope}[xshift=3.25cm,yshift=3.00cm]
  \draw[->] (-2.65,0)--(2.72,0) node[right,polelabel] {\(\operatorname{Re}\ell^0\)};
  \draw[->] (0,-2.55)--(0,2.55) node[above,polelabel] {\(\operatorname{Im}\ell^0\)};
  \draw[contour] (-2.40,0)--(2.40,0);
  \draw[contour] (2.40,0) arc[start angle=0,end angle=180,radius=2.40];
  \draw[photon] (-2.08,0.14)--(-1.90,0.32);
  \draw[photon] (-2.08,0.32)--(-1.90,0.14);
  \draw[photon] (1.90,-0.32)--(2.08,-0.14);
  \draw[photon] (1.90,-0.14)--(2.08,-0.32);
  \fill[simple] (-0.62,0.23) circle (2pt);
  \draw[higher] (0.68,-0.30) rectangle +(0.14,0.14);
  \node[polelabel,above] at (-1.99,0.33) {\(\gamma,-\)};
  \node[polelabel,below] at (1.99,-0.33) {\(\gamma,+\)};
  \node[polelabel,above] at (-0.62,0.31) {\(a\)};
  \node[polelabel,below] at (0.75,-0.34) {\(b^{(d-3)}\)};
  \node[font=\small] at (0,-2.80) {(b) \((\eta_a,\eta_b)=(-,-)\)};
\end{scope}
\begin{scope}[xshift=-3.25cm,yshift=-3.00cm]
  \draw[->] (-2.65,0)--(2.72,0) node[right,polelabel] {\(\operatorname{Re}\ell^0\)};
  \draw[->] (0,-2.55)--(0,2.55) node[above,polelabel] {\(\operatorname{Im}\ell^0\)};
  \draw[contour] (-2.40,0)--(2.40,0);
  \draw[contour] (2.40,0) arc[start angle=0,end angle=-180,radius=2.40];
  \draw[photon] (-2.08,0.14)--(-1.90,0.32);
  \draw[photon] (-2.08,0.32)--(-1.90,0.14);
  \draw[photon] (1.90,-0.32)--(2.08,-0.14);
  \draw[photon] (1.90,-0.14)--(2.08,-0.32);
  \fill[simple] (-0.62,0.23) circle (2pt);
  \draw[higher] (0.68,0.16) rectangle +(0.14,0.14);
  \node[polelabel,above] at (-1.99,0.33) {\(\gamma,-\)};
  \node[polelabel,below] at (1.99,-0.33) {\(\gamma,+\)};
  \node[polelabel,above] at (-0.62,0.31) {\(a\)};
  \node[polelabel,above] at (0.75,0.34) {\(b^{(d-3)}\)};
  \node[font=\small] at (0,-2.80) {(c) \((\eta_a,\eta_b)=(-,+)\)};
\end{scope}
\begin{scope}[xshift=3.25cm,yshift=-3.00cm]
  \draw[->] (-2.65,0)--(2.72,0) node[right,polelabel] {\(\operatorname{Re}\ell^0\)};
  \draw[->] (0,-2.55)--(0,2.55) node[above,polelabel] {\(\operatorname{Im}\ell^0\)};
  \draw[contour] (-2.40,0)--(2.40,0);
  \draw[contour] (2.40,0) arc[start angle=0,end angle=180,radius=2.40];
  \draw[photon] (-2.08,0.14)--(-1.90,0.32);
  \draw[photon] (-2.08,0.32)--(-1.90,0.14);
  \draw[photon] (1.90,-0.32)--(2.08,-0.14);
  \draw[photon] (1.90,-0.14)--(2.08,-0.32);
  \fill[simple] (-0.62,-0.23) circle (2pt);
  \draw[higher] (0.68,-0.30) rectangle +(0.14,0.14);
  \node[polelabel,above] at (-1.99,0.33) {\(\gamma,-\)};
  \node[polelabel,below] at (1.99,-0.33) {\(\gamma,+\)};
  \node[polelabel,below] at (-0.62,-0.31) {\(a\)};
  \node[polelabel,below] at (0.75,-0.34) {\(b^{(d-3)}\)};
  \node[font=\small] at (0,-2.80) {(d) \((\eta_a,\eta_b)=(+,-)\)};
\end{scope}
\end{tikzpicture}
\caption{Feynman \(\ell^0\)-plane contours for the four assignments in
\eqref{eq:feynman_pole_locations}.  Red crosses are the photon poles
\(\ell^0_{\gamma,-}=-L+i\epsilon\) and
\(\ell^0_{\gamma,+}=L-i\epsilon\); the filled circle and square are the
\(a\)- and order-\((d-3)\) \(b\)-particle poles, respectively.  For
\(\eta_b=+1\), panels (a) and (c), the contour closes clockwise below and encloses
\(\ell^0_{\gamma,+}\); it also encloses \(\ell^0_a\) only in panel (a).  For
\(\eta_b=-1\), panels (b) and (d), the contour closes counterclockwise above and
encloses \(\ell^0_{\gamma,-}\); it also encloses \(\ell^0_a\) only in panel (b).
The order-\((d-3)\) pole \(\ell^0_b\) is excluded in every panel.}
\label{fig:feynman_l0_contours}
\end{figure}

We choose the contour so that the order-\((d-3)\) pole at \(\ell^0_b\) is not enclosed.
For \(\eta_b=+1\) we close in the lower half-plane; for \(\eta_b=-1\) we close in the
upper half-plane, as shown in Figure~\ref{fig:feynman_l0_contours}.  Equation
\eqref{eq:feynman_pole_locations} then shows that:
\begin{enumerate}
\item one Feynman photon pole is always enclosed;
\item the simple scalar pole at \(\ell^0_a\) is enclosed only if
\(\eta_a\eta_b=1\).
\end{enumerate}
The enclosed contributions are distinguished by their pole origin.  We write
\be\label{eq:feynman_pole_split}
\mathcal K^{\rm F}_{ab}
=\mathcal K^{\rm matt}_{ab}
+\mathcal K^{\rm ph}_{ab},
\ee
Under the contour choice in Figure~\ref{fig:feynman_l0_contours},
\(\mathcal K^{\rm matt}_{ab}\) is the scalar-pole residue, when present, and
\(\mathcal K^{\rm ph}_{ab}\) is the enclosed Feynman photon-pole residue.

\subsection{Matter-pole contribution and comparison with the classical integral}

For two outgoing lines, \(P_i=\mathfrak m_i v_i\), the Feynman matter-pole residue is
the negative of the first term in \eqref{eq:Jab_collinear_log}, after extracting
\(\mathfrak m_a\mathfrak m_b^{d-3}\).  The Feynman photon-pole residue is similarly
the negative of either one of the two retarded photon-pole terms.  In even dimensions
the latter two terms cancel under \(q\to-q\), leaving only the matter-pole term in
the full outgoing classical result \eqref{eq:Jab_even_result}.  Consequently,
after extracting the mass factor and accounting for the relative contour sign,
\(\mathcal K^{\rm matt}_{ab}\) is fully determined in terms of
\(\mathcal J^{\rm class}_{ab}\).  The same conclusion follows from the incoming
result \eqref{eq:Jab_even_in_result}.  Explicitly,
\bea\label{eq:feynman_matter_relation_even}
\left.
\mathcal K^{\rm matt}_{ab}(P_a,P_b)
\right|_{\substack{d=\mathrm{even}\\ \mathrm{out},\,\ln\\
\eta_a=\eta_b=+1}}
&=&
-\f{1}{\mathfrak m_a\mathfrak m_b^{d-3}}\,
\mathcal J^{\rm class}_{ab}(v_a,v_b)
\big|_{\substack{d=\mathrm{even}\\ \mathrm{out},\,\ln}},
\non\\[2mm]
\left.
\mathcal K^{\rm matt}_{ab}(P_a,P_b)
\right|_{\substack{d=\mathrm{even}\\ \mathrm{in},\,\ln\\
\eta_a=\eta_b=-1}}
&=&
-\f{1}{\mathfrak m_a\mathfrak m_b^{d-3}}\,
\mathcal J^{\prime\,\rm class}_{ab}(v'_a,v'_b)
\big|_{\substack{d=\mathrm{even}\\ \mathrm{in},\,\ln}},
\non\\[2mm]
\left.
\mathcal K^{\rm matt}_{ab}(P_a,P_b)
\right|_{\substack{d=\mathrm{even},\,\ln\\ \eta_a\eta_b=-1}}
&=&0 .
\eea
Here \(v_i=P_i/\mathfrak m_i\) for outgoing lines and
\(v'_i=-P_i/\mathfrak m_i\) for incoming lines.  Substituting
\eqref{eq:Jab_even_result} and \eqref{eq:Jab_even_in_result} into
\eqref{eq:feynman_matter_relation_even} fixes the even-dimensional coefficient.
For odd \(d\), the same residue is fixed independently by
\eqref{eq:Jab_odd_in_contour}--\eqref{eq:Jab_odd_in_result}.  The conversion to
all-outgoing momenta uses \(P_i=\eta_i\mathfrak m_i v_i\), together with
\be\label{eq:odd_gamma_product}
\Gamma\left(\f{4-d}{2}\right)
\Gamma\left(\f{d-2}{2}\right)
=(-1)^{(d-3)/2}\pi ,
\qquad d\ \text{odd}.
\ee
The even- and odd-dimensional calculations therefore give
\bea\label{eq:feynman_matter_result}
\mathcal K^{\rm matt}_{ab}\big|_{\ln}
&\simeq&
-\f{1+\eta_a\eta_b}{2}\,
\mathcal C_d\,
\f{\mathfrak m_a^{d-4}}{\Delta_{ab}^{(d-3)/2}}\ln\omega ,
\eea
where, explicitly,
\be\label{eq:feynman_matter_coefficient_app}
\mathcal C_d\equiv
\begin{cases}
\displaystyle
\f{(-1)^{(d-4)/2}}
{(4\pi)^{(d-2)/2}\Gamma\left(\f{d-2}{2}\right)},
&d\ \text{even},\\[4mm]
\displaystyle
\f{i\,\Gamma\left(\f{4-d}{2}\right)}
{2^{d-2}\pi^{d/2}},
&d\ \text{odd}.
\end{cases}
\ee

For odd \(d\), Figure~\ref{fig:retarded_l0_contours}(a) and
\eqref{eq:Jab_odd_finite_part} show that the outgoing classical integral contains the
matter pole and two retarded photon poles.  The two photon-pole terms in
\eqref{eq:Jab_collinear_log} are exchanged by \(q\to-q\) and therefore give equal
contributions for odd \(d\).  For \(\eta_a=\eta_b=+1\), comparison with
\eqref{eq:feynman_photon_residue_eta_plus} shows that, after extracting
\(\mathfrak m_a\mathfrak m_b^{d-3}\), the Feynman photon-pole residue is the negative
of either retarded photon-pole contribution.  Their sum is consequently represented
by \(2\mathcal K^{\rm ph}_{ab}\), with the overall sign and mass factor displayed
explicitly below.  The matter-pole residues obey the same relative normalization.
The incoming contour in Figure~\ref{fig:retarded_l0_contours}(b) instead contains
only the matter pole, as in \eqref{eq:Jab_odd_in_contour}, so
\eqref{eq:Jab_odd_in_result} is related directly to
\(\mathcal K^{\rm matt}_{ab}\).  Thus
\bea\label{eq:feynman_retarded_odd_relation}
\left.
\left(\mathcal K^{\rm matt}_{ab}+2\mathcal K^{\rm ph}_{ab}\right)
\right|_{\substack{d=\mathrm{odd}\\ \mathrm{out},\,\ln\\
\eta_a=\eta_b=+1}}
&=&
\left.
-\f{\mathcal J^{\rm class}_{ab}(v_a,v_b)}
{\mathfrak m_a\mathfrak m_b^{d-3}}
\right|_{\substack{d=\mathrm{odd}\\ \mathrm{out},\,\ln}}
\non\\
&\simeq&
\f{\ln\omega}{\mathfrak m_a\mathfrak m_b^{d-3}}
\left[
-\f{\mathcal C_d}{(s_{ab}^2-1)^\f{d-3}{2}}
+\f{i\,\mathcal H_d(s_{ab})}
{2^{d-2}\pi^{d/2}\Gamma\left(\f{d-2}{2}\right)}
\right],
\non\\[2mm]
\left.
\mathcal K^{\rm matt}_{ab}
\right|_{\substack{d=\mathrm{odd}\\ \mathrm{in},\,\ln\\
\eta_a=\eta_b=-1}}
&=&
\left.
\f{\mathcal J^{\prime\,\rm class}_{ab}(v'_a,v'_b)}
{\mathfrak m_a\mathfrak m_b^{d-3}}
\right|_{\substack{d=\mathrm{odd}\\ \mathrm{in},\,\ln}}
\non\\
&\simeq&
-\f{\ln\omega}{\mathfrak m_a\mathfrak m_b^{d-3}}\,
\f{\mathcal C_d}{(s_{ab}^2-1)^\f{d-3}{2}} .
\eea
These are the precise connections with the full classical results
\eqref{eq:Jab_odd_result} and \eqref{eq:Jab_odd_in_result}.  For a mixed pair,
\(\mathcal K^{\rm matt}_{ab}=0\) by \eqref{eq:feynman_matter_result}.  The
odd-dimensional mixed tail integral \eqref{eq:Jab_mix_result} is generated entirely
by the two retarded photon poles.  For an incoming source \(a\) and an outgoing
trajectory \(b\), their exact relation to the Feynman photon residue is
\bea\label{eq:feynman_retarded_odd_mixed_relation}
\left.
2\mathcal K^{\rm ph}_{ab}
\right|_{\substack{d=\mathrm{odd}\\\eta_a=-1,\,\eta_b=+1\\\ln}}
&=&
\left.
\f{\mathcal J^{{\rm class},\,{\rm mix}}_{ab}(v'_a,v_b)}
{\mathfrak m_a\mathfrak m_b^{d-3}}
\right|_{\substack{d=\mathrm{odd}\\\mathrm{mix},\,\ln}} .
\eea
Indeed, setting \(d\) odd and
\((\eta_a,\eta_b)=(-1,+1)\) in \eqref{eq:feynman_photon_result} and then comparing
with \eqref{eq:Jab_mix_result} gives
\eqref{eq:feynman_retarded_odd_mixed_relation} directly.  In even dimensions the two
retarded photon-pole contributions cancel under \(q\to-q\), so the mixed retarded
integral vanishes.  For an incoming same-branch pair, conversion to all-outgoing
momenta changes the factors \((P_a.\ell)^{-1}\) and
\((P_b.\ell)^{-(d-3)}\) by the combined sign \((-1)^{d-2}\), while the Feynman
photon propagator has the opposite sign from the retarded propagator at the matter
pole.  Their product gives the factor \((-1)^{d-1}\) responsible for the incoming
signs in \eqref{eq:feynman_matter_relation_even} and
\eqref{eq:feynman_retarded_odd_relation}.  The distinct outgoing, incoming and mixed
odd-dimensional coefficients are illustrated for \(d=5\) in
\eqref{eq:odd_d5_out_in_comparison}.
\par\medskip
The odd-dimensional outgoing and mixed relations above use the lower-half-plane
representation.  Since the integrands fall as \((\ell^0)^{-d}\), the large semicircle
vanishes and the same retarded integrals may instead be evaluated by closing in the
upper half-plane, which contains no retarded photon poles.  This closure encloses the
order-\((d-3)\) pole of particle \(b\) for an outgoing same-branch pair, the simple
pole of particle \(a\) for an incoming same-branch pair, and both matter poles for the
allowed incoming-to-outgoing ordering \eqref{eq:Jab_mix_def}.  For the reverse mixed
ordering it encloses no pole, in agreement with the vanishing result preceding
\eqref{eq:Jab_in_def}.  The upper closure therefore reproduces
\eqref{eq:Jab_even_result}, \eqref{eq:Jab_even_in_result},
\eqref{eq:Jab_odd_result}, \eqref{eq:Jab_odd_in_result} and
\eqref{eq:Jab_mix_result} entirely through matter-particle residues.

In the upper-half-plane closure of the corresponding Feynman integral, the
matter-pole locations are the same and the negative-energy Feynman photon pole is
also enclosed.  After the all-outgoing substitutions
\eqref{eq:quantum_eta_convention}, including the opposite proper-time orientation of
the mixed incoming source displayed before \eqref{eq:odd_I_mix_covariant}, the matter
residues reproduce the classical current of section~\ref{S:log_correction} and hence
the explicit soft factors \eqref{eq:quantum_classical_part} and
\eqref{eq:quantum_classical_part_odd}; the additional Feynman photon residue is the
intrinsically quantum correction.  The
lower-half-plane representation in
\eqref{eq:feynman_retarded_odd_relation} and
\eqref{eq:feynman_retarded_odd_mixed_relation} merely rewrites part of the same
classical coefficients in terms of retarded photon residues.
\par\medskip

\subsection{Photon-pole contribution}

For \(\eta_b=+1\), the enclosed photon pole is \(\ell^0=L-i\epsilon\) and the contour is
clockwise.  Using \eqref{eq:feynman_pole_factorization}, the residue gives
\be\label{eq:feynman_photon_residue_eta_plus}
\left.\mathcal K^{\rm ph}_{ab}\right|_{\substack{\eta_b=+1\\ \ln}}
=
\f{i(-1)^d\eta_a}
{2\mathfrak m_a\gamma_a(\mathfrak m_b\gamma_b)^{d-3}}
\int_\omega^\Lambda
\f{d^{d-2}\vec\ell_\perp\,dq}{(2\pi)^{d-1}}
\f{1}
{L(L-\beta_aq)(L-\beta_bq)^{d-3}} .
\ee
For \(\eta_b=-1\), the enclosed photon pole is \(\ell^0=-L+i\epsilon\) and the contour is
counterclockwise.  Evaluating the remaining denominators at this pole gives
\be\label{eq:feynman_photon_factor_eta_minus}
\left(\ell^0-L+i\epsilon\right)^{-1}
\to
-\f{1}{2L},
\ee
\be\label{eq:feynman_matter_factor_eta_minus}
\left(\ell^0-\beta_aq+i\eta_a\epsilon\right)^{-1}
\to
-\left(L+\beta_aq\right)^{-1},
\ee
\be\label{eq:feynman_b_factor_eta_minus}
\eta_b^{d-3}\left(\ell^0-\beta_bq-i\eta_b\epsilon\right)^{-(d-3)}
\to
\left(L+\beta_bq\right)^{-(d-3)},
\qquad \eta_b=-1,
\ee
so the factor \(\eta_b^{d-3}\) from \eqref{eq:feynman_pole_factorization} is cancelled
by the sign of the order-\((d-3)\) denominator evaluated at \(\ell^0=-L+i\epsilon\).
The second residue is therefore
\be\label{eq:feynman_photon_residue_eta_minus}
\left.\mathcal K^{\rm ph}_{ab}\right|_{\substack{\eta_b=-1\\ \ln}}
=
-\f{i(-1)^d\eta_a}
{2\mathfrak m_a\gamma_a(\mathfrak m_b\gamma_b)^{d-3}}
\int_\omega^\Lambda
\f{d^{d-2}\vec\ell_\perp\,dq}{(2\pi)^{d-1}}
\f{1}
{L(L+\beta_aq)(L+\beta_bq)^{d-3}} .
\ee
After changing \(q\to-q\) in the \(\eta_b=-1\) expression, the spatial integral becomes
identical to the \(\eta_b=+1\) case, and because \(\eta_b=-1\) the two cases combine
into
\bea\label{eq:feynman_photon_residue}
\mathcal K^{\rm ph}_{ab}\big|_{\ln}
&=&
\f{i(-1)^d\eta_a\eta_b}
{2\mathfrak m_a\gamma_a(\mathfrak m_b\gamma_b)^{d-3}}
\int_\omega^\Lambda
\f{d^{d-2}\vec\ell_\perp\,dq}{(2\pi)^{d-1}}
\f{1}
{L(L-\beta_aq)(L-\beta_bq)^{d-3}} .
\eea
Set \(x=q/L\).  The spatial measure is
\be\label{eq:feynman_spatial_measure}
d^{d-2}\vec\ell_\perp\,dq
=
\f{2\pi^\f{d-2}{2}}{\Gamma\left(\f{d-2}{2}\right)}
L^{d-2}dL\,(1-x^2)^\f{d-4}{2}dx .
\ee
Every denominator in \eqref{eq:feynman_photon_residue} is linear in \(L\), so the
power \(L^{d-2}\) from the measure leaves the radial factor \(dL/L\).  Therefore
\bea\label{eq:feynman_photon_angular}
\mathcal K^{\rm ph}_{ab}\big|_{\ln}
&=&
\f{i(-1)^d\eta_a\eta_b}
{2^{d-1}\pi^{d/2}\Gamma\left(\f{d-2}{2}\right)}
\f{1}{\mathfrak m_a\mathfrak m_b^{d-3}}
\non\\
&&\times
\int_\omega^\Lambda\f{dL}{L}\,
\f{1}{\gamma_a\gamma_b^{d-3}}
\int_{-1}^{1}dx\,
\f{(1-x^2)^{(d-4)/2}}
{(1-\beta_ax)(1-\beta_bx)^{d-3}} .
\eea
The coefficient of the logarithmic radial integral is Lorentz invariant.  A boost
changes the homogeneous radial cutoffs only by angular factors and hence changes the
finite part, but not the coefficient of \(\ln\omega\).  The angular integral in
\eqref{eq:feynman_photon_angular} is therefore a function only of
\(s_{ab}=\gamma_a\gamma_b(1-\beta_a\beta_b)\), defined in
\eqref{eq:quantum_pair_invariants}.  In fact, after
\((\beta'_a,\beta_b,z)\to(\beta_a,\beta_b,s_{ab})\), it is exactly the integral
encountered in the mixed retarded calculation
\eqref{eq:odd_mix_H_function}.  In that calculation the two retarded photon poles
give equal contributions for odd \(d\), whereas the Feynman contour used here
encloses one photon pole.  This common angular coefficient is the origin of the
factor of two in \eqref{eq:feynman_retarded_odd_mixed_relation}.

For completeness, in the rest frame of particle \(b\), one has
\(\beta_b=0\), \(s_{ab}=\gamma_a\), and
\(\beta_a^2=1-s_{ab}^{-2}\).  The part odd under \(x\to-x\) integrates to zero, and
the remaining Euler beta-function integral gives
\bea\label{eq:feynman_H_collinear_identity}
&&\f{1}{\gamma_a\gamma_b^{d-3}}
\int_{-1}^{1}dx\,
\f{(1-x^2)^{(d-4)/2}}
{(1-\beta_ax)(1-\beta_bx)^{d-3}}
=
\mathcal H_d(s_{ab}),
\non\\
&&\mathcal H_d(s_{ab})
\equiv
\f{1}{s_{ab}}\,
\mathrm B\left(\f12,\f{d-2}{2}\right)
{}_2F_1\left(
1,\f12;\f{d-1}{2};1-\f{1}{s_{ab}^2}
\right).
\eea
Thus \(\mathcal H_d\) in \eqref{eq:feynman_H_collinear_identity} is the same
function as in appendix~\ref{app:acceleration_integral}, rather than an independent
quantum coefficient.  The hypergeometric representation is valid in both even and odd dimensions;
for odd \(d\), its finite polynomial form is displayed in
\eqref{eq:odd_mix_H_function}.
Finally,
\(\int_\omega^\Lambda dL/L\simeq-\ln\omega\), since the
hard-scale term is discarded under the \(\simeq\) prescription.  Substituting
\eqref{eq:feynman_H_collinear_identity} into
\eqref{eq:feynman_photon_angular} therefore gives
\bea\label{eq:feynman_photon_result}
\mathcal K^{\rm ph}_{ab}\big|_{\ln}
&\simeq&
-\f{i(-1)^d\eta_a\eta_b}
{2^{d-1}\pi^{d/2}\Gamma\left(\f{d-2}{2}\right)}
\f{\mathcal H_d(s_{ab})}
{\mathfrak m_a\mathfrak m_b^{d-3}}\ln\omega .
\eea

\subsection{Total logarithmic contribution}
Adding \eqref{eq:feynman_photon_result} and
\eqref{eq:feynman_matter_result} gives
\bea\label{eq:feynman_master_result_app}
\mathcal K^{\rm F}_{ab}\big|_{\ln}
&\simeq&
\ln\omega\Bigg[
-\f{1+\eta_a\eta_b}{2}\,
\mathcal C_d\,
\f{\mathfrak m_a^{d-4}}{\Delta_{ab}^{(d-3)/2}}
-\f{i(-1)^d\eta_a\eta_b}
{2^{d-1}\pi^{d/2}\Gamma\left(\f{d-2}{2}\right)}
\f{\mathcal H_d(s_{ab})}{\mathfrak m_a\mathfrak m_b^{d-3}}
\Bigg].
\eea
Equation \eqref{eq:feynman_master_result_app} is written after imposing
\(P_i^2=-\mathfrak m_i^2\).  Although the regulated integral
\eqref{eq:feynman_master_app} depends on the radial scales \(\omega\) and \(\Lambda\),
the coefficient of \(\ln\omega\) contains no independent mass parameter.  For
timelike \(P_a\) and \(P_b\), with their time orientations held fixed, Lorentz
invariance and the separate homogeneities of the master integral therefore determine
its continuation away from the mass shell: the on-shell norms are replaced by
\(\sqrt{-P_i^2}\), and \(s_{ab}\) by the corresponding normalized inner product.
Thus the off-shell logarithmic coefficient required for differentiation is
\bea\label{eq:feynman_master_off_shell}
\left.\mathcal K^{\rm F}_{ab}\right|_{\ln,\,\mathrm{off}}
&\simeq&
\ln\omega\Bigg[
-\f{1+\eta_a\eta_b}{2}\,
\mathcal C_d\,
\f{(-P_a^2)^{(d-4)/2}}{\Delta_{ab}^{(d-3)/2}}
\non\\
&&\hspace{1.5cm}
-\f{i(-1)^d\eta_a\eta_b}
{2^{d-1}\pi^{d/2}\Gamma\left(\f{d-2}{2}\right)}
\f{\mathcal H_d(\widehat s_{ab})}
{\sqrt{-P_a^2}\,(-P_b^2)^{(d-3)/2}}
\Bigg].
\eea
Here
\be\label{eq:feynman_off_shell_s_def}
\widehat s_{ab}\equiv
-\eta_a\eta_b\f{P_a.P_b}{\sqrt{(-P_a^2)(-P_b^2)}} .
\ee
The contractions that make the order of operations explicit are
\bea\label{eq:feynman_off_shell_derivative_contractions}
\left(P_a^\rho P_{b\sigma}-P_a.P_b\,\delta^\rho_\sigma\right)
\f{\p(-P_b^2)}{\p P_{b\sigma}}
&=&-2\left(P_a^\rho P_b^2-P_a.P_b\,P_b^\rho\right),
\non\\
\left(P_a^\rho P_{b\sigma}-P_a.P_b\,\delta^\rho_\sigma\right)
\f{\p\widehat s_{ab}}{\p P_{b\sigma}}
&=&-\f{\widehat s_{ab}}{P_b^2}
\left(P_a^\rho P_b^2-P_a.P_b\,P_b^\rho\right),
\non\\
\left(P_a^\rho P_{b\sigma}-P_a.P_b\,\delta^\rho_\sigma\right)
\f{\p\Delta_{ab}}{\p P_{b\sigma}}
&=&-2P_a^2
\left(P_a^\rho P_b^2-P_a.P_b\,P_b^\rho\right).
\eea
The off-shell expression \eqref{eq:feynman_master_off_shell} has degree \(-1\) in
\(P_a\) and degree \(-(d-3)\) in \(P_b\), as required by
\eqref{eq:feynman_master_app}, and reduces to
\eqref{eq:feynman_master_result_app} on shell, where
\(\widehat s_{ab}=s_{ab}\).  It is \eqref{eq:feynman_master_off_shell} that must be
differentiated.  Replacing the norms by the masses before differentiation would
incorrectly treat the explicit \(P_b^2\) dependence and the norm entering
\(\widehat s_{ab}\) as constants; replacing \(\Delta_{ab}\) by its on-shell form
would likewise obscure its \(P_b\) derivative.  The first two identities in
\eqref{eq:feynman_off_shell_derivative_contractions} produce the intermediate
combination \((d-3)\mathcal H_d(s_{ab})+s_{ab}\mathcal H'_d(s_{ab})\), whose
even- and odd-dimensional reductions give the photon-pole terms in
\eqref{eq:quantum_soft_factor_even_d} and
\eqref{eq:quantum_soft_factor_odd_d}.  The third identity produces the common
matter-pole factor
\(\mathfrak m_a^{d-2}\mathfrak m_b^2/\Delta_{ab}^{(d-1)/2}\).

Equation \eqref{eq:feynman_master_result_app} is the complete coefficient of
\(\ln\omega\) in the master integral for distinct, non-collinear massive lines in
both even and odd spacetime dimensions.
The function \(\mathcal H_d\) is given explicitly in
\eqref{eq:feynman_H_collinear_identity}, while the appropriate even- and
odd-dimensional values of \(\mathcal C_d\) are given in
\eqref{eq:feynman_matter_coefficient_app}.  Equation
\eqref{eq:feynman_master_result_app} is the result quoted in
\eqref{eq:quantum_feynman_master_result}.

For \(d=4\), using \(\mathcal C_4=1/(4\pi)\) from
\eqref{eq:feynman_matter_coefficient_app} and setting \(d=4\) in
\eqref{eq:feynman_H_collinear_identity}, one finds
\be\label{eq:feynman_H_four_dimensions}
\mathcal H_4(s)=\f{2\,\operatorname{arccosh}s}{\sqrt{s^2-1}}.
\ee
Equation
\eqref{eq:feynman_master_result_app} reduces to the evaluation of the integral
\(\mathcal I_{ab}\) defined in equation (5.22) of
\cite{Sahoo:2018lxl}.  That reference uses the all-incoming momentum convention,
which is related to the present one by \(p_i=-P_i\); with this replacement,
\(\mathcal K^{\rm F}_{ab}\) is precisely \(\mathcal I_{ab}\).  Equivalently, if the
momentum symbols are held fixed, interchanging the \(i\epsilon\) prescriptions on
\(P_a.\ell\) and \(P_b.\ell\) relates \(\mathcal K^{\rm F}_{ab}\) to
\(\mathcal I_{ba}\).
Using \(\ln\omega^{-1}=-\ln\omega\), the matter- and photon-pole terms then
reproduce equations (5.26) and (5.27) of that reference, respectively.

\end{document}